\documentclass[fleqn,usenatbib]{mnras}
\usepackage{lmodern}
\usepackage{anyfontsize}
\usepackage[T1]{fontenc}

\DeclareRobustCommand{\VAN}[3]{#2}
\let\VANthebibliography\thebibliography
\def\thebibliography{\DeclareRobustCommand{\VAN}[3]{##3}\VANthebibliography}
\usepackage{graphicx}	% Including figure files
\usepackage{newtxtext,newtxmath}
\usepackage{amsmath}
\usepackage{xurl} % URL links
\usepackage{threeparttable} % Allows table footnotes
\usepackage{aas_macros} % Allows parsing of journal codes in .bib files exported from ADS
\usepackage{xcolor}
\usepackage{hyperref}

\title[Validated New Mass-Radius Relations]{A Validated Low-to-Intermediate Mass Planetary Interior Structure Model and New Mass-Radius Relations}

\author[B. N. Skinner, R. E. Pudritz, \& R. Cloutier]{Bennett Neil Skinner$^{1,2}$\thanks{Contact e-mail: \href{mailto:skinnb1@mcmaster.ca}{skinnb1@mcmaster.ca}}, Ralph E. Pudritz$^{1,2}$, and Ryan Cloutier$^{1}$
\\
$^1$Department of Physics \& Astronomy, McMaster University, 1280 Main St W, Hamilton, ON, L8S 4L8, Canada\\
$^2$Origins Institute, McMaster University, 1280 Main St W, Hamilton, ON, L8S 4L8, Canada}

\date{Accepted XXX. Received YYY; in original form ZZZ}
\pubyear{\the\year{}}
\begin{document}
\label{firstpage}
\pagerange{\pageref{firstpage}--\pageref{lastpage}}
\maketitle

\begin{abstract}
  The increasing precision of planetary mass and radius observations is bringing major questions about the structure and formation of planets--such as the nature of the radius valley and origin of super-Mercuries--within reach\textcolor{black}{, demanding} the development of interior structure models with more physics to more accurately determine \textcolor{black}{planetary radii for a} given composition. Here, we present a new model that includes state-of-the-art equations of state following the latest experimental and computational results, a physically-motivated mineralogy allowing multiple species to coexist within planetary layers, a \textcolor{black}{non-adiabatic temperature profile}, melting, and other features. This model replicates Earth\textcolor{black}{'s radius and moment of inertia coefficient} to within $0.2\%$\textcolor{black}{, }Mars and the Moon's to within $0.5\%$, and Mercury\textcolor{black}{, Venus,} and Europa's to within $1\%$ or 3$\sigma$. We use this model to calculate mass-radius relationships for H/He-enveloped, water-rich, Earth-like, and iron-rich bodies with masses between $0.01$--$100\, M_\oplus$. \textcolor{black}{We calculate mass-radius tables and fit piece-wise power-laws to them for ${<}8M_\oplus$ planets, finding that the exponent in $M=bR^a$ increases with mass and core mass fraction.} \textcolor{black}{We find radii generally smaller than in literature mass-radius relations at low instellations and larger at high instellations, with our improvement on the literature comparable to observational uncertainties. State-of-the-art interior structure models are thus required to interpret observational data.} Our mass-radius curves \textcolor{black}{comprising 32,97\textcolor{black}{5} model planets are publicly available.}
\end{abstract}

\section{Introduction} \label{sec:real_intro}
\subsection{Interior Structure Models}\label{sec:models}
The inference of the composition of a planet from its observed mass and radius requires a planetary interior structure model, which solves for the internal density and temperature structure of a planet. This inference is often performed using forward-modeled mass-radius curves that give the expected radii of planets for a given mass and composition.

Many planetary interior structure models exist in the literature \citep[e.g.][]{FortneyMarley2007,SeagerKuchner2007,SotinGrasset2007,ValenciaSasselov2007,ZengSasselov2013,LopezFortney2014,DornKhan2015,DornVenturini2017,vandenBergYuen2019,ZengJacobsen2019,BoujibarDriscoll2020,AcunaDeleuil2021,UnterbornDesch2023,PlotnykovValencia2024,RiceHuang2025,AguichineBatalha2025,HaldemannDorn2024}. As is necessary for any model they include \textcolor{black}{several} simplifying assumptions, a number of which are increasingly challenged by better experimental, computational, and observational findings.

As an example, planetary interiors are often assumed to consist of a number of distinct differentiated layers, each composed of only a single species. This ignores the fact that Earth's mantle and core are both composed of multiple species: iron in Earth's mantle makes it denser and light elements in Earth's core makes it less dense \citep{Mcdonough2014,PalmeOneill2014}. Planet interiors are often assumed to be purely adiabatic, but this neglects temperature jumps that occur at the top of the mantle and core due to heat trapped from planet formation \citep{Stixrude2014}, making \textcolor{black}{model} planets \textcolor{black}{denser than reality}. High-pressure phases of mantle materials predicted to occur beyond the pressures of Earth's mantle are neglected, making \textcolor{black}{model} planets less dense \citep{UmemotoWentzcovitch2017}. Solid and liquid cores are often not included simultaneously, resulting in density overestimates if cores are assumed to be purely solid and underestimates if they are assumed to be purely liquid \citep{RiceHuang2025}. Planetary radii are taken at the outer boundary pressure of \textcolor{black}{a} model rather than at the $\tau=\frac{2}{3}$ optical depth of a grazing ray of light that exoplanet transit surveys probe, systematically leading to radius overestimates \citep{HaldemannDorn2024}. Finally, older equations of state are still in use that do not take into account recent experimental and computational advances or use poorly-extrapolating formulations \citep{HakimRivoldini2018}. These simplifications and others induce systematic changes in estimated radii, but these are not all in the same direction. It is thus unclear a priori whether the literature overestimates or underestimates planetary radii. As these effects and others will strongly influence our understanding of how planets are formed, we present here a more physical model of planetary interiors that relaxes or removes these assumptions.

The main task accomplished in this paper is to develop new mass-radius relations for exoplanets based on a greatly improved interior structure model. These improvements are developed along four main axes: new equations of state \citep[e.g.][]{HakimRivoldini2018,HaldemannAlibert2020,HowardGuillot2023}, a more comprehensive mineralogy \citep[e.g.][]{StixrudeLithgow-Bertelloni2024}, the inclusion of additional phase changes \citep[e.g.][]{DorogokupetsDymshits2017,UmemotoWentzcovitch2017}, and a more physical thermal structure \citep[e.g.][]{ParmentierGuillot2014,Stixrude2014,ParmentierGuillot2015}.

We have benchmarked this model against solar system objects for which there \textcolor{black}{are} a variety of measured constraints on their interior structures. Of prime importance is comparing a forward model of Earth's interior to the \textcolor{black}{seismically-derived} one-dimensional Reference Earth Model (REM1D) \citep{MoulikEkstrom2025a,MoulikEkstrom2025b}. \textcolor{black}{Our model Earth's radius and moment of inertia coefficient agree with REM1D's to} within $0.2\%$, with a strikingly close correspondence to REM1D throughout the entire interior. We perform similar benchmarking for Mars, the Moon, Venus, Mercury, and Europa, obtaining radii and moment of inertia coefficients with typical errors compared to reality of less than 0.5\% while replicating their observed mantle/core boundaries.

\subsection{Exoplanet Applications Requiring Updated Interior Structure Models}\label{sec:outstanding_questions}
One of the most \textcolor{black}{notable} planet demographics findings of the past decade is the discovery of the bimodality of the radii of planets in close-in orbits referred to as the Radius Valley \citep{FultonPetigura2017,FultonPetigura2018,CloutierMenou2020}. The lower-radius super-Earth peak is composed of planets with densities consistent with Earth-like compositions, while the higher-radius sub-Neptune peak is composed of planets with lower bulk densities, implying a significant inventory of a lower-density substance, either H/He or water \citep{DressingCharbonneau2015,Rogers2015}. 

A relatively small mass fraction of H/He or a relatively large mass fraction of water would produce similar bulk densities, so observations of planetary masses and radii alone are insufficient to distinguish between H/He-enveloped and water-rich sub-Neptunes on an individual level \citep{Miller-RicciSeager2009,DressingCharbonneau2015,Rogers2015}. A population-level study, therefore, requires extreme precision in planetary radii as the degeneracy in H/He-enveloped and water-rich planetary radii means that even a $\%$-level inaccuracy in a planetary interior structure model can make the difference between a planet appearing H/He-enveloped or water-rich. We show here that effects often ignored in the literature--such as the non-ideal mixing of hydrogen and helium, the presence of steam atmospheres, and the discrepancy between the radius of a planet at a specific pressure and the radius of a planet as measured \textcolor{black}{during} transit--produce errors of this magnitude.

The consequences for our interpretation of observations are striking. If the lower-density substance is H/He, the radius valley separates planets that have retained a primordial H/He envelope from those that have lost a primordial H/He envelope via mass loss. This could be driven by stellar X-ray and UV radiation, residual heat in the planetary core, or some combination of the two \citep{OwenWu2013,LopezFortney2014,LopezRice2018,GuptaSchlichting2019,OwenSchlichting2024}. If the lower-density substance is water, the radius valley separates planets that accreted nearly all their material within the iceline and thus never acquired much water from those that accreted a significant mass fraction of water outside the iceline and subsequently migrated inwards \citep{VenturiniGuilera2020b,IzidoroSchlichting2022,ChakrabartyMulders2024,VenturiniRonco2024}. \textcolor{black}{The water in these water worlds is expected to be steam at sufficiently high instellations, with \citet{BurnMordasini2024b} finding that the radius valley cannot be replicated without taking the phase transition from condensed to gas-phase water into account.}

Around sun-like stars, the positive slope of the location of the radius valley with instellation \citep{VanEylenAgentoft2018,MartinezCunha2019} and its weakening in very young systems have been used as observational evidence for the prevalence of mass loss of H/He envelopes \citep{ChristiansenZink2023,VachZhou2024,FernandesBergsten2025}. However, the radius-installation slope of the radius valley \textcolor{black}{flattens or} reverses around \textcolor{black}{early} M dwarf stars, potentially indicating the prevalence of inward-migrating water worlds \citep{GaidosAli2024,HoRogers2024,CloutierMenou2020}. \textcolor{black}{The fading of the radius valley around mid-to-late M dwarfs \citep{GillisCloutier2026} is another hint at the prevalence of steam-rich water worlds, as it potentially reflects the disappearance of steam atmospheres as lower stellar temperatures lead to lower planetary temperatures. Testing of the water world hypothesis for the radius valley thus requires the inclusion of steam, as in this model.}

Another quandary concerns Super-Mercuries, planets with high bulk densities and thus core mass fractions implying formation either via giant impacts that stripped a significant fraction of their mantles \citep[e.g.][]{ScoraValencia2020,ScoraValencia2022} or by condensation at high temperatures \citep[e.g.][]{DornHarrison2019} (or some combination thereof). One super-Mercury--GJ 367 b--has a well-measured bulk density near that of a pure iron planet of its size \citep{GoffoGandolfi2023}.  Even relatively minor changes in model radii could render this body and others like it (many of which have been provisionally detected but do not yet have well-constrained masses and radii) denser than worlds made of pure iron and therefore unphysical, providing a lower radius limit to compare interior structure models to \citep{DaiMasuda2019,AdamsJackson2021,GoffoGandolfi2023}\footnote{This is not a\textcolor{black}{n absolute} limit as observations of radius are not made to infinite precision so it is possible that a physical planet would have a measured core mass fraction above 100\% due to uncertainty in mass and radius.}. Our model--unlike some in the literature--respects this lower radius limit.

This paper is structured as follows: in section \ref{sec:model} we describe our new interior structure model in physical detail; the general reader interested only in the main results can read subsections \ref{sec:methods_overview} and \ref{sec:effects_summary} for an overview of the section. In section \ref{sec:validation}, we compare our model to the observed interior structures of solar system objects for validation. Section \ref{sec:mr} presents our new isocomposition mass-radius curves and in section \ref{sec:discussion} we compare our model to empirical mass-radius relations, compare the assumptions of our model to those in the literature, and discuss the implications of missing physics. Finally, section \ref{sec:conclusion} summarizes our results and looks towards future work.

\section{A New Interior Structure Model} 
\label{sec:model}

We begin this section with an overview of the basic ingredients used in constructing our models. We then discuss the forms of equations of state used in our model before addressing the specific equations of state used as one proceeds from the outermost layer of a planet to its core. Finally, we discuss additional physics included in our model before providing a brief summary.  A list of all parameters input into our interior structure model is given in Table \ref{tab:free_params}.

\begin{table*}
    \centering
    \begin{tabular}{ccc}
        Variable&Parameter&Source\\\hline
        $m_\mathrm{p}$ & Planet mass & Independent Variable\\
        $F_\mathrm{p}$ & Planet instellation & Independent Variable\\
        $w_\mathrm{H/He}$ & Envelope mass fraction & Independent Variable\\
        $w_\mathrm{H\textsubscript{2}O}$ & Water mass fraction & Independent Variable\\
        $w_\mathrm{mantle}$ & Mantle mass fraction & Independent Variable\\
        $w_\mathrm{core}$ & Core mass fraction & Independent Variable\\
        $x^{s}_\mathrm{Fe}$ & Fe \textcolor{black}{molar fraction} in solid core & Independent Variable ($x^{s}_\mathrm{Fe,\oplus}=0.885$)\\
        $x^{s}_\mathrm{S}$ & S \textcolor{black}{molar fraction} in solid core & Independent Variable ($x^{s}_\mathrm{S,\oplus}=0.115$) \\
        \textcolor{black}{$x^{s}_\mathrm{O}$} & \textcolor{black}{O \textcolor{black}{molar fraction} in solid core} & \textcolor{black}{Set to zero ($x^{s}_\mathrm{O,\oplus}\leq0.01$, 0 assumed)}\\
        $x^{l}_\mathrm{Fe}$ & Fe \textcolor{black}{molar fraction} in liquid core & Independent Variable ($x^{l}_\mathrm{Fe,\oplus}=0.83$)\\
        $x^{l}_\mathrm{S}$ & S \textcolor{black}{molar fraction} in liquid core & Independent Variable ($x^{l}_\mathrm{S,\oplus}=0.09$)\\
        $x^{l}_\mathrm{O}$ & FeO \textcolor{black}{molar fraction} in liquid core & Independent Variable ($x^{l}_\mathrm{O,\oplus}=0.08$)\\
        $x_\mathrm{MgO}$ & MgO endmember \textcolor{black}{molar fraction} in mantle & Independent Variable ($x_\mathrm{MgO,\oplus}=0.512$)\\$x_\mathrm{SiO\textsubscript{2}}$ & SiO\textsubscript{2} endmember \textcolor{black}{molar fraction} in mantle & Independent Variable ($x_\mathrm{SiO_2,\oplus}=0.425$)\\
        $x_\mathrm{FeO}$ & FeO endmember \textcolor{black}{molar fraction} in mantle & Independent Variable ($x_\mathrm{FeO,\oplus}=0.063$)\\
        $w_\mathrm{H}$ & H weight fraction in envelope & Set as solar (0.725)\\
        $w_\mathrm{He}$ & He weight fraction in envelope & Set as solar (0.275)\\
        $t$ & Age of planet & Set as Earth-like (4.5 Gyr)\\
        $A_B$ & Bond Albedo of planet & Set as Earth-like (0.3)\\
        $T_*$ & Stellar Effective Temperature & Set as sun-like (6000 K)\\
        $Z_\mathrm{Atm}$ & Atmospheric metallicity & Set following observations (50$Z_\odot$)\\
        $P_\mathrm{rot}$ & Rotation period of planet & Effect set to zero ($P_\mathrm{rot}=\infty$)
    \end{tabular}
    \caption{All parameters input into the interior structure model. Note the restrictions $w_\mathrm{H/He}+w_\mathrm{H_2O}+w_\mathrm{mantle}+w_\mathrm{core}=1$, \textcolor{black}{$w_\mathrm{H}+w_\mathrm{He}=1$}, $x_\mathrm{MgO}+x_\mathrm{SiO\textsubscript{2}}+x_\mathrm{FeO}=1$, \textcolor{black}{$x^{s}_\mathrm{Fe}+x^{s}_\mathrm{S}+x^{s}_\mathrm{O}=1$, and $x^{l}_\mathrm{Fe}+x^{l}_\mathrm{S}+x^{l}_\mathrm{O}=1$}, reducing the number of free parameters by 5 from the total number of variables. In total, our model has \textcolor{black}{17} free parameters, 5 of which (planet mass, planet instellation, and relative masses of the planetary layers) we vary in our generation of mass-radius curves, \textcolor{black}{6} of which (compositional parameters in the mantle and core) are varied but generally set to be Earth-like, \textcolor{black}{3} of which (H/He mass fractions, bond albedo, and stellar effective temperature) we hold to solar or Earth values, 1 of which (atmospheric metallicity) we set following exoplanet observations, \textcolor{black}{1 of which (age) we set to be Earth-like throughout most of the publication but vary for one test case,} and 1 of which (rotation) \textcolor{black}{is set following observations in our validation sample but disabled for most mass-radius curves.} We emphasize that all parameters can be freely varied within our framework.}
    \label{tab:free_params}
\end{table*}

\subsection{Overview and Key Assumptions}
\label{sec:methods_overview}
We employ the standard assumption that planets are objects in hydrostatic equilibrium composed of multiple layers each with a homogeneous elemental composition, although not necessarily homogeneous in chemical composition or phase. These layers--from the outermost layers of a planet to its core--are 1) a H/He envelope, 2)  a pure water layer, 3) a mantle composed of FeO, SiO\textsubscript{2}, MgO and minerals formed from the combination of those endmembers (most prominently MgSiO\textsubscript{3}), 4) a liquid core composed of Fe, S, and O, and 5) a solid core composed of Fe\textcolor{black}{, S, and O}. The envelope, water, mantle, and total core mass fractions are set a priori while the size of the liquid and solid core are determined by solving the standard equations of planetary structure below. \textcolor{black}{The water within the water layer can be of any state. The gaseous fraction of the H/He envelope and water layer compose the planetary atmosphere, which is included in our model with no explicit \textcolor{black}{differentiation} between the atmosphere and layers below the atmosphere \textcolor{black}{besides} the EOS in use and thermal structure profile (see section \ref{sec:thermal}).}

We calculate a planet's structure following the three ordinary differential equations: mass conservation
\begin{equation}
    \frac{\partial r}{\partial m}=\frac{1}{4\pi r^2 \rho(P,T)}\mathrm{,}
    \label{eq:mass_cons}
\end{equation}
hydrostatic equilibrium
\begin{equation}
    \frac{\partial P}{\partial m}=-\frac{Gm}{4\pi r^2}\mathrm{,}
    \label{eq:hydro_equil}
\end{equation}
and thermal transport
\begin{equation}
    \frac{\partial T}{\partial m}=\frac{\partial P}{\partial m}\frac{T}{P}\nabla(P,T)\mathrm{,}
    \label{eq:thermal_struct}
\end{equation}
where $r(m)$ is the distance from the planetary centre at which the variables are evaluated, $m$ is the enclosed mass, $\rho(P,T)$ is the density, $P(m)$ is pressure, $T(m)$ is temperature, $G$ is the gravitational constant, and $\nabla\equiv\frac{P}{T}\frac{\partial T}{\partial P}$ is the dimensionless temperature gradient \citep{Prialnik2009}.

Our boundary conditions are the pressure, temperature, and radius of the planet at its centre and outermost layer (set as $P_\mathrm{out}=100$ Pa throughout this work, not to be confused with the condensed surface of the planet \textcolor{black}{or its transit radius}). At the core of the planet, $r(0)=0$ is known, while at the surface of the planet, $P(r_\mathrm{out})=P_\mathrm{out}$ and $T(r_\mathrm{out})=\textcolor{black}{T_\mathrm{out}}$ are known, where $P_\mathrm{out}$ is a small pressure representing the edge of the planet and \textcolor{black}{$T_\mathrm{out}$ is a function of the effective surface temperature of the planet (see section \ref{sec:thermal}).} As not all three dependent variables are known at the same location, we guess the unknown variables at the surface and core, then integrate the ODEs from the surface inward and the core outward simultaneously to some $m_\mathrm{fit}$, a location within the planet at which the inwards and outwards integration meet that carries no physical meaning. If the values of $r(m_\mathrm{fit})$, $P(m_\mathrm{fit})$, and $T(m_\mathrm{fit})$ are similar for the surface\textcolor{black}{--}inward and core\textcolor{black}{--}outward integrations, then our guesses were correct. Guesses are iterated via the Newton-Raphson method \citep{PressVetterling1996}. We seed initial guesses by the \citet{ZengJacobsen2019} parametric fit to planetary radii. We integrate using the Cash-Karp Runge-Kutta method, a fifth-order Runge-Kutta scheme in which step sizes are adaptively varied to keep error--estimated by comparing fifth- and fourth-order solutions--below a chosen threshold \citep{PressVetterling1996}. Further numerical details are provided in appendix \ref{sec:num_details}.

We make the assumption (with the exception of the low-pressure mantle region covered by HeFESTo, see section \ref{sec:mantle}) that all species are solely in one phase at a given pressure and temperature, even though in reality some phases may coexist. We justify this assumption by the already imprecise nature of many of the phase transitions of interest. If lower-\textcolor{black}{pressure} phases coexist with higher-pressure phases beyond where we expect a phase transition to occur, our model tends to overestimate densities, conversely, if higher-\textcolor{black}{pressure} phases coexist with lower-\textcolor{black}{pressure} phases below where we expect a phase transition to occur, our model tends to underestimate densities.

Our assumption of an Earth-like inventory of mantle and core elements is justified by the relatively consistent midplane temperatures in the protoplanetary discs of Sun-like stars. This results in Mg, Si, and Fe almost universally condensing at the radii where planets are formed, causing planets to inherit abundance ratios from their host stars, which themselves exhibit relatively minor abundance deviations \citep{BedellBean2018,UnterbornSchaefer2020,HinkelTimmes2014}. Indeed, measurements of the chemical composition of polluted white dwarf atmospheres assumed to come from the infall of exoplanetary material indicate Earth-like compositions \citep[e.g.][]{DoyleYoung2019}. We emphasize that our model only assumes an Earth-like inventory of elements and that the relative abundances of MgO, SiO\textsubscript{2}, and FeO in the mantle and Fe, FeS, and FeO in the core can be freely varied.

Measurements of Earth's bulk mantle composition from mid-ocean ridge basalts indicate that the end members that we include compose $\gtrsim90\%$ of Earth's mantle, further justifying our simplification of the mantle to those components \citep{WorkmanHart2005}. Al\textsubscript{2}O\textsubscript{3} and CaO comprise the largest fraction of missing materials from our model in the case of Earth, however our (see section \ref{sec:earth}) and other interior structure models represent Earth well without them \citep{WorkmanHart2005,UnterbornSchaefer2020}. Carbon may constitute a more significant component in exoplanetary systems than \textcolor{black}{in} our own and is the subject of future work \citep{BerginKempton2023,LiBergin2026,LinSeager2025}.

Our core inventory is slightly less complete, with greater than $85\%$ of the weight of Earth's core \textcolor{black}{likely} being Fe, O, or S, with Fe being by far the most abundant element \citep{Mcdonough2014}. Ni and/or Si likely represent much of the remainder of the core but we justify their exclusion by (1) our good match to observations within the Solar System discussed in section \ref{sec:validation}, (2) Mars' core being much more S-rich than Earth's indicating a diversity in core S compositions \citep{GendreBadro2022,LeMaistreRivoldini2023}, and (3) the possibility of FeS enhancement in planetary cores as mantle materials oxidize the core, removing bare Fe from the core while oxidation-protected FeS is relatively enhanced in the remaining core \citep[e.g][]{JohansenRonnet2023}. The inclusion of lower-density materials within the core reduces its bulk density and thus previous models assuming pure Fe cores underpredict core radii, especially when those cores are assumed to be purely solid rather than at least partially liquid, as is the case for Earth \citep{UnterbornSchaefer2020}.

In addition to these compositional layers, the planet is divided into an irradiated outermost atmospheric layer (atmosphere) that receives all incident stellar flux and an interior layer \textcolor{black}{does not directly receive any} stellar flux. The outermost layer has a temperature profile following the radiative transfer results of \citet{ParmentierGuillot2014,ParmentierGuillot2015}, while the interior layer has a temperature profile that is radiative or adiabatic. 

In section \ref{sec:eos} we provide background on the \textcolor{black}{formulations of} equations of state (EOS) used in our model. In sections \ref{sec:envelope}-\ref{sec:core}, we discuss the mineralogy, phases, and EOS used in our model, from the H/He envelope down to the core. In sections \ref{sec:thermal}-\ref{sec:rotation}, we discuss further details of \textcolor{black}{our} model.

\subsection{Equations of State} \label{sec:eos}
Equations Of State (EOS) are thermodynamic equations that relate the properties of a substance, with the relevant properties in our formalism being pressure $P$, temperature $T$, density $\rho$, and entropy $S$ (necessary for calculating $\nabla_\mathrm{ad}$, see section \ref{sec:combination}), i.e EOS give $\rho(P,T)$ and $S(P,T)$. They are thus necessary to calculate Equations \ref{eq:mass_cons} and \ref{eq:thermal_struct}. The entirety of any planetary interior structure model thus hinges on EOS, which we now give a brief overview of before discussing the particular EOS \textcolor{black}{that} we use in sections \ref{sec:envelope}-\ref{sec:core}.

In the majority of EOS used, $V$ instead of $\rho$ is the variable related to the size of the substance, giving the volume of a mole of that material. $V$ can be converted to $\rho$ via $V=\frac{M}{\rho}$ where $M$ is the molar mass of the material. EOS can either be provided as tables in a publication or fit to one of several equations. All fit EOS \textcolor{black}{used in this study} have Helmholtz free energies composed of an isothermal term (i.e $\rho(P)$, $S(P)$) and up to two temperature-dependent \textcolor{black}{(i.e $\rho(P,T)$, $S(P,T)$)} terms, with the first related to harmonic oscillations in the material and the second related to anharmonic oscillations and electrons, whose forms we discuss in sections \ref{sec:isothermal_background} \ref{sec:thermal_background}, and \ref{sec:anharmonic_electronic}, respectively. The thermal and anharmonic/electronic pressures are generally (but not universally) positive, so neglecting them leads to lower pressures at similar densities than would otherwise be calculated. As we calculate densities from pressures, this results in a bias towards higher densities. Intuitively, ignoring these terms ignores thermal expansion, resulting in higher densities (for example, our reference Earth's radius decreases by 0.04\% if the mantle EOS is always calculated using $T=300$ K).

In this subsection, we provide the functional forms of all EOS used and a brief description of the physics used to derive them.  The purpose of this overview is to give the reader an intuition for where our EOS arise and over what pressure ranges certain EOS are appropriate, justifying the forms \textcolor{black}{that} we employ in our model as discussed in sections \ref{sec:envelope}-\ref{sec:core}. We progress from EOS suited to low to \textcolor{black}{EOS suited to} high pressures. The reader is directed towards the references provided if interested in complete derivations.

When discussing phase transitions, in this publication \textcolor{black}{we use} the subscript notation $X^{Y}_{Z_1/Z_2}$, where $X$ is $P$ or $T$, $Y$ is the material and $Z_1$ and $Z_2$ are the higher and lower-pressure phases (even if $X$ is $T$), respectively.

\subsubsection{Isothermal EOS} \label{sec:isothermal_background}
\textcolor{black}{Isothermal EOS come in several analytical formulations, with differences between materials expressed as constants within these formulations. Analytical EOS formulations ultimately rely on results from simplified toy models and their validity is solely justified by their fit to the available data.}

The third-order finite strain Birch-Murnagham (BM3) EOS formulation given by Eq.~\ref{eq:BM3} is derived by expanding a material's Helmholtz free energy in terms of the change in surface area of a compressed cube \citep{Murnaghan1937,Murnaghan1944,Birch1947,KatsuraTange2019}\textcolor{black}{:}
\begin{equation}
    P = \frac{3}{2}K_0\left[ \left( \frac{\rho}{\rho_0} \right)^\frac{7}{3}-\left( \frac{\rho}{\rho_0} \right)^\frac{5}{3} \right]\left[ 1+\frac{3}{4}(K_0^\prime-4)\left( \left( \frac{\rho}{\rho_0}\right )^\frac{2}{3}-1\right) \right] \mathrm{,}
    \label{eq:BM3}
\end{equation}
where $\rho_0$, $K_0$, and $K_0^\prime$ are fitting constants representing the density in fiducial conditions, the isothermal bulk modulus ($K_T$) at reference conditions, and its pressure derivative, respectively.

The Rose-Vinet\footnote{Synonymous with Vinet or Vinet-Rydberg.} formulation in Eq.~\ref{eq:rose_vinet} is derived by taking the volume derivative of the Helmholtz free energy of a Rydberg potential, an approximation of atomic binding energies \citep{Rydberg1932,RoseSmith1983,VinetFerrante1987,Holzapfel2002}:
\begin{equation}
P = 3K_0(\frac{\rho}{\rho_0})^\frac{2}{3}(1-(\frac{\rho}{\rho_0})^{-\frac{1}{3}})e^{\frac{3}{2}(K_0^\prime-1)(1-(\frac{\rho}{\rho_0})^{-\frac{1}{3}})}\mathrm{.}
\label{eq:rose_vinet}
\end{equation}

The Holzapfel formulation of Eq.~\ref{eq:holzapfel} is derived by modifying the Rose-Vinet equation such that it matches the high-compression free-electron Fermi gas limit (i.e $K^\prime$ approaches $\frac{5}{3}$ as pressure approaches $\infty$) while remaining integrable in closed form \citep{Holzapfel1996,Holzapfel1998,Holzapfel2002,HakimRivoldini2018}\textcolor{black}{:}
\begin{multline}
P = 3K_0(\frac{\rho}{\rho_0})^\frac{5}{3}(1-(\frac{\rho}{\rho_0})^{-\frac{1}{3}})e^{c_0(1-(\frac{\rho}{\rho_0})^{-\frac{1}{3}})}\\
\cdot(1+c_2(\frac{\rho}{\rho_0})^{-\frac{1}{3}}(1-(\frac{\rho}{\rho_0})^{-\frac{1}{3}}))\mathrm{,}
\label{eq:holzapfel}
\end{multline}
where $c_0$ and $c_2$ are fitting constants. All fitting constants are derived via simulation or experimentation.

The Keane formulation of Eq.~\ref{eq:keane} is a consequence of Keane's rule, which states that the pressure derivatives of the bulk modulus at zero pressure, $K_0^\prime$, and infinite pressure, $K_\infty^\prime$, must satisfy $\frac{K_0^\prime}{2}<K_\infty^\prime<K_0^\prime-1$ \citep{Keane1954,StaceyDavis2004,SakaiDekura2016}:
\begin{equation}
    P = K_0(\frac{K_0^\prime}{{K_\infty^\prime}^2}((\frac{\rho}{\rho_0})^{K_\infty^\prime}-1)-(\frac{K_0^\prime}{K_\infty^\prime}-1)\ln{(\frac{\rho}{\rho_0})})\mathrm{.}
    \label{eq:keane}
\end{equation}
The Keane formulation is only used in one EOS in our sample, where it\textcolor{black}{s fit} lies between the BM3 and Rose-Vinet equations and is the best fit to the data up to 300 GPa among those three \citep{SakaiDekura2016}.

\textcolor{black}{The Holzapfel formulation is the only one among these four EOS that recovers the infinite-pressure theoretical limit of a free-electron Fermi gas (as it was designed to do). This is reflected in the Holzapfel EOS achieving better fits to available data at pressures above ${\sim}$1 TPa than other EOS} \citep{HamaSuito1996,CohenGulseren2000,SeagerKuchner2007,HakimRivoldini2018}. BM3 or Rose-Vinet EOS are thus not applicable to the high-pressure interiors of super-Earths or sub-Neptunes, motivating the necessity of EOS updates. \textcolor{black}{Additionally, if $K_0^\prime<4$, as is the case for some materials (e.g high spin B1 FeO \citet{FischerCampbell2011}), equation \ref{eq:BM3} undergoes an unphysical turnover at extremely high pressures wherein density decreases with increasing pressure.} We thus prefer the Rose-Vinet EOS over BM3 when applied to high pressures \textcolor{black}{and therefore preferentially use literature sources that provide Rose-Viet EOS}. \textcolor{black}{We do not refit any data with new EOS in this study.} Although the Holzapfel EOS has the best high\textcolor{black}{-}pressure limit, we caution that the Rose-Vinet \textcolor{black}{and BM3} EOS have been found to match experiments better at lower pressures, motivating an approach as considered here where the Holzapfel EOS is applied only to the highest pressure phases \citep{HamaSuito1996,CohenGulseren2000}.

We note that all of these EOS are of the form $P(\rho)$ while we desire $\rho(P)$. We generated tables of these EOS in the desired form using the Newton-Raphson method of iteration and interpolated over these tables while running our interior structure model to save the computational expense of repeated iteration.

\subsubsection{Thermal Contributions to the EOS} \label{sec:thermal_background}
The temperature-dependent portion of the EOS is itself split into two parts, a quasi-harmonic contribution and an anharmonic and/or electronic contribution.

All quasi-harmonic contributions to the EOS considered here arise from the Mie–Grüneisen framework of Eq.~\ref{eq:mie-gruneisen},
\begin{equation}
    \Delta P_{th}=\gamma_{th}\frac{\Delta E_{th}}{V}\mathrm{,}
    \label{eq:mie-gruneisen}
\end{equation}
where $\Delta$ is the change between a temperature $T$ and some reference temperature $T_0$ and $\gamma_{th}$ is the thermal Grüneisen parameter \citep{Mie1903,Gruneisen1912,Goodstein1985,Poirier2000,Heuze2012}. The EOS formulations differ in their expressions for $E_{th}$.

In the Einstein Model, a material is made up of simple harmonic oscillators vibrating at some characteristic frequency $\omega_E$ corresponding to a characteristic temperature $\theta_E=\frac{\hbar \omega_E}{k_B}$ and the resulting partition function is used to derive an energy as shown in Eq.~\ref{eq:einstein} \citep{Einstein1906,Goodstein1985,Poirier2000}\footnote{This expression leaves out the zero-point energy, which makes no difference on the final result as we are only concerned with $\Delta E$, not its absolute value.}:
\begin{equation}
    E=3nR(\frac{\theta_E}{e^{\frac{\theta_E}{T}}-1})\mathrm{.}
    \label{eq:einstein}
\end{equation}

In the Debye model, rather than assuming a uniform characteristic frequency, all frequencies below some frequency $\omega_D$ corresponding to a temperature $\theta_D=\frac{\hbar \omega_D}{k_B}$ contribute to the internal energy of the material \citep{Debye1912,Goodstein1985,Poirier2000}. The total internal energy is then an integral in frequency space of the energy per oscillator per mode, $\hbar\omega$, times the density of modes $g(\omega)$, times the number of quanta of vibration occupying that mode (phonons, c.f photons with $E=\hbar\omega$), resulting in Eq.~\ref{eq:debye} \citep{Debye1912,Goodstein1985,Poirier2000}.

\begin{align}
E&=\int_0^{\omega_D}<n>\hbar \omega g(\omega)d\omega \nonumber\\
&=9nRT(\frac{T}{\theta_D})^3\int_{0}^{\frac{\theta_D}{T}}\frac{x^3}{e^x-1}dx
\label{eq:debye}
\end{align}

At $T>>>\theta_D$, the entire portion of Eq.~\ref{eq:debye} to the right of \textcolor{black}{$9nRT$} (also known as the Debye Function) approaches a limit of $\frac{1}{3}$, so the Debye EOS can be re-expressed in the Linear form of Eq.~\ref{eq:linear} \textcolor{black}{also known as the Dulong-Petit law} \citep{DulongPetit1818,IchikawaTsuchiya2014,IchikawaTsuchiya2020,KuwayamaMorard2020}:
\begin{equation}
    E=3nRT\mathrm{.}
    \label{eq:linear}
\end{equation}
\textcolor{black}{Eq.~\ref{eq:linear} can also be derived from equipartition under the assumption that every atom has three translational and three vibrational degrees of freedom, each contributing $\frac{1}{2}RT$ of energy.}

The three formulations converge at higher temperatures, with the Debye and Einstein models in agreement for $T>>\theta_D$ and at even higher temperatures all energies converging to Eq.~\ref{eq:linear}, as can be derived by Taylor expanding $e^{\frac{\theta_D}{T}}$ in Eq.~\ref{eq:einstein} \citep{Dorogokupets2010}.

Thus, as one goes from the Debye model of Eq.~\ref{eq:debye} to the Einstein model of Eq.~\ref{eq:einstein} to the Linear model of Eq.~\ref{eq:linear}, the complexity of the equation decreases at the cost of a worse representation of reality, with this cost decreasing at higher temperatures.

If $\theta_D$ and $\theta_E$ were constant, the model would be harmonic; however, as a material compresses, its characteristic temperature $\theta_D$ changes, hence the ``quasi-'' in quasi-harmonic \citep{Anderson2005}. Under the quasi-harmonic approximation used in all EOS considered here\textcolor{black}{,} $\theta_D$ has no temperature dependence \citep{Anderson2005}. \textcolor{black}{Under this model, $\gamma_{th}$ represents} the change in $\theta_D$ with volume\textcolor{black}{. This} can be combined with Eq.~\ref{eq:mie-gruneisen} to express $\gamma_{th}$ in terms of the constant volume heat capacity $C_V$, the thermal expansion coefficient $\alpha$, and $K_T$, as in Eq.~\ref{eq:gamma} \citep{Gruneisen1912,Poirier2000,Anderson2005}:
\begin{equation}
    \gamma_{th} = \frac{P_{th}}{E_{th}} V= -\frac{d(\ln{(\theta)})}{d(\ln{(V)})} = \frac{\alpha V K_T}{C_V}\mathrm{.}
    \label{eq:gamma}
\end{equation}
The three definitions of $\gamma_{th}$ are only fully equivalent if the Mie–Grüneisen EOS is valid and can vary by up to tens of \% in low-density, high-temperature environments \citep{OganovDorogokupets2004}. \textcolor{black}{Throughout this work, $\gamma_\mathrm{th}$ is assumed to be temperature-independent.}

The volume dependence of $\gamma_{th}$ has been theoretically shown in the Thomas-Fermi atom to approach a constant at low volumes and experimentally determined to follow a power law at higher volumes \citep{Gilvarry1956,BoehlerRamakrishnan1980}, leading to the Al'tshuler form of $\gamma_{th}$ of Eq.~\ref{eq:al'tshuler} \citep{Al'tshulerBrusnikin1987}:
\begin{equation}
    \gamma_{th} = \gamma_{\infty}+(\gamma_0-\gamma_\infty)(\frac{\rho}{\rho_0})^{-\beta}\mathrm{.}
    \label{eq:al'tshuler}
\end{equation}
For many materials under Earth-like conditions, $\gamma_{\infty}\to0$, leading to an assumption of $\gamma_{\infty}=0$ or in some cases \textcolor{black}{$\beta\approx0$}, resulting in a constant $\gamma$ assumption \citep{Anderson1979}. \citet{Al'tshulerBrusnikin1987}'s original form had $\beta\equiv(\gamma_0)/(\gamma_0-\gamma_\infty)$, but most of the EOS used here treat $\beta$ as a free parameter. As one moves from a fixed $\gamma$ to $\gamma_{\infty}=0$ to the Al'tshuler Form to the Al'tshuler Form (Varied $\beta$), one increases the number of free parameters. This can result in a better fit to the data, but risks overfitting, especially as the number of free parameters approaches the order of magnitude of the number of datapoints.

In any case, the definition of $\gamma_{th}$ in Eq.~\ref{eq:gamma} combined with Eq.~\ref{eq:al'tshuler} leads to a formulation for $\theta_D$ or $\theta_E$ (numerically, $\theta_E\approx0.75\theta_D$, so scaling laws that apply to one apply to both \citep{Anderson2019}) presented in Eq.~\ref{eq:theta}:
\begin{equation}
    \theta_D = \theta_0(\frac{\rho}{\rho_0})^{\gamma_\infty}\exp{(\frac{\gamma_0-\gamma_\infty}{\beta}(1-(\frac{\rho}{\rho_0})^{-\beta}))}\mathrm{.}
    \label{eq:theta}
\end{equation}
\textcolor{black}{As $\theta_D$ depends on $\rho_0$, the thermal portion of the equation of state is linked to the isothermal portion of the equation of state. Throughout this work, the $\rho_0$ used in the thermal EOS is the $\rho_0$ from the source publication of the thermal EOS, even if we use a different source publication for our isothermal EOS for that material. This is the same methodology used by e.g. \citet{HakimRivoldini2018}}.

Despite the numerous approximations used in the derivation of the thermal energy, the fact that thermodynamic properties are integrals over an entire spectrum of vibrational states makes them relatively insensitive to the exact form of the spectrum, explaining the good agreement between this simplified theory, experiment, and the properties of Earth's interior \citep{StixrudeLithgow-Bertelloni2005}.

\subsubsection{Anharmonic and Electronic EOS}
\label{sec:anharmonic_electronic}
Real material structures are not composed solely of harmonic oscillators; application of perturbation theory to the potential of a weakly anharmonic oscillator in the high-temperature limit reveals an additional contribution to the internal energy proportional to $T^2$ \citep{OganovDorogokupets2004,DewaeleLoubeyre2006}. Electrons within the structure contribute an additional energy term; solving the Thomas-Fermi model for the electrons in an atom results in a contribution to the internal energy proportional to $T^2$ \citep{Thomas1927,Fermi1928,Gilvarry1954a,Gilvarry1954b,Al'TshulerKormer1960}. Experimental data reveals that the electronic energy varies with $\frac{V}{V_0}$ to a constant power \citep{Latter1955,Gilvarry1956,Al'TshulerKormer1960}. Similarly, the anharmonic energy varies with $\frac{V}{V_0}$ to a constant power, decreasing with pressure (and thus increasing with $V$) \citep{ZharkovKalinin1971,OganovDorogokupets2004,Oganov2015}.

As \textcolor{black}{the} anharmonic and electronic contributions to internal energy have identical temperature scaling, all EOS of interest in this paper combine them into one term or use a formulation involving only an electronic or anharmonic term but not the other \citep{BelonoshkoDorogokupets2008,BouchetMazevet2013}. Although the volume scaling of the electronic and anharmonic terms can differ, the impact is minor enough that this complication does not \textcolor{black}{e}ffect the quality of the fit \citep{BelonoshkoDorogokupets2008,BouchetMazevet2013}. The anharmonic-electronic internal energy is thus given in Eq.~\ref{eq:ae}\footnote{\label{n_abs}\citet{BouchetMazevet2013,MusellaMazevet2019} absorb $n$ into $e_0$}, with $e_0$ and $g$ being fit terms:
\begin{equation}
E_{ae} = \frac{3}{2}nRe_0(\frac{V}{V_0})^{g}T^2\mathrm{.}
\label{eq:ae}
\end{equation}

\textcolor{black}{The EOS of \citet{IchikawaTsuchiya2020,KuwayamaMorard2020,XieFu2025} are explicitly defined with their $E_{th}$ term including anharmonic-electronic internal energy and thus for these EOS we apply Eq.~\ref{eq:mie-gruneisen} in the form of $\Delta(P_{th}+P_{ae})=\gamma_{th}\frac{\Delta (E_{th}+E_{ae})}{V}$. In all other cases,} Eq.~\ref{eq:mie-gruneisen} is not valid for non quasi-harmonic terms and pressure is instead derived via the thermodynamic relation $P=-(\frac{\partial F}{\partial V})_T$ for a volume-independent $S$, resulting in Eq.~\ref{eq:P_ae}\footnote{See Footnote \ref{n_abs}} \citep{Al'TshulerKormer1960,Oganov2015,DorogokupetsDymshits2017}:
\begin{equation}
P_{ae} = \frac{3R}{2V}ne_0g(\frac{V}{V_0})^gT^2\mathrm{.}
\label{eq:P_ae}
\end{equation}

\textcolor{black}{\subsubsection{Adiabatic Gradients} \label{sec:ad_grad_discussion}}
\textcolor{black}{The values of the adiabatic gradient $\nabla_\mathrm{ad}$ are calculated in three distinct ways: method 0, wherein it is directly provided by a table that we interpolate over; method 1, relying on derivatives of the entropy; and method 2, relying on the rightmost definition of $\gamma_{th}$ in Eq.~\ref{eq:gamma}. Method 0 is used for the water layer; method 1 is used for the H/He envelope, the solid mantle at pressures below the transition of post-perovskite into high-pressure phases, the liquid mantle, and the solid core; whereas method 2 is used in the solid mantle at pressures above the transition of post-perovskite into high-pressure phases and the liquid core. We prefer method 1, but use method 0 when the provided entropy $S$ table is unreliable (see section \ref{sec:water}) and use method 2 when the simplification of Eq.~\ref{eq:linear} forces $(\frac{\partial \log{(T)}}{\partial \log{(P)}})_S=0$ and thus $\nabla_\mathrm{ad}=0$--even though $\nabla_\mathrm{ad} \neq 0$. Additionally, in some cases we have no explicit thermal EOS from which to calculate entropy and thus must use thermodynamics quantities that are provided in tables, as is the case for the high-pressure mantle \citep{TsuchiyaTsuchiya2011}. }

\textcolor{black}{Method 1 requires calculating} the entropy $S(V,T)$ of a material via Eq.~\ref{eq:S_from_E}:
\begin{equation}
    S(V,T) = \int_0^T\frac{\frac{\partial E(V,T^\prime)}{\partial T^\prime}}{T^\prime}dT^\prime\mathrm{.}
    \label{eq:S_from_E}
\end{equation}
\noindent
\textcolor{black}{Similarly to $\rho$ (see section \ref{sec:isothermal_background}), we desire $S(P,T)$ even though our equation specifies $S(V,T)$ (and thus $S(\rho,T)$). We pre-compute $S$ over the same $P$-$T$ grid for which we computed $\rho$ and thus already have the function $\rho(P,T)$. Our $S$ is thus directly calculated as $S(\rho(P,T),T)$. At runtime, both $\rho$ and $S$ are calculated for a combination of $P$ and $T$ via interpolation.}

Having determined $S$, we calculate the adiabatic gradient \textcolor{black}{in method 1} via Eq.~\ref{eq:adgrad},
\begin{equation}
    \nabla_\mathrm{ad} \equiv (\frac{\partial \log{T}}{\partial \log{P}})_S= -\frac{(\frac{\partial \log{S}}{\partial \log{P}})_T}{(\frac{\partial \log{S}}{\partial \log{T}})_P}\mathrm{,}
    \label{eq:adgrad}
\end{equation}
a consequence of the definition of $\nabla_\mathrm{ad}$ used here and setting $dS=0$ in $dS=(\frac{\partial S}{\partial T})_PdT+(\frac{\partial S}{\partial P})_TdP$.

\textcolor{black}{In method 2,} we instead calculate $\nabla_\mathrm{ad}$ using $\alpha$, $K_T$, and $C_V$. $\alpha\equiv\frac{1}{V}(\frac{\partial V}{\partial T})_P$ and $K_T\equiv-V(\frac{\partial P}{\partial V})_T$ are determined by numerically differentiating our EOS \textcolor{black}{while} $C_V\equiv(\frac{\partial E}{\partial T})_V$ is determined by taking derivatives of Equations \ref{eq:einstein}-\ref{eq:linear} \citet{Anderson2005,Glasser2013}. Combining the rightmost definition of $\gamma_\mathrm{th}$ in Eq.~\ref{eq:gamma} with the thermodynamic relations and Eq.~\ref{eq:adgrad} yields Eq.~\ref{eq:adgrad_different} \citep{HakimRivoldini2018,StaceyHodgkinson2019,HaldemannDorn2024}:
\begin{equation}
    \nabla_\mathrm{ad}=\frac{\gamma_{th}P}{K_T(1+\alpha\gamma_{th}T)} = \frac{\alpha V P}{C_V+\alpha^2VK_TT}\mathrm{.}
    \label{eq:adgrad_different}
\end{equation}

\subsubsection{Multiple Species in a Layer: Combining EOS} \label{sec:combination}
When multiple species are present in one layer, their EOS are combined using the fact that the extensive variables ($V$, the volume of a mole of a substance, and $S$, entropy, \textcolor{black}{in method 1 and $C_V$, $\alpha V$, and $V/K_T$ in method 2--notably, $\nabla_\mathrm{ad}$ is not extensive}) of different substances add for constant intensive variables ($P$ and $T$) \citep{FontaineGraboske1977,SaumonChabrier1995}, as expressed in Eq.~\ref{eq:additive_volume}:
\begin{equation}
    X = \sum_{i=1}^Nx_iX_i\mathrm{,}
    \label{eq:additive_volume}
\end{equation}
where $X$ is some extensive property and $X_i$ is that extensive property for some species $i$ comprising a molar fraction fraction $x_i$ of the mixed material\footnote{\citet{ChabrierMazevet2019} use $w_i$ instead of $x_i$, this is because all of their properties are not extensive but rather extensive variables per unit mass. Converting their variables back into extensive variables would involve multiplying by their masses, changing $x_i$ into $w_i$.}.

\textcolor{black}{Eq.~\ref{eq:additive_volume} is only strictly valid for ideal gases but has has been experimentally verified for a subset of materials up to 3 TPa and computationally verified for Earth's core \citep{BradleyLoomis2018,HuangBadro2019,UmemotoHirose2020}. Nonetheless, non-ideal effects resulting from the mixing of two materials do arise and are discussed in section \ref{sec:envelope} in the case of the H/He envelope. We do not include non-ideal effects for other layers due to a lack of available experimental or computational data.}

We apply Eq.~\ref{eq:additive_volume} directly for $S$ \textcolor{black}{when calculating $\nabla_\mathrm{ad}$ in method 1 (see section \ref{sec:ad_grad_discussion})}, but in the case of $V$, we desire the density rather than specific volume of substances. We convert between the two using the molar mass of the species, which yields Eq.~\ref{eq:density_derivation} for $\rho$\textcolor{black}{, also known as the additive volume law}:
\begin{equation}
    \rho = \left(\sum_{i=1}^N\frac{w_i}{\rho_i}\right)^{-1}\mathrm{.}
    \label{eq:density_derivation}
\end{equation}

\textcolor{black}{When calculating $\nabla_\mathrm{ad}$ via method 2 (see section \ref{sec:ad_grad_discussion}), we calculate} values of \textcolor{black}{$C_V$, $\alpha V$, and $V/K_T$} for a mixture using Eq.~\ref{eq:additive_volume}. \textcolor{black}{We calculate V \textcolor{black}{by}} dividing the sum of molar weights of a substance by $\rho$ from Eq.~\ref{eq:density_derivation} \citep{StixrudeLithgow-Bertelloni2005,StixrudeLithgow-Bertelloni2011}.

We calculate the melting temperature of a layer that is composed of multiple species by modifying the melting temperature of the dominant species in that layer following \citet{Stixrude2014} via Eq.~\ref{eq:melting_T}\footnote{Eq.~\ref{eq:melting_T} breaks down for very low $x_\mathrm{species|layer}$. In this study, the only occurrence of such a break-down is in the \textcolor{black}{iron}-rich mantle of Europa (see section \ref{sec:europa}). We thus set $x_\mathrm{MgSiO\textsubscript{3}|mantle}$ to Earth-like in this equation for Europa.},
\begin{equation}
    T^\mathrm{layer}_\mathrm{m} = \frac{T^\mathrm{species}_\mathrm{m}}{1-\ln{(x_\mathrm{species|layer})}}\mathrm{,}
    \label{eq:melting_T}
\end{equation}
where $T^\mathrm{layer}_\mathrm{m}$ is the layer melting temperature, $T^\mathrm{species}_\mathrm{m}$ is the melting temperature of some species derived via its EOS, and $x_\mathrm{species|layer}$ is the molar fraction of that species in the layer. 

We do not include partial melting in our model, which could be enhanced by volatiles within the mantle not included in our model \citep{KatzSpiegelman2003,HirschmannTenner2009,UnterbornSchaefer2020}. As the liquid mantle is generally less dense than the solid mantle, this results in an overestimation of mantle densities. Although simple, Eq.~\ref{eq:melting_T} reproduces the expected result that melting temperatures \textcolor{black}{for a mixture} containing MgSiO\textsubscript{3} mixed with a few weight percent of other substances are lower than melting temperatures \textcolor{black}{for} pure MgSiO\textsubscript{3} \citep{FiquetAuzende2010,AndraultBolfan-Casanova2011,NomuraHirose2014,UnterbornSchaefer2020}.

The conversion between elemental abundances and species abundances is covered in appendix \ref{sec:stoichiometry}.

\subsection{H/He Envelope} \label{sec:envelope}

\citet{ChabrierMazevet2019} calculate $P$-$T$ tables for pure H and He that we combine using Eq.~\ref{eq:additive_volume} with the addition of \textcolor{black}{the} $V_\mathrm{mix}(P,T)$ and $S_\mathrm{mix}(P,T)$ terms derived by \citet{HowardGuillot2023} using the results of \citet{ChabrierDebras2021} to account for non-ideal mixing. \textcolor{black}{Explicitly, we retrieve the $\rho$ and $S$ columns from the tables TABLE\_H\_TP\_v1 and TABLE\_HE\_TP\_v1 from \citet{ChabrierMazevet2019} and mixing $V$ and mixing $S$ from the table HG23\_Vmix\_Smix from \citet{HowardGuillot2023}.} \textcolor{black}{Following \citet{ChabrierMazevet2019}, we impose minimum and maximum values of $\nabla_\mathrm{ad}$ in the H/He envelope of 0.1 and 0.4, respectively.} Our model can account for any H/He ratio, but we assume a solar H/He ratio throughout this publication.

Below ${\sim}$10 GPa, non-ideal effects systematically increase the density of H/He \citep{HowardGuillot2023}. This results in planets with masses significantly smaller than Jupiter--constituting \textcolor{black}{most} of our sample--having systematically smaller radii than estimated ignoring non-ideal effects \citep{HowardGuillot2023}. Compounding this trend, the widely-used but out-of-date EOS of \citet{SaumonChabrier1995} (which uses semi-analytical rather than numerical calculations) is underdense for nearly all $P\leq$2 TPa and thus radii calculated using \citet{ChabrierMazevet2019,ChabrierDebras2021,HowardGuillot2023} are smaller than those calculated using \citet{SaumonChabrier1995} for planets in our sample by up to 10\%. %The EOS of \citet{HowardGuillot2023,ChabrierDebras2021,ChabrierMazevet2019} generally agrees with the older widely-used EOS of \citet{SaumonChabrier1995} below ${\sim}$100 GPa, but above are systematically denser, with densities up to ${\sim}$10\% greater. The inclusion of non-ideal mixing generally increases the density of planets less massive than Jupiter (as we are concerned with) but decreases the density of planets of comparable or greater mass than Jupiter \citep{HowardGuillot2023}.

% These systematically higher densities can have consequences in the interpretation of interior structure models, as is the case for interpretation of Juno data. The core mass of Jupiter estimated from \citet{ChabrierMazevet2019} alone is 14.1$M_\oplus$ whereas the core mass of Jupiter estimated from \citet{ChabrierMazevet2019,ChabrierDebras2021,HowardGuillot2023} is 20.8 $M_\oplus$, closer to the pebble isolation mass \citep{LambrechtsJohansen2014b,BitschMorbidelli2018}. However, we note that (1) Jupiter may have accreted significant solids after core formation and thus Jupiter's current metallicity may not reflect its mass before beginning runaway gas accretion and (2) data from the Juno spacecraft point toward a diluted core of Jupiter, so these numbers are not exact and merely indicate that these EOS increase Jupiter's inferred metallicity \citep{WahlHubbard2017,HelledStevenson2022}.

\subsection{Water} \label{sec:water}
We used the AQUA model $P$-$T$ tables of \citet{HaldemannAlibert2020} to calculate the EOS of the water layer. These tables combine the previous EOS from \citet{FeistelWagner2006} \textcolor{black}{(Ice Ih)}, \citet{JournauxBrown2020} \textcolor{black}{(Ice II-VI)}, \citet{FrenchRedmer2015} \textcolor{black}{(Ice VII, VII*, and X)}, \citet{WagnerPruss2002} \textcolor{black}{(Liquid, gas, and supercritical fluid <1200 K and <1 GP)}, \citet{Brown2018} \textcolor{black}{(Liquid and supercritical fluid >1 GPa)}, \citet{GordonMcbride1994,McbrideGordon1996} \textcolor{black}{(Gas >1200 K)}, and \citet{MazevetLicari2019} \textcolor{black}{(Supercritical fluid and high-pressure superionic)}, with the high\textcolor{black}{-}pressure results of \citet{MazevetLicari2019} forming the dominant contributor for planets with significant water mass fractions \citep{HaldemannAlibert2020}.

 \textcolor{black}{Explicitly, we retrieve the $\rho$ and $\nabla_\mathrm{ad}$ columns from the table aqua\_eos\_pt\_v1\_0 from \citet{HaldemannAlibert2020}. We retrieve $\nabla_\mathrm{ad}$ directly rather than calculating it via $S$ because the published $S$ table of \citet{MazevetLicari2019} has a typo in the calculation of $S$ (but not its derivatives) that is present in v1 of the AQUA EOS of \citet{HaldemannAlibert2020} \citep{AguichineBatalha2025}. As we are not mixing the water with any other substance and are not evolving the planetary structure, we have no explicit need for $S$.}
 
\textcolor{black}{AQUA's comprehensive phase inventory includes gas phases and thus a steam atmosphere is self-consistently recovered for sufficiently high instellations. As steam is less dense than condensed water, this leads to planets in this model with significant water mass fractions and high surface temperatures having systematically higher radii than planets in models that do not account for steam atmospheres.}

\textcolor{black}{\citet{HaldemannAlibert2020} find that a}pplying this EOS to isothermal pure water planets produces radii ${\sim}$3\% smaller than the older EOS used by \citet{ZengSasselov2016,ZengJacobsen2019}; a similar trend is found using a different set of water EOS reported by \citet{GrandeHuang2019,HuangRice2021}. In both cases, the fact that \citet{ZengSasselov2016,ZengJacobsen2019}'s model is not isothermal at high pressures makes a direct comparison difficult, although \citet{HaldemannDorn2024} find a similar increased water layer density compared to \citet{ZengSasselov2016,ZengJacobsen2019} with a non-isothermal temperature profile.

\subsection{Mantle} \label{sec:mantle}
\begin{figure*}
\centering     \includegraphics[width=\textwidth]{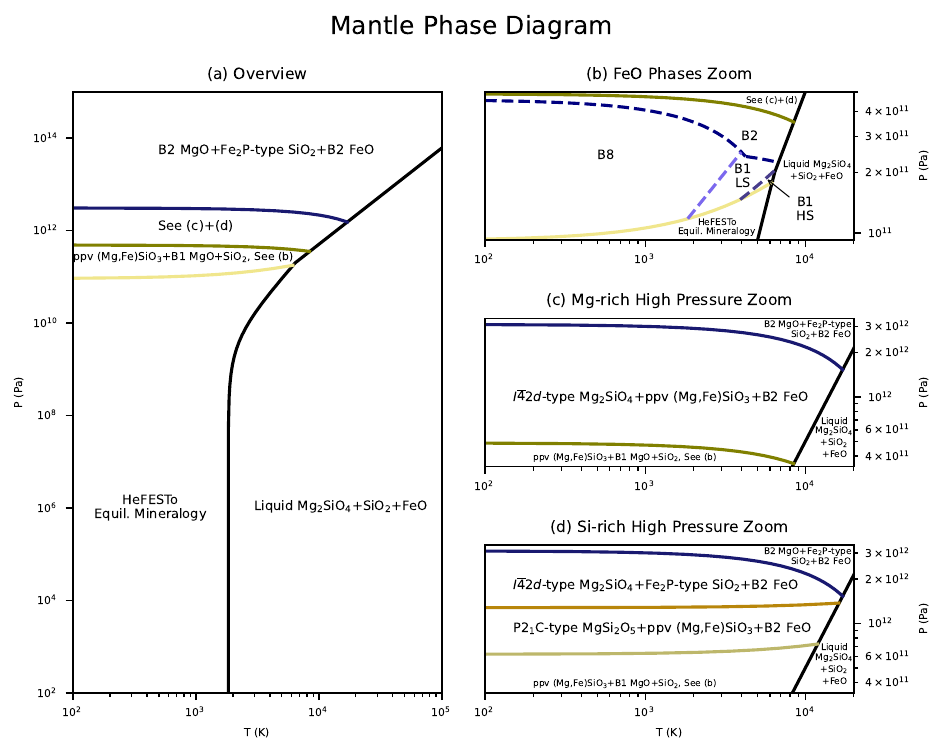}
\vspace{-2em}\caption{The species and phases of those species in our mantle. Solid lines represent changes in chemical composition, dashed lines represent solely changes in phase of a species. Colours are arbitrary but are consistent across panels. Equil. is an abbreviation for equilibrium. \textcolor{black}{A + sign} indicates the coexistence of multiple species. Note that many phase and chemical transitions take place in the region labelled HeFESTo Equi. Mineralogy. The case shown is for $x_\mathrm{MgSiO\textsubscript{3}}=1$ in the region with post-perovskite, the melting curve would move to lower temperatures with the addition of other species. See text for details. (a) The entire parameter space of the mantle. (b) Zoom-in to the region where post-perovskite is present, with FeO phase transitions indicated. (c) Zoom-in to ${\sim}$TPa pressures for a mantle with Mg/Si $>1$. (d) Zoom-in to ${\sim}$TPa pressures for a mantle with Mg/Si $\le 1$.}
\vspace{-2em}\label{fig:mantle_phase_diagram}
\end{figure*}

An overview of the species and phases present in our mantle is shown in Fig.~\ref{fig:mantle_phase_diagram}.

We calculate the melting temperature of the mantle using the MgSiO\textsubscript{3} melting temperature equations of \citet{BelonoshkoSkorodumova2005} for low pressures, \citet{Stixrude2014} for intermediate pressures, and \citet{FeiSeagle2021} for high pressures, switching at their intersections, as shown in Eq.~\ref{eq:MgSiO3_melting_T}:
\begin{equation}
    T^\mathrm{MgSiO\textsubscript{3}}_\mathrm{Melt} = \begin{cases} 
      1831\mathrm{ K}(1+\frac{P}{4.6\mathrm{ GPa}})^{0.33} & P\leq199.5\mathrm{ GPa}\\
      5400\mathrm{ K}(\frac{P}{140\mathrm{ GPa}})^{0.48} & 358.7>\frac{P}{\mathrm{GPa}}> 199.5\\
      6295\mathrm{ K}(\frac{P}{140\mathrm{ GPa}})^{0.317} & P\geq 358.7\mathrm{ GPa}\\
   \end{cases}\mathrm{.}
   \label{eq:MgSiO3_melting_T}
\end{equation}
Our reason for this three-piece piecewise curve is that \citet{BelonoshkoSkorodumova2005} is required to reproduce the melting temperature at \textcolor{black}{zero} pressure \textcolor{black}{while \citet{FeiSeagle2021} is required to fit high-pressure experimental data,} but \citet{FeiSeagle2021} and \citet{BelonoshkoSkorodumova2005} intersect at pressures too low ($8.4$ GPa) for \citet{FeiSeagle2021} to be appropriate. We thus utilize \citet{Stixrude2014} as the intermediate curve between the two extremes. \citet{DongMardaru2025,DongFischer2025} used a machine learning technique to determine the slope of the MgSiO\textsubscript{3} melting temperature, giving results similar but not identical to this (within 12\% from 3,200--10,000 K). We do not use this melting curve because of concerns with extrapolating a constant melting curve slope to high pressures with sparse data.
We calculate $x_\mathrm{MgSiO\textsubscript{3}|mantle}$ for Eq.~\ref{eq:melting_T} following the analytical formulae for the lower mantle given in \citet{HaldemannDorn2024}.
% Removed: For Earth-like planets, using solely \citet{Stixrude2014} at high pressures leads to an unphysical pattern in which planetary cores at increasing masses first solidify and then re-liquify while using \citet{FeiSeagle2021} at Earth's pressures leads to temperatures far too high to explain Earth's solid inner core in our methodology (see sections \ref{sec:core}, \ref{sec:thermal}, \ref{sec:earth}, and \ref{sec:mr}).

Above the melting temperature, we use the equations of state listed in Table \ref{tab:liquid_mantle}. \citet{StewartDavies2020}'s Mg$_2$SiO$_4$ EOS is provided in $\rho$-$T$ space and we use linear interpolation between adjacent data points in $\rho$ and $T$ to generate a table in $P$-$T$ space. We use this EOS rather than one for MgO as we are unaware of any liquid MgO EOS covering the desired parameter space \citep{HaldemannDorn2024}. As there is no EOS for Mg in the mantle not in Mg$_\mathrm{2}$SiO$_\mathrm{4}$\textcolor{black}{\textcolor{black}{, o}ur model is limited to} planets with $\frac{\mathrm{Mg}}{\mathrm{Si}}>0.5$, which is the case for every star in \textcolor{black}{a} study of sun-like stars in the solar neighbourhood by \citet{BedellBean2018}. % Their fig.~7

\citet{StewartDavies2020} find a melting point different from that obtained using Equations \ref{eq:melting_T} and \ref{eq:MgSiO3_melting_T} and we do not have access to separate solid and liquid EOS, so we allow for solid Mg$_2$SiO$_4$ within our liquid state. We avoid this phase transition introducing an artificial jump in the adiabatic gradient by keeping $(\frac{\partial S}{\partial T})_P$ constant for pressures inside or within one datapoint of the solid-liquid transition. This constant value of $(\frac{\partial S}{\partial T})_P$ is that of the highest-pressure liquid for the given temperature.

We use the publicly available FEOS code of \citet{FaikTauschwitz2018} to calculate $\rho$-$T$ SiO\textsubscript{2} EOS grids and use linear interpolation between adjacent data points in $\rho$ and $T$ to generate a table in $P$-$T$ space.
We use \citet{FaikTauschwitz2018} for all pressures; although there are concerns with \citet{FaikTauschwitz2018}'s EOS' applicability to pressures below the critical point \citep{HaldemannDorn2024}, we prioritize avoiding an artificial discontinuity in the SiO\textsubscript{2} EOS, especially one in which higher pressure leads to lower densities. Planetary mantle pressures are generally in excess of the critical point (in our model Earth, ${\sim}$99\% of the mass of the mantle is at pressures greater than the SiO\textsubscript{2} critical point), so the effects of non-applicability are minimal.

Below the melting temperature, we use different sets of EOS depending on the pressure.
The Perovskie--Post-Perovskite (pv--ppv) transition is given by \citet{DongMardaru2025,DongFischer2025} as in Eq.~\ref{eq:ppv_pv}:
\begin{equation}
    P^\mathrm{MgSiO\textsubscript{3}}_{\mathrm{ppv/pv}}=\left(120+0.014(\frac{T}{\mathrm{K}}-2000)\right)\mathrm{ GPa}\mathrm{.}
    \label{eq:ppv_pv}
\end{equation}
Below the Perovskite--Post\textcolor{black}{-}Perovskite transition, we implement the HeFESTo system of equations of state presented in \citet{StixrudeLithgow-Bertelloni2005,StixrudeLithgow-Bertelloni2011,StixrudeLithgow-Bertelloni2024} using Perple\_X 7.1.7 \citep{Connolly2009}. Perple\_X is a geochemical software that solves for the equilibrium mineralogy of mantle materials using Gibbs free energy minimization \citep{Connolly2009,Connolly2017}. The chemical composition of the upper mantle thus varies freely to its equilibrium state while its elemental composition remains uniform.

HeFESTo includes its own EOS for MgSiO\textsubscript{3} post-perovskite, but it is in the poorly-extrapolating BM3 form \citep{StixrudeLithgow-Bertelloni2005,StixrudeLithgow-Bertelloni2011,StixrudeLithgow-Bertelloni2024}. We thus stop using HeFESTo above the pv-to-ppv transition\textcolor{black}{. By switching the EOS used at a pre-existing phase transition and thus density jump, we avoid inducing an artificial density jump caused solely by a change in the EOS used.}

For pressures above the Perovskite--Post-Perovskite transition we calculate the equations of state for the species MgSiO\textsubscript{3}, FeSiO\textsubscript{3}, MgO, SiO\textsubscript{2}, and FeO, with each species' relative abundance determined from the input mantle-wide molar fractions of MgO, SiO\textsubscript{2}, and FeO\footnote{\citet{StixrudeLithgow-Bertelloni2024}'s model takes $x_\mathrm{Fe}$ and $x_\mathrm{O}$ as separate free parameters, we take $x_\mathrm{O}=x_\mathrm{Fe}=x_\mathrm{FeO}/2$} following \citet{HaldemannDorn2024}'s equations that ensure the \textcolor{black}{molar fraction} of perovskite is as high as possible. The EOS that we use in the solid mantle are shown in Table \ref{tab:solid_mantle}.

\citet{SakaiDekura2016} report multiple fits for ppv MgSiO\textsubscript{3}'s EOS, we use Fit 8--which is derived via simulation--rather than Fit 7--which is derived via experiment--because it satisfies Keane's rule (see section \ref{sec:isothermal_background}) while Fit 7 does not \citep{Keane1954,GuignotAndrault2007}. \citet{SakaiDekura2016} do not recommend use of their fits 1--6. At high (${\sim}$300 GPa) pressures, fit 8 is $\lesssim$1\% less dense than fit 7, indicating good agreement.

\textcolor{black}{We combine the isothermal EOS of \citet{GreenbergNazarov2023} with the thermal EOS of \citet{FischerCampbell2011,MorardAntonangeli2022}. We use the thermal parameters for B8 FeO from \citet{FischerCampbell2011} for B2 FeO because \citet{FischerCampbell2011,MorardAntonangeli2022} do not investigate B2 FeO. We select B8 FeO rather than B1 FeO for this purpose because B8 FeO is likely to be the phase present immediately above B2 FeO in super-Earth mantles.}

The FeO B8-B1 transition begins at the triple point from \citet{FeiMao1994} and follows the slope of \citet{OzawaHirose2010}, resulting in a phase transition given by Eq.~\ref{eq:feo_b8_b1_transition}:
\begin{equation}
    P^\mathrm{FeO}_\mathrm{B8/B1} = \left(66.4+0.063(\frac{T}{\mathrm{K}}-1020)\right)\textrm{ GPa}\mathrm{.}
    \label{eq:feo_b8_b1_transition}
\end{equation}
The FeO B2-B8 transition of Eq.~\ref{eq:feo_b2_b8_transition},
\begin{equation}
    P^\mathrm{FeO}_\mathrm{B2/B8} = \left(329-0.052(\frac{T}{\mathrm{K}}-2496)\right)\textrm{ GPa}\mathrm{,}
    \label{eq:feo_b2_b8_transition}
\end{equation}
is from \citet{ZhangSun2023}. The FeO B1-B8 transition of Eq.~\ref{eq:feo_b2_b1_transition},
\begin{equation}
    P^\mathrm{FeO}_\mathrm{B2/B1} = \left(242-0.0062(\frac{T}{\mathrm{K}}-3842)\right)\textrm{ GPa}\mathrm{,}
    \label{eq:feo_b2_b1_transition}
\end{equation}
is from \citet{OzawaTakahashi2011}, with the slope reported in the text and the intercept extracted from their fig.~1.

FeO can be in a low-spin (LS) or high-spin (HS) state \citep{Badro2014,GreenbergNazarov2023}. In the parameter space relevant to planetary interiors, B8 FeO and B2 FeO are always in the LS state \citep{GreenbergNazarov2023}. In contrast, B1 undergoes a transition from HS to LS as pressure increases following Eq.~\ref{eq:feo_b1_hs_ls_transition}, 
\begin{equation}
    P^\mathrm{FeO}_\mathrm{B1(LS)/B1(HS)} = \left(93-0.023(\frac{T}{\mathrm{K}}-1571)\right)\textrm{ GPa}\mathrm{,}
    \label{eq:feo_b1_hs_ls_transition}
\end{equation}
which we obtain from \citet{GreenbergNazarov2023}'s fig.~1. This transition can cause FeO density jumps nearing 10\% \citep{GreenbergNazarov2023}, so studies that assume FeO is in the HS state--as it is in ambient conditions--systematically underestimate FeO's density. \textcolor{black}{\citet{GreenbergNazarov2023} provide no thermal EOS for LS FeO B1 and thus we use the HS FeO B1 thermal EOS from \citet{FischerCampbell2011,MorardAntonangeli2022} for LS FeO B1, cautioning that \citet{OzawaTakahashi2011} find that LS B8 FeO has a greater thermal expansivity than HS B8 FeO and thus the usage of one thermal EOS for HS and LS is likely not strictly valid.}

The MgO B2-B1 transition curve of Eq.~\ref{eq:mgo_transition},
\begin{equation}
    P^\mathrm{MgO}_\mathrm{B2/B1} = \left(490-0.016(\frac{T}{\mathrm{K}}-300)\right)\textrm{ GPa}\mathrm{,}
    \label{eq:mgo_transition}
\end{equation}
is identical to the chemical transition curve of Eq.~\ref{eq:Mg2SiO4_comb_P}, we choose this to align our transitions. This is extremely close ($T=0$ point differs by 9 GPa and slope by 2 MPa) to the findings of \citet{DongMardaru2025,DongFischer2025}.

\citet{UmemotoWentzcovitch2011,UmemotoWentzcovitch2017} report that theoretically ppv MgSiO\textsubscript{3} undergoes a high-pressure phase transition into $I\overline{4}2d$ Mg$_2$SiO$_4$ and P2$_1$/c-type MgSi$_2$O$_5$. This transition has never been experimentally confirmed due to its high pressure, however, the $I\overline{4}2d$ or the similar $I\overline{4}3d$ phase \textcolor{black}{has} been experimentally observed in Mg$_2$GeO$_4$ and Fe$_3$O$_4$, analogues for Mg$_2$SiO$_4$ that undergo similar phase transitions at lower pressures \citep{UmemotoWentzcovitch2019,DuttaTracy2022,ZurkowskiYang2022}.

The exact sequence \textcolor{black}{of phase transitions} depends on whether there is free MgO or free SiO\textsubscript{2} in the mantle (i.e if Mg/Si is $>$ or $\leq$1). If there is free MgO in the mantle, past the pressure of Eq.~\ref{eq:Mg2SiO4_comb_P},
\begin{equation}
    P_\mathrm{MgSiO\textsubscript{3}+MgO}^\mathrm{Mg\textsubscript{2}SiO\textsubscript{4}+MgSiO_3} = \left(0.49*10^{12}-16*10^6\frac{T}{\mathrm{K}}\right)\mathrm{ Pa}\mathrm{,}
    \label{eq:Mg2SiO4_comb_P}
\end{equation}
as much MgO as possible combines with MgSiO\textsubscript{3} to form Mg$_2$SiO$_4$ \citep{UmemotoWentzcovitch2017}.
If there is free SiO\textsubscript{2} in the mantle, past the pressure of Eq.~\ref{eq:MgSi2O5_comb_P},
\begin{equation}
P_\mathrm{MgSiO\textsubscript{3}+SiO\textsubscript{2}}^\mathrm{MgSi\textsubscript{2}O\textsubscript{5}+MgSiO\textsubscript{3}} = \left(0.62*10^{12}\textcolor{black}{-}9*10^6\frac{T}{\mathrm{K}}\right)\mathrm{ Pa}\mathrm{,}
    \label{eq:MgSi2O5_comb_P}
\end{equation}
as much SiO\textsubscript{2} as possible combines with MgSiO\textsubscript{3} to form MgSi$_2$O$_5$ \citep{UmemotoWentzcovitch2017}. At the higher pressures of Eq.~\ref{eq:MgSi2O5_dissoc_P},
\begin{equation}
    P_\mathrm{MgSi\textsubscript{2}O\textsubscript{5}+MgSiO\textsubscript{3}}^\mathrm{Mg\textsubscript{2}SiO\textsubscript{4}+SiO\textsubscript{2}} = \left(1.31*10^{12}+6*10^6(\frac{T}{\mathrm{K}}-5000)\right)\mathrm{ Pa}\mathrm{,}
    \label{eq:MgSi2O5_dissoc_P}
\end{equation}
MgSi$_2$O$_5$ dissociates into Mg$_2$SiO$_4$ \citep{UmemotoWentzcovitch2017}.
Regardless of the value of Mg/Si, above the pressures of Eq.~\ref{eq:dissoc_P},
\begin{equation}
    P_\mathrm{Mg\textsubscript{2}SiO\textsubscript{4}+SiO\textsubscript{2}\textrm{ or }MgO}^\mathrm{MgO+SiO\textsubscript{2}\textrm{ or }MgO} = \left(3.1*10^{12}-92*10^6\frac{T}{\mathrm{K}}\right)\mathrm{ Pa}\mathrm{,}
    \label{eq:dissoc_P}
\end{equation}
all Mg and Si-bearing species in the mantle are completely dissolved into oxides \citep{UmemotoWentzcovitch2017}.

This sequence was specifically reported for molar ratios of MgO or SiO\textsubscript{2} to MgSiO\textsubscript{3} of 1:1. We do not include all transitions on the path between MgO or SiO\textsubscript{2} and MgSiO\textsubscript{3} to MgO and SiO\textsubscript{2} because to the authors' knowledge there is no publicly available precise curve for these transitions, only approximate curves as in \citet{vandenBergYuen2019}. We assign all excess atoms to post-perovskite MgSiO\textsubscript{3} because it is the only molecule at these pressures existing in a form that has been directly observed. Thus, for Mg/Si$>1$ we have Reactions \ref{eq:mgo_step_1}-\ref{eq:mgo_step_2}:
\begin{equation}
    \mathrm{xMgSiO_3+yMgO\rightarrow(x-y)MgSiO_3+yMg_2SiO_4}\mathrm{,}
    \label{eq:mgo_step_1}
\end{equation}
\begin{equation}
        \mathrm{(x-y)MgSiO_3+yMg_2SiO_4\rightarrow(x+y)MgO+xSiO_2}\mathrm{,}
    \label{eq:mgo_step_2}
\end{equation}
while for Mg/Si$<1$ we have Reactions \ref{eq:sio2_step_1}-\ref{eq:sio2_step_3}:
\begin{equation}
    \mathrm{xMgSiO_3+ySiO_2\rightarrow(x-y)MgSiO_3+yMgSi_2O_5}\mathrm{,}
    \label{eq:sio2_step_1}
\end{equation}
\begin{equation}
    \mathrm{(x-y)MgSiO_3+yMgSi_2O_5\rightarrow\frac{x}{2}Mg_2SiO_4+(\frac{x}{2}+y)SiO_2}\mathrm{,}
    \label{eq:sio2_step_2}
\end{equation}
\begin{equation}
    \mathrm{\frac{x}{2}Mg_2SiO_4+(\frac{x}{2}+y)SiO_2\rightarrow xMgO+(x+y)SiO_2}\mathrm{.}
    \label{eq:sio2_step_3}
\end{equation}

\citet{NiuOganov2015} report a different sequence with a temperature dependence, we favor the sequence of \citet{UmemotoWentzcovitch2017} due to its relative simplicity.

\citet{TsuchiyaTsuchiya2011} and \citet{WuUmemoto2011} report EOS for Fe\textsubscript{2}P-type SiO\textsubscript{2}. \citet{WuUmemoto2011} report no thermal parameters while \citet{TsuchiyaTsuchiya2011} do not report a $K_0^\prime$. We thus combine their EOS. We use the values of $V_0$, $K_0$, $\gamma$, and $\theta_D$ (all assumed constant) from \citet{TsuchiyaTsuchiya2011} at 0.7 TPa. Our $K_0$ value is for a $T_0$ of 300 K, the lowest temperature reported, and our $\gamma$ value is for a temperature of 4000 K, which is representative both of the temperatures at which this phase exists in planetary interiors and of the typical value of $\gamma$ across temperatures (as we assume $\gamma$ is temperature-\textcolor{black}{in}dependent throughout this work). We take the value of $K_0^\prime$ from \citet{WuUmemoto2011} at 0.8 TPa as extrapolated from a BM3 fit. As our reference conditions are at high rather than ambient pressures, we add the constant term $P_0=0.7$ TPa to Eq.~\ref{eq:rose_vinet}. We caution that this approach is not consistent as their parameters were \textcolor{black}{were not designed to be used together}.

\citet{DuttaTracy2023} find that $I\overline{4}2d$ Mg$_2$SiO$_4$ disorders at high temperatures, causing it to have lower densities than predicted by the EOS we use here. We do not include this effect due to the lack of a closed-form EOS for \textcolor{black}{it}.

We additionally caution that the tabular EOS reported by \citet{UmemotoWentzcovitch2011} \textcolor{black}{for} P2$_1$/c-type MgSi$_2$O$_5$ and by \citet{UmemotoWentzcovitch2017} for I$\overline{4}2d$\textcolor{black}{-type} Mg$_2$SiO$_4$ have poor resolution (200 GPa and 1000 K). Despite these weaknesses in the EOS and the unconfirmed nature of these high-pressure phases in the mantle, we include them to avoid a systematic underprediction of the mantle densities of massive planets. Even if these phase transitions are not accurate, the phase transitions in reality would have the same systematic effect of increasing densities as our treatment here.

Another high-pressure phase of SiO\textsubscript{2}, cotunnite, has been reported at high temperatures \citep{UmemotoWentzcovitch2006,Oganov2015,Gonzalez-CataldoDavis2016}, we do not include it here as it occurs at temperatures higher than occur in the mantles of the vast majority of planets in our isocomposition curves and it has very similar parameters to Fe\textsubscript{2}P-type SiO\textsubscript{2} (\citet{WuUmemoto2011} find it has the same $K_0^\prime$ and a $K_0$ within 0.5\% of Fe\textsubscript{2}P-type SiO\textsubscript{2}).

Further phase transitions such as Fe\textsubscript{2}P-type SiO\textsubscript{2} into I4/mmm at 10--14 TPa (encountered in mantles of Earth-like planets $\gtrsim20M_\oplus$) have been predicted, but we avoid implementing unobserved phase transitions beyond the dissociation of the mantle into oxides \citep{LylePickard2015,WangLv2024}. \citet{UmemotoWentzcovitch2017} found that further changes in chemical composition deeper in the mantle are unlikely, indicating our model is likely largely accurate up to very high pressures.

The \textcolor{black}{main} net effect\textcolor{black}{s} of this suite of mantle EOS \textcolor{black}{are} \textcolor{black}{(1)} denser planets than those assumed by models of pure MgSiO\textsubscript{3} due to the inclusion of FeSiO\textsubscript{3}\textcolor{black}{--}the densest silicate stable in the lower mantle of the Earth \citep{YangSong2024}\textcolor{black}{, (2)} denser \textcolor{black}{massive} planets than those derived from extrapolations from Earth's structure due to the inclusion of high-pressure phase transitions not present in Earth's interior\textcolor{black}{, and (3) less dense hot planets than those assumed by models that do not account for mantle melting.}

\defcitealias{MorardAntonangeli2022}{MA22}
\defcitealias{StewartDavies2020}{SD20}
\defcitealias{FaikTauschwitz2018}{FK18}
\begin{table*}
    \caption{The EOS used within the liquid mantle. AE is short for anharmonic \& electronic. Vinet is short for Vinet-Rydberg. All references are abbreviated to the first letters of the last names of the first two authors, as follows: \citet{MorardAntonangeli2022} is \citetalias{MorardAntonangeli2022}, \citet{StewartDavies2020} is \citetalias{StewartDavies2020}, \citet{FaikTauschwitz2018} is \citetalias{FaikTauschwitz2018}.}
    \begin{center}
    \centerline{
    \begin{threeparttable}
\begin{tabular}{ccccccc}
    Compound&Phase&Isothermal EOS&Thermal EOS&AE EOS&$\gamma_{th}$&Reference\\\hline
        FeO & Liquid & Vinet & Einstein & Quadratic & $\gamma_{\infty}=0$ & \citetalias{MorardAntonangeli2022}\\
        Mg\textsubscript{2}SiO\textsubscript{4} & Liquid & \multicolumn{4}{c}{Tabular} & \citetalias{StewartDavies2020}\\
        SiO\textsubscript{2} & Liquid & \multicolumn{4}{c}{Tabular} & \citetalias{FaikTauschwitz2018}
    \end{tabular}
    \end{threeparttable}}
    \end{center}
    \label{tab:liquid_mantle}
\end{table*}

\defcitealias{SakaiDekura2016}{SD16}
\defcitealias{StixrudeLithgow-Bertelloni2024}{SL24}
\defcitealias{FischerCampbell2011}{FC11}
\defcitealias{GreenbergNazarov2023}{GN23}
\defcitealias{YePrakapenka2017}{YP17}
\defcitealias{DorogokupetsDewaele2007}{DD07}
\defcitealias{MusellaMazevet2019}{MM19}
\defcitealias{TsuchiyaTsuchiya2011}{TT11}
\defcitealias{UmemotoWentzcovitch2017}{UW17}
\defcitealias{UmemotoWentzcovitch2011}{UW11}

\begin{table*}
    \caption{The EOS used within the solid mantle at pressures beyond the perovskite-post-perovskite transition. AE is short for anharmonic \& electronic. Vinet is short for Vinet-Rydberg. Ppv is short for post-perovskite. BM3 is short for third-order Birch-Murnaghan. Debye is short for Mie-Grüneisen-Debye. All references are abbreviated to the first letters of the last names of the first two authors, as follows: \citet{SakaiDekura2016} is \citetalias{SakaiDekura2016}, \citet{StixrudeLithgow-Bertelloni2024} is \citetalias{StixrudeLithgow-Bertelloni2024}, \citet{MorardAntonangeli2022} is \citetalias{MorardAntonangeli2022}, \citet{FischerCampbell2011} is \citetalias{FischerCampbell2011}, \citet{GreenbergNazarov2023} is \citetalias{GreenbergNazarov2023}, \citet{YePrakapenka2017} is \citetalias{YePrakapenka2017}, \citet{DorogokupetsDewaele2007} is \citetalias{DorogokupetsDewaele2007}, \citet{MusellaMazevet2019} is \citetalias{MusellaMazevet2019}, \citet{TsuchiyaTsuchiya2011} is \citetalias{TsuchiyaTsuchiya2011}, \citet{FaikTauschwitz2018} is \citetalias{FaikTauschwitz2018}, \citet{UmemotoWentzcovitch2017} is \citetalias{UmemotoWentzcovitch2017}, \citet{UmemotoWentzcovitch2011} is \citetalias{UmemotoWentzcovitch2011}. \textcolor{black}{If multiple citations are provided, the first citation is for the isothermal EOS and the second citation is for the thermal EOS.}}
    \begin{center}
    \centerline{
    \begin{threeparttable}
\begin{tabular}{ccccccc}
    Compound&Phase&Isothermal EOS&Thermal EOS&AE EOS&$\gamma_{th}$&Reference\\\hline
        MgSiO\textsubscript{3} & ppv & Keane & Debye & None & Al'tshuler & \citetalias{SakaiDekura2016}\\
        FeSiO\textsubscript{3} & ppv & BM3 & Debye & None & $\gamma_\infty=0$ & \citetalias{StixrudeLithgow-Bertelloni2024}\\
        FeO & B1 (HS) &  BM3 & Debye & None & $\gamma_\infty=0$ & \citetalias{MorardAntonangeli2022,FischerCampbell2011}\\
        FeO & B1 (LS) &  BM3 & Debye & None & $\gamma_\infty=0$ & \citetalias{GreenbergNazarov2023,MorardAntonangeli2022,FischerCampbell2011}\\
        FeO & B8 &  BM3 & Debye & None & $\gamma_\infty=0$ & \citetalias{GreenbergNazarov2023,FischerCampbell2011}\\
        FeO & B2 &  BM3 & Debye & None & $\gamma_\infty=0$ & \citetalias{GreenbergNazarov2023,FischerCampbell2011}\\
        MgO & B1 & Vinet & Debye & Quadratic & $\beta$ varied & \citetalias{YePrakapenka2017,DorogokupetsDewaele2007}\\
        MgO & B2 & Holzapfel & Einstein & Quadratic & $\beta$ varied & \citetalias{MusellaMazevet2019}\tnote{\textsuperscript{a}}\\
        SiO\textsubscript{2} & Fe\textsubscript{2}P-type & Vinet & Debye & None & Fixed & \citetalias{TsuchiyaTsuchiya2011}; This work\\
        SiO\textsubscript{2} & Various &\multicolumn{4}{c}{Tabular}& \citetalias{FaikTauschwitz2018}\\
        Mg\textsubscript{2}SiO$_4$ & $I\overline{4}2d$ &\multicolumn{4}{c}{Tabular}& \citetalias{UmemotoWentzcovitch2017}\\
        MgSi\textsubscript{2}O\textsubscript{5} & $P2_1/c$-type &\multicolumn{4}{c}{Tabular}& \citetalias{UmemotoWentzcovitch2011}
    \end{tabular}
    \begin{tablenotes}
  \item{\textsuperscript{a}}{$\beta$ is incorrectly listed as negative in \citet{MusellaMazevet2019}'s original publication. Using the same value of $\beta$ reported with a positive sign replicates their fig.~1.}
    \end{tablenotes}
    \end{threeparttable}}
    \end{center}
    \label{tab:solid_mantle}
\end{table*}

\subsection{Core} \label{sec:core}
\begin{figure*}
\centering     \includegraphics[width=\textwidth]{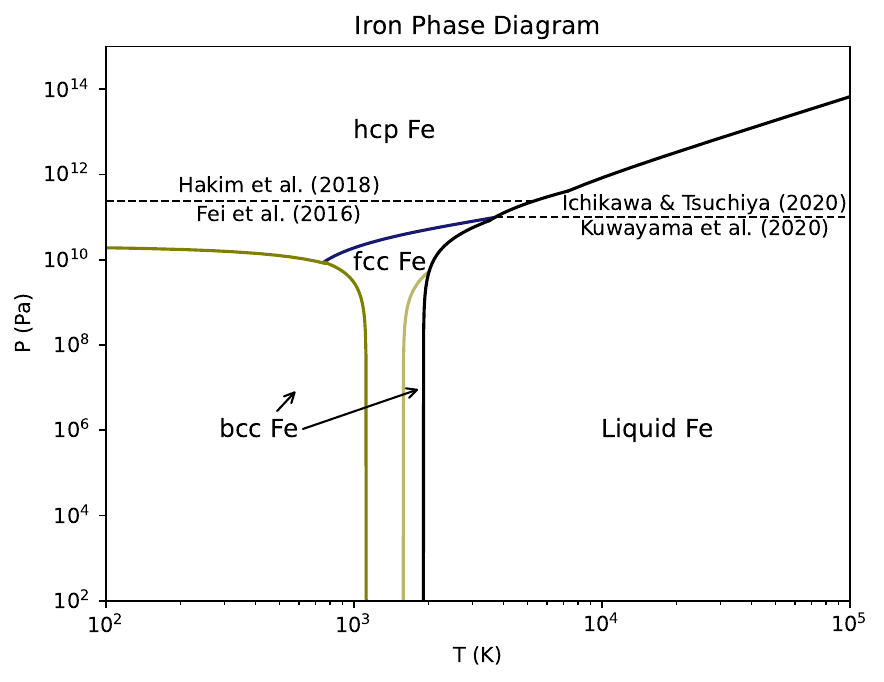}
\vspace{-2em}\caption{The phase diagram of iron employed in our model. Solid lines indicate phase transitions while dashed lines represent transitions between using different EOS for the same phase. EOS labels indicate the isothermal portion only, see text and Table \ref{tab:solid_core} for details. Colours are arbitrary. Note that solid Fe of all phases coexists with solid FeS VI/VII and liquid Fe coexists with liquid FeS and liquid FeO. The case shown is for $x^\mathrm{Solid}_\mathrm{S}=0$, the \textcolor{black}{black} melting curve would move to lower temperatures with greater $x^\mathrm{Solid}_\mathrm{S}$. See text for details.}
\vspace{-2em}\label{fig:iron_phase_diagram}
\end{figure*}

The phase diagram we employ for iron is shown in Fig.~\ref{fig:iron_phase_diagram}.

Seismology indicates that Earth's core density is lower than that of pure iron and thus the inclusion of lower-density core materials is essential to recreate the radii of Earth-like planets \citep{Birch1952,Birch1964,McQueenMarsh1966,Ahrens1979,Jeanloz1979,UnterbornSchaefer2020}. The main low-density elements within the Earth's core are Si, O, and S \citep{Mcdonough2014}. Si and/or O likely form the majority by weight of the lower-density element in Earth's core, however the partitioning of Si and O into the core rather than the mantle is uncertain at higher pressures: \citet{FischerNakajima2015} find that Si and O continue to partition into planetary cores up to ${\sim}800$ GPa while \citet{SchaeferJacobsen2017} find that Si and O stop partitioning into planetary cores at ${\sim}$100 GPa.

The light element composition of planetary cores depends on the pressures at which the metals and silicates of planets equilibrated \citep{WadeWood2005}, which requires an understanding of the microphysics of metals in planetary primordial magma oceans and \textcolor{black}{of} the history of giant melting impacts still in development for Earth \citep{SchaeferJacobsen2017,LichtenbergSchaefer2023}, making an exact determination of the light elements within the cores of super-Earths beyond the scope of this work. We thus use reasonable parameters for the light elements in the core, approximating all light elements as FeO--for which we already have EOS for the mantle--and FeS, which is \textcolor{black}{generally agreed to exist in Earth's core} (albeit at a lower concentration than in our model\textcolor{black}{, see below).} \textcolor{black}{The likely presence of FeS in exoplanetary cores is further supported by \citet{KamaShorttle2019}--who find that the spectra of the accretion-contaminated photospheres of young stars imply that most sulphur is in a refractory form (e.g. FeS) within the protoplanetary disk--and \citet{RogersBonsor2024}, who find a correlation between Fe and S abundances across several polluted white dwarfs.}

A key constraint on the composition of the core comes from the observed density jump between Earth's liquid and solid cores of $4.5\pm0.5\%$ \citep{DziewonskiAnderson1981,AlfeGillan2002,IchikawaTsuchiya2020}. This jump is larger than expected solely from the transition between liquid and solid ($1.5\%$ in our model for Earth\textcolor{black}{'s} core with 0.15 \textcolor{black}{molar fraction} S), implying that lighter elements preferentially partition into the liquid core, as also found via experiment and simulation \citep{AlfeGillan2002,ZhangCsanyi2020,SakaiHirose2023}. We thus allow light elements to be present in different abundances in the liquid and solid core, to our knowledge the first time that this effect has been considered in an exoplanetary interior structure model \textcolor{black}{that mixes materials with each having its own unique EOS. Previous literature \citep[e.g.][]{ValenciaO'Connell2006,ValenciaSasselov2007} has used different compositions for the solid and liquid core, but employed a single EOS for the combined species (in contrast with our methodology outlined in subsection \ref{sec:combination}). This can be problematic as different species in the core will have different compressibilities, so using one EOS for a mixture of species will lead to a different high-pressure extrapolation than mixing the different EOS of multiple species.} 

We emphasize that the relative abundances of light elements in the inner and outer core \textcolor{black}{are} determined by partitioning, which is dependent on the pressures in the planetary core. Thus, the light element abundances in the liquid and solid cores are not constant across different masses for the same planetary composition as assumed here. Our prescription is nevertheless a first step towards the inclusion of this effect in planetary interior structure models.

Using the size of the inner-outer core density jump, \citet{AlfeGillan2002} find that a \textcolor{black}{molar fraction} of $8\pm2.5\%$ of O in the liquid core, negligible O in the solid core, $10\pm2.5\%$ S in the liquid core, and $8.5\pm2.5\%$ S in the solid core reproduces Earth's density jump well\footnote{\citet{AlfeGillan2007} revise their estimates for a density jump of 6.5\% rather than 4.5\% following the revised Earth interior density jump of \citet{MastersGubbins2003}, we refer to their results derived from a comparison with PREM here as the upwards-revised density jump is not universally agreed, see \citet{KoperDombrovskaya2005}.}. \citet{HiroseLabrosse2013} conducted a literature review of proposed light core elements, combining density constraints and geochemical constraints, and preferred weight fractions of Si of ${\sim}6\%$, of O of ${\sim}3\%$, and of S of $1$--$2$\%. This corresponds to a molar fraction of O of ${\sim}9\%$ and a non-Fe-non-O mass fraction of ${\sim}8\%$, corresponding to a S molar fraction of ${\sim}13\%$ assuming that all non-Fe-non-O mass fraction in the core is S (i.e we approximate \textcolor{black}{silicon} and all other less abundant elements as \textcolor{black}{sulphur}, justified by \textcolor{black}{sulphur's} close molar mass to \textcolor{black}{silicon}). \citet{UmemotoHirose2020} give best fits to PREM for a Fe-Ni-H-Si-O-S-C system for various assumed solid-inner core boundary temperatures, with the most likely $T_\mathrm{ICB}$ of 5400 K giving $x_\mathrm{O}=0.08$ and $x_\mathrm{S}=0.01$ and the somewhat higher $T_\mathrm{ICB}$ of 6000 K giving $x_\mathrm{O}=0.22$ and $x_\mathrm{S}=0.005$. \citet{HiroseWood2021} combine density, geochemical, and cosmochemical constraints to get $x^\mathrm{Liquid}_\mathrm{O}=0.03$--$0.1\textcolor{black}{8}$ and $x^\mathrm{Liquid}_\mathrm{S}=0.03$ with up to $x^\mathrm{Liquid}_\mathrm{Si}=0.0\textcolor{black}{8}$, $x^\mathrm{Liquid}_\mathrm{H}=0.1\textcolor{black}{4}$, and $x^\mathrm{Liquid}_\mathrm{C}=0.01$ for the outer core and up to $x^\mathrm{Solid}_\mathrm{O}=0.003$, $x^\mathrm{Solid}_\mathrm{S}=0.02$, $x^\mathrm{Solid}_\mathrm{Si}=0.0\textcolor{black}{5}$, $x^\mathrm{Solid}_\mathrm{H}=0.1\textcolor{black}{3}$, and $x^\mathrm{Solid}_\mathrm{C}=0.06$ for the inner core. \textcolor{black}{Given the agreement in the literature that $x^\mathrm{Solid}_\mathrm{O}$ is low ($\lesssim$0.01), we set it to zero throughout this work to reduce the number of free parameters and avoid overfitting the core.}

We find that $x^\mathrm{Liquid}_\mathrm{O}=0.08$ and $x^\mathrm{Liquid}_\mathrm{S}=0.09$ in the liquid core and $x^\mathrm{Solid}_\mathrm{S}=0.115$ in the solid core best \textcolor{black}{reproduces} Earth's interior (see section \ref{sec:earth}) and thus select these values as fiducial. These values are within the uncertainties of those reported by \citet{AlfeGillan2002} and \citet{HiroseLabrosse2013} (note that the liquid core is much larger than the solid core and thus the bulk core composition is near to the liquid core composition) but lower than those reported by \citet{UmemotoHirose2020,HiroseWood2021}. These discrepancies are to be expected because (1) we use a different set of EOS from previous attempts to identify the light element in Earth's core and thus have no reason to replicate their results and (2) we neglect several elements included in their calculations and thus our values represent effective values resulting from fitting a simplified mineralogy to Earth's core. Additionally, planetary cores are volumetrically small and thus have relatively little impact on planetary radii (see section \ref{sec:validation}), minimizing the impact of the assumption of the light element in the core. \textcolor{black}{We note that S in Earth's core is stratified and not uniformly mixed in each layer as assumed here \citep{AlfeGillan2002,Ganguly2025}.}

We also emphasize that these values are selected to fit the density structure within Earth's \textcolor{black}{core{}} and not to reproduce Earth's radius, so our replication of Earth's radius shown in section \ref{sec:earth} is not a product of actively selecting core parameters to replicate Earth's radius. Due to our close replication of Earth in the rest of our model, alterations to $x^\mathrm{Liquid}_S$, $x^\mathrm{Liquid}_\mathrm{O}$, and $x^\mathrm{Solid}_S$ could be employed to replicate Earth's radius to arbitrary \textcolor{black}{accuracy}. This high precision would merely be the product of canceling out any errors in the mantle with an arbitrary core structure and thus would have no reason to extrapolate to other bodies.

We determine the melting temperature using Eq.~\ref{eq:Fe_melting},
\begin{equation}
    T^\mathrm{Fe}_\mathrm{Melt} = \begin{cases}
    2100+19.5(\frac{P}{\mathrm{GPa}}-10)\mathrm{ K} & P<82.\textcolor{black}{7}\mathrm{ GPa}\\
      3950+11.6(\frac{P}{\mathrm{GPa}}-120)\mathrm{ K} & 82.\textcolor{black}{7}<\frac{P}{\mathrm{GPa}}<409.8\\
    6469(1+(\frac{P-300\mathrm{ GPa}}{434.82\mathrm{ GPa}})^\frac{1}{1.839})\mathrm{ K} & P\geq 409.8\mathrm{ GPa}
   \end{cases}\mathrm{,}
   \label{eq:Fe_melting}
\end{equation}
combined with Eq.~\ref{eq:melting_T}, with $x^\mathrm{Solid}_\mathrm{Fe}$ being the species of interest. The lowest-pressure piece corresponds to the melting temperature of fcc \textcolor{black}{iron} (see below for solid phases) while the higher two pressure pieces correspond to the melting temperature of hcp \textcolor{black}{iron}. Note that as our phase changes and melting temperatures come from different sources, there is no fcc-hcp-liquid triple point (see Fig.~\ref{fig:iron_phase_diagram}). We thus allow some hcp \textcolor{black}{iron} to melt following the fcc \textcolor{black}{iron melting} curve to ensure continuity in the melting temperature.

The first two pieces of Eq.~\ref{eq:Fe_melting} come from \citet{DongMardaru2025,DongFischer2025}'s machine learning classification of literature phase experimental or computational observations, constraining the slope of $T^\mathrm{Fe}_\mathrm{fcc/Liquid}$ to 16--23$\frac{\mathrm{K}}{\mathrm{GPa}}$ and the slope of $T^\mathrm{Fe}_\mathrm{hcp/Liquid}$ to 5--12$\frac{\mathrm{K}}{\mathrm{GPa}}$. \citet{DongMardaru2025,DongFischer2025} do not report separate melting slopes for bcc \textcolor{black}{iron} and thus we assume that bcc \textcolor{black}{iron} has the same melting curve as fcc \textcolor{black}{iron} (the lowest-pressure piece of \textcolor{black}{Eq.~\ref{eq:Fe_melting}}). We choose our slope in the middle piece of \textcolor{black}{Eq.~\ref{eq:Fe_melting}} to reproduce the radius of Earth's liquid-solid core boundary (see section \ref{sec:earth}), which we caution may not accurately reflect reality as the crystallization of the Earth's core is a temporal process not captured by the static model presented here\textcolor{black}{. O}ur slope in the low-pressure piece of Eq.~\ref{eq:Fe_melting} is chosen to be the middle of the provided distribution of plausible values. We caution that values of the slope of Eq.~\ref{eq:Fe_melting} consistent with \citet{DongMardaru2025,DongFischer2025} both allow for a completely liquid and completely solid Earth core (changing a model Earth's radius by ${\sim}$0.5\%) and thus our slope is strongly dependent on geological constraints that might not be universally applicable outside of Earth. The highest-pressure piece of Eq.~\ref{eq:Fe_melting} comes from \citet{Gonzalez-CataldoMilitzer2023}, who perform simulations at high pressures. \textcolor{black}{We swap between the results of \citet{DongMardaru2025,DongFischer2025} and \citet{Gonzalez-CataldoMilitzer2023} when they intersect. We note that this is not a smooth transition.}
We calculate the phase of solid Fe using Equations \ref{eq:epsilon-gamma}-\ref{eq:epsilon-alpha},
\begin{equation}
    T^\mathrm{Fe}_\mathrm{hcp/fcc}=575+18.7\left(\frac{P}{\textrm{GPa}}\right)+0.213\left(\frac{P}{\textrm{GPa}}\right)^2-0.000817\left(\frac{P}{\textrm{GPa}}\right)^3\textrm{ K}\mathrm{,}
    \label{eq:epsilon-gamma}
\end{equation}
\begin{equation}
    T^\mathrm{Fe}_\mathrm{fcc/bcc1}=1120-300\left(\frac{P}{\textrm{GPa}}\right)\textrm{ K}\mathrm{,}
    \label{eq:gamma-alpha}
\end{equation}
\begin{equation}
    T^\mathrm{Fe}_\mathrm{fcc/bcc2}=1580+418\left(\frac{P}{\textrm{GPa}}\right)\textrm{ K}\mathrm{,}
    \label{eq:gamma-alpha2}
\end{equation}
\begin{equation}
    T^\mathrm{Fe}_\mathrm{hcp/bcc}=820-520\left(\frac{P-7.3\textrm{ GPa}}{8.5\textrm{ GPa}}\right)\textrm{ K}\mathrm{,}
    \label{eq:epsilon-alpha}
\end{equation}
from \citet{DorogokupetsDymshits2017,HaldemannDorn2024}. There are two fcc/bcc transitions in Equations \ref{eq:gamma-alpha} and \ref{eq:gamma-alpha2} because bcc is the most stable phase of Fe in two non-contiguous regions of $P$-$T$ parameter space (see Fig.~\ref{fig:iron_phase_diagram}).

The EOS for the liquid core are shown in Table \ref{tab:liquid_core}. We use two separate EOS for liquid Fe as \citet{IchikawaTsuchiya2020} and \citet{KuwayamaMorard2020} reported EOS fits over different pressure regimes. We caution that these two EOS do not \textcolor{black}{smoothly intersect}, with \citet{IchikawaTsuchiya2014}'s result ${\sim}$4\% denser at 5000 K and 100 GPa than \citet{KuwayamaMorard2020}\textcolor{black}{'s}. However, as the higher-density EOS is found at greater pressures, this does not result in an unphysical density decrease and thus we find it acceptable. We do not use the more recent isothermal Fe EOS from \citet{XieFu2025} because they are fit with BM3 \textcolor{black}{and have $K_0^\prime<4$, leading to inaccurate results when extrapolating to high pressures (see section \ref{sec:isothermal_background})}, however we do augment \citet{IchikawaTsuchiya2020}'s isothermal pressure with \citet{XieFu2025}'s thermal pressure. \textcolor{black}{We do not use the liquid iron EOS reported by \citet{LuoDorn2024} as it is only for Fe and hydrated Fe, whereas we want to use one source for our liquid Fe, FeO, and FeS EOS. This is because EOS parameters can vary significantly between studies and we want any differences in the core structure as we add light elements to the core to be caused primarily by the changing composition itself rather than the changing source publication.}

The liquid FeS EOS from \citet{IchikawaTsuchiya2020} is for Fe$_\mathrm{0.81}$S$_\mathrm{0.19}$, placing an upper bound on $x^\mathrm{Liquid}_\mathrm{S}$ of 0.19, as above this value there is not sufficient Fe to form the alloy whose EOS we use. Similarly, we cannot have $x^\mathrm{Liquid}_\mathrm{O}>0.5$ as it would require an EOS for free \textcolor{black}{O} in the liquid core that we do not have. As the latter would likely correspond to a denser mantle than core, it is unphysical. In contrast, the restriction on $x^\mathrm{Liquid}_\mathrm{S}$ rules out plausible values (see the required low values to accommodate a greater $x^\mathrm{Liquid}_\mathrm{O}$ in section \ref{sec:mars}).

We do not use the same liquid FeO EOS for the mantle and core because the low-pressure EOS of \citet{MorardAntonangeli2022} was derived for $P\lesssim120$ GPa and does not \textcolor{black}{smoothly intersect} with the EOS of \citet{IchikawaTsuchiya2020} derived for higher pressures. The structure of Earth's liquid core cannot be recovered by extrapolating \citet{MorardAntonangeli2022}'s low-\textcolor{black}{pressure} FeO EOS to high pressures (its $\frac{\partial r}{\partial \rho}$ is too large), motivating our switch between the EOS at the mantle-core boundary. As is the case for liquid Fe, the two EOS do not \textcolor{black}{smoothly intersect}, causing a density jump of ${\sim}7\%$ at 100 GPa and 5000 K.

\defcitealias{IchikawaTsuchiya2020}{IT20}
\defcitealias{XieFu2025}{XF25}
\defcitealias{KuwayamaMorard2020}{KM20}
\defcitealias{HaldemannDorn2024}{HD24}

\begin{table*}
    \caption{The EOS used within the liquid core. AE is short for anharmonic \& electronic. Vinet is short for Vinet-Rydberg. All references are abbreviated to the first letters of the last names of the first two authors, as follows: \citet{IchikawaTsuchiya2020} is \citetalias{IchikawaTsuchiya2020}, \citet{XieFu2025} is \citetalias{XieFu2025}, \citet{KuwayamaMorard2020} is \citetalias{KuwayamaMorard2020}, and \citet{HaldemannAlibert2020} is \citetalias{HaldemannDorn2024}. \textcolor{black}{If multiple citations are provided, the first citation is for the isothermal EOS and the second citation is the thermal EOS (in the case of liquid Fe) or the publication which refit the data (in the case of liquid FeO).}}
    \begin{center}
    \centerline{
    \begin{threeparttable}
\begin{tabular}{ccccccc}
    Compound&Phase&Isothermal EOS&Thermal EOS&AE EOS&$\gamma_{th}$&Reference\\\hline
        Fe & Liquid (P > 100 GPa) & Vinet & Linear & Quadratic & Constant & \citetalias{IchikawaTsuchiya2020,XieFu2025}\\
        Fe & Liquid (P < 100 GPa) & Vinet & Linear & Quadratic & $\gamma_\infty=0$ & \citetalias{KuwayamaMorard2020}\\
        FeO & Liquid & Vinet & Einstein & Quadratic & $\gamma_{\infty}=0$ & \citetalias{IchikawaTsuchiya2020,HaldemannDorn2024}\\
        FeS & Liquid & Vinet & Linear & Quadratic & Constant & \citetalias{IchikawaTsuchiya2020}
    \end{tabular}
    \end{threeparttable}}
    \end{center}
    \label{tab:liquid_core}
\end{table*}

The EOS used for the solid core are shown in Table \ref{tab:solid_core}. At high pressures within Earth and the cores of more massive planets, the density difference between Fe phases is $\lesssim$1\% and decreases with pressure, so our results are insensitive to the exact phase transitions assumed \citep{CottenierProbert2011,HakimRivoldini2018}. This justifies our non-inclusion of the reported phase transition of hcp \textcolor{black}{iron} back into fcc and bcc \textcolor{black}{iron} at $P\gtrsim8$ TPa \citep{PickardNeeds2009,CottenierProbert2011,Stixrude2012}. \citet{HakimRivoldini2018} calculate their Fe EOS without thermal effects and thus use the results of \citet{BouchetMazevet2013} to calculate thermal pressures\textcolor{black}{. In contrast,} we use the more recent results of \citet{ZhangZhang2025} to calculate the thermal pressure of Fe, which has \textcolor{black}{$\theta_D$} $\gtrsim$20 times larger than \citet{BouchetMazevet2013} due to the inclusion of experimental data in \citet{ZhangZhang2025} and approximations in the theory of \citet{BouchetMazevet2013}. We use Fit \#5 of \citet{ZhangZhang2025} because it is the lowest-residual fit using an isothermal Vinet formulation, more closely matching the isothermal formulation used in our model than the BM3 Fit \#1 with a slightly lower residual.

\citet{ZhangZhang2025} do not include an anharmonic/electronic term. We have found that including the anharmonic/electronic term of \citet{BouchetMazevet2013} with the thermal term of \citet{ZhangZhang2025} leads to $\nabla_\mathrm{ad}$ values near the pressure-temperature regime of Earth's core that are higher than the $\nabla_\mathrm{ad}$ values for liquid Fe at those same pressures and temperatures, leading to our solutions being \textcolor{black}{unstable around the iron melting curve}. This is because as our solutions move toward the planetary center, the crossing of the liquid-solid transition increases the temperature gradient, causing higher temperatures that return the solution to the liquid state. We thus do not use the anharmonic/electronic term of \citet{BouchetMazevet2013}. We found that combining the anharmonic/electronic term of \citet{DorogokupetsDymshits2017} with the thermal term of \citet{ZhangZhang2025} causes densities to increase with temperature at $P\gtrsim2$ TPa because \citet{DorogokupetsDymshits2017}'s $g$ is negative. As neither of these anharmonic/electronic terms are compatible with our formalism and to our knowledge no EOS with anharmonic/electronic terms derived using data up to a TPa exists besides that of \citet{BouchetMazevet2013}, we include no anharmonic/electronic term for \textcolor{black}{hcp Fe}.

The closed-form EOS provided by \citet{HakimRivoldini2018} is only valid above 234.4 GPa, necessitating that the EOS of \citet{FeiMurphy2016} be used below that pressure. We use the thermal expression from \citet{ZhangZhang2025} and no anharmonic/electronic expression for \citet{FeiMurphy2016}'s data for consistency with our update to \citet{HakimRivoldini2018}'s data. \citet{HakimRivoldini2018}'s EOS was designed to cleanly match \citet{FeiMurphy2016}'s EOS, but our update to the thermal terms slightly breaks this clean meeting, introducing an artificial density jump (${\sim}$0.5\% at 5500 K). As the artificial density jump is positive, our model still produces a gravitationally stable interior structure.

We use a single EOS for FeS as we prioritize the superior high-pressure extrapolation of the Holzapfel form over the accuracy of including a phase transition that occurs at ${\sim}$180 GPa, a lower pressure than the ${\sim}$330 GPa at which Earth's core becomes solid and thus a pressure unlikely to be encountered in the solid cores of super-Earths or sub-Neptunes \citep{OhfujiSata2007,SataOhfuji2008}. Only six data points in the FeS fit are for the higher\textcolor{black}{-}pressure Fe\textcolor{black}{S} VII, so our fit is dominated by the lower\textcolor{black}{-}pressure Fe\textcolor{black}{S} VI even though the core will be composed mostly of Fe\textcolor{black}{S} VII, however, the authors are unaware of a study with sufficient measurements of Fe\textcolor{black}{S} VII to remedy this issue \citep{SataHirose2010}. The FeS VI/VII data were only collected at one temperature and thus have no thermal fit \citep{OhfujiSata2007,SataOhfuji2008,SataHirose2010}, so we calculate $\nabla_\mathrm{ad}$ for the solid core as if it is pure Fe. This is motivated by the fact that although the density of Earth's solid core is highly discrepant with that expected for pure hcp \textcolor{black}{iron}, its $K_S$ is in line with that of a pure hcp \textcolor{black}{iron} core \citep{DorogokupetsDymshits2017}.

\textcolor{black}{}

\defcitealias{HakimRivoldini2018}{HR18}
\defcitealias{ZhangZhang2025}{ZZ25}
\defcitealias{FeiMurphy2016}{FM16}
\defcitealias{DorogokupetsDymshits2017}{DD17}
\defcitealias{SataHirose2010}{SH10}
\defcitealias{SataOhfuji2008}{SO08}
\defcitealias{OhfujiSata2007}{OS07}
\begin{table*}
    \caption{The EOS used within the solid core. AE is short for anharmonic \& electronic. Vinet is short for Vinet-Rydberg. BM3 is short for third-order Birch-Murnaghan. Debye is short for Mie-Grüneisen-Debye. hcp is short for hexagonal close-packed. bcc is short for body-centered cubic. fcc is short for face-centered cubic. FeS undergoes a phase transition from VI to VII at ${\sim}$180 GPa \citep{OhfujiSata2007,SataOhfuji2008,SataHirose2010}, but the EOS used fits all FeS data with one form. All references are abbreviated to the first letters of the last names of the first two authors, as follows: \citet{HakimRivoldini2018} is \citetalias{HakimRivoldini2018}, \citet{ZhangZhang2025} is \citetalias{ZhangZhang2025}, \citet{DorogokupetsDymshits2017} is \citetalias{DorogokupetsDymshits2017}, \citet{SataHirose2010} is \citetalias{SataHirose2010}, \citet{OhfujiSata2007} is \citetalias{OhfujiSata2007}, \citet{SataOhfuji2008} is \citetalias{SataOhfuji2008}. \textcolor{black}{If multiple citations are provided, the rightmost citation is for the thermal EOS and all citations to the left of it are for the isothermal EOS.}}
    \begin{center}
    \centerline{
    \begin{threeparttable}
\begin{tabular}{ccccccc}
    Compound&Phase&Isothermal EOS&Thermal EOS&AE EOS&$\gamma_{th}$&Reference\\\hline
        Fe & hcp (P > 234.4 GPa) & Holzapfel & Debye & None & $\gamma_\infty=0$ & \citetalias{HakimRivoldini2018,ZhangZhang2025}\\
        Fe & hcp (P < 234.4 GPa) & BM3 & Debye & None & $\gamma_\infty=0$& \citetalias{FeiMurphy2016,ZhangZhang2025}\\
        Fe & bcc & Vinet & Einstein & Quadratic & $\gamma_\infty=0$ & \citetalias{DorogokupetsDymshits2017}\\
        Fe & fcc & Vinet & Debye & Quadratic & $\gamma_\infty=0$ & \citetalias{DorogokupetsDymshits2017}\\
        FeS & VI/VII & Holzapfel & Einstein & None & $\gamma_\infty=0$ & \citetalias{OhfujiSata2007,SataOhfuji2008,SataHirose2010,HakimRivoldini2018,ZhangZhang2025}\\
        FeO & \multicolumn{6}{c}{See Table \ref{tab:solid_mantle}}
    \end{tabular}
    \end{threeparttable}}
    \end{center}
    \label{tab:solid_core}
\end{table*}

In the high core pressure regimes of planets with masses greater than Earth, the EOS of \citet{HakimRivoldini2018} dominates. \citet{HakimRivoldini2018} finds high-pressure densities using density functional theory that are higher than predicted from extrapolations from low-pressure experimental results and uses the EOS formulation of \citet{Holzapfel1996} that is more suited to high pressures. These effects combine to generate radii ${\sim}6\%$ smaller than \citet{ZengJacobsen2019} for pure Fe bodies. However, the introduction of light elements into the core counteracts this effect, increasing planetary radii. Allowing the core to melt also increases planetary radii compared to the assumption of a solid interior. Our inclusion of low-pressure Fe phases increases the radii of small planets with large cores but has no effect on larger planets ($\gtrsim 1 M_\oplus$) whose entire solid cores' iron are in the hcp state.

\subsection{Thermal Structure} \label{sec:thermal}
Our model divides the planet into two thermal layers: an irradiated layer in which a temperature profile derived from fits to radiative transfer equations \textcolor{black}{is} used and a non-irradiated layer in which the temperature profile is radiative or convective. The irradiated layer extends from the outer boundary of the planet to the high optical depth ($\tau \gg 1$) region where no more stellar flux is absorbed, which we approximate as an optical depth of $\tau=1000$.  \textcolor{black}{The irradiated layer is similar to an atmospheric layer, as it is composed entirely of non-condensed H/He or water at the outer boundary of the planet. However, gas at an optical depth $\tau>1000$ is within the non-irradiated layer. This model makes no explicit differentiation between an atmosphere and an interior--the entire structure, interior and atmosphere included, is treated within the same framework.}

We calculate $\tau$ using Eq.~\ref{eq:tau_from_kappa},
\begin{equation}
    \frac{\partial \tau}{\partial m} = -\frac{\kappa(P,T,Z)}{4\pi r^2}\mathrm{,}
    \label{eq:tau_from_kappa}
\end{equation}
which arises from the definition of $\tau$ ($d\tau\equiv-\rho\kappa dz$) \citep{Guillot2010} \textcolor{black}{combined with Eq.~\ref{eq:mass_cons}} where $\kappa$ is the \textcolor{black}{Rosseland mean} opacity. \textcolor{black}{We calculate $\frac{d\tau}{dm}$ as a function of $r$ and do not assume that a universal surface gravity applies throughout the atmosphere. If a region of \textcolor{black}{solid} matter is reached while integrating inwards before $\tau$ reaches $1000$, $\tau$ is set to $1001$, indicating that the irradiated layer immediately ends.}% \textcolor{black}{We employ the} plane-parallel assumption\textcolor{black}{,} justified by the irradiated atmosphere composing a relatively small fraction of the total planetary radii for \textcolor{black}{most planets without significant H/He envelopes}, for example, the irradiated atmosphere only comprises 0.005\% of the radius of an Earth-mass planet comprised of half water and half Earth-like composition with an equilibrium temperature of 1000 K. 

Within the irradiated atmospheric layer, we calculate the temperature profile using the fits provided by \citet{ParmentierGuillot2014,ParmentierGuillot2015}'s model D fit to radiative transfer models \textcolor{black}{for a plane-parallel atmosphere}, which rely on values for $\kappa$, the equilibrium temperature from stellar irradiation \textcolor{black}{$T_\mathrm{eq0}\equiv(F_p/\sigma_\mathrm{SB})^{0.25}$ where $\sigma_\mathrm{SB}$ is the Stefan–Boltzmann constant}, the intrinsic luminosity of the planet $L_\mathrm{int}$, and the bond albedo of the planet $A_B$. The globally-averaged effective temperature of a planet \textcolor{black}{(a parameterization of the irradiated atmosphere's energy budget)} is given by $T_\mathrm{eff}=((1-A_B)T_\mathrm{eq0}^4+T_\mathrm{int}^4)^{0.25}$, where $T_\mathrm{int}$ is the intrinsic temperature arising from the intrinsic luminosity given by the Stefan-Boltzmann law as $T_\mathrm{int}=(L_\mathrm{int}/(4\pi r^2\sigma_\mathrm{SB}))^{0.25}$.  \textcolor{black}{$T_\mathrm{eq0}$ is the outer boundary temperature planet if the planet had no intrinsic luminosity or albedo ($L_\mathrm{int}=0$, $A_B=0$) and corresponds to the $T_\mathrm{eq}$ typically reported in observation papers.  However, we emphasize that it is merely a convenient parameterization of the incident stellar flux and its use does not imply that the planet has achieved thermal equilibrium.}

\textcolor{black}{The outer boundary temperature $T_\mathrm{out}$ (see section \ref{sec:methods_overview}) is the skin temperature given by \citet{ParmentierGuillot2014,ParmentierGuillot2015} as a function of the above parameters as calculated at an optical depth of 0. If a planet is purely condensed and has no irradiated layer, its outer boundary temperature is $T_\mathrm{eff}$.}

\textcolor{black}{Objects with extended atmospheres and low masses have diverging $r_\mathrm{out}$ (the radius of the outermost layer of the model, not to be confused with the transit radius of the planet, see section \ref{sec:transit}) with decreasing mass, leading to irradiated atmospheres that comprise a significant fraction of the total planetary radius, invalidating our plane-parallel assumption. This excludes the application of our model to planets that simultaneously have both low masses ($\lesssim2M_\oplus$) and high H/He mass fractions ($\gtrsim5$\%). That said, no well-characterized planets lie in this region of mass-radius space, so this cutoff does not in any way limit the science application of this model.}

\textcolor{black}{T}hese fits are only valid for surface gravities between 2.5 and 250 m/s\textsuperscript{2} and effective temperatures between 100 and 3000 K.  \textcolor{black}{Our models for atmospheres are thus valid for planetary conditions within these two regimes. The gravity limits do not impact the validity of our model, as the low gravity boundary is already imposed by the plane-parallel assumption (see above) and the high gravity boundary is not encountered for any object under consideration.} We caution that these analytical solutions were derived around a sun-like star and thus are not strictly valid for planets with host stars with significantly different radiation profiles \citep{ParmentierGuillot2014,ParmentierGuillot2015}.

We assume that $\kappa$ arises from a \textcolor{black}{chemical} equilibrium composition of metals \textcolor{black}{for the local pressure and temperature, $P(m)$ and $T(m)$.} \textcolor{black}{We use the $P-T-Z$ Rosseland mean opacity grids provided by \citet{FreedmanLustig-Yaeger2014} to calculate $\kappa(P,T,Z)$ within our planetary atmosphere, both in the water (steam) and H/He layers.} \textcolor{black}{The spectrum of the incident light in the calculation of the Rosseland mean opacity is assumed to completely arise from the host star, assumed to be a blackbody with $T_*=6000$ K, the closest temperature to the \textcolor{black}{S}un provided in the tables of \citet{FreedmanLustig-Yaeger2014}.} \textcolor{black}{Water is the dominant contributor to opacity at temperatures $\lesssim$1000 K, but neutral atomic alkalis become increasingly important at higher temperatures until TiO and VO begin to dominate at $\gtrsim$2000 K \citep{FreedmanLustig-Yaeger2014}. Although the opacity of a steam atmosphere should be expected to be different from the opacity of a H/He atmosphere due to the presence of a different dominant background gas, considerations of variations in $\kappa$ due to differing chemical compositions in the atmosphere is left to future work.} \textcolor{black}{At pressures or temperatures beyond the table limits of \citet{FreedmanLustig-Yaeger2014}, we use the values at the maximum pressure or temperature within the table, that its, $\kappa(P>P_\mathrm{max})=\kappa(P_\mathrm{max})$ and $\kappa(T>T_\mathrm{max})=\kappa(T_\mathrm{max})$. $T_\mathrm{max}=4000$ K and $P_\mathrm{max}=0.03$ GPa for the high temperatures inside planetary interiors.}

Observations of exoplanetary upper atmospheric metallicity as probed by H\textsubscript{2}O have revealed substantial scatter from ${\sim}$0.3 to ${\sim} 100 \times$ solar while solar system C/H spectroscopy and extrasolar interior structure models have given mass-metallicity relations implying $Z\gtrsim50$ for the $M<25M_\oplus$ planets \textcolor{black}{composing the bulk} of our sample \citep{KreidbergBean2014,ThorngrenFortney2016,EdwardsChangeat2023,GuillotFletcher2023,SwainHasegawa2024}. Keeping in mind the large uncertainties in observed and theoretical planetary atmospheric metallicities \textcolor{black}{as well as the fact that bulk metallicities are not necessarily reflective of envelope metallicities}, we assume \textcolor{black}{a constant} $Z=50Z_\odot$ ($[M/H]=1.7$) \textcolor{black}{for all models presented in this work}, the highest metallicity provided by the \citet{FreedmanLustig-Yaeger2014} tables and roughly in line with Uranus and Neptune's atmospheres. \textcolor{black}{$Z$ is unlikely to be truly constant throughout a planetary atmosphere, with metallicity increasing towards the interior \citep[e.g.][]{OrmelVazan2021,Piaulet-GhorayebThorngren2025,SurTejadaArevalo2025}, but this is a consideration for future work.}

\textcolor{black}{W}e calculate $L_\mathrm{int}$ following the analytical fits provided by \citet{Mordasini2020}\footnote{By incorporating \textcolor{black}{internal energy} into our outer temperature boundary condition, we forego the need to explicitly include the energy conservation equation in our solution, i.e we set $\epsilon=0$ in $\frac{\partial L}{\partial m}=\epsilon$.}. These analytical fits include intrinsic heating generated by radiogenic luminosity as well as the cooling and contraction of the core and envelope \citep{MordasiniAlibert2012,LinderMordasini2019,Mordasini2020}. \textcolor{black}{The input parameters to these fits are the mass, composition, and age of the planet \citep{Mordasini2020}. This study thus accounts for the contraction of cooling planets with time through the decrease of $L_\mathrm{int}$ with planetary age without the use of an explicit dynamical evolution model.}

\textcolor{black}{A 13$M_\oplus$ planet with a 99\% Earth-like and 1\% H/He by mass composition with a zero-albedo equilibrium temperature of 1565 K (an incident flux of $1000F_\oplus$) has a radius \textcolor{black}{23.9\%} larger at an age of 100 Myr than at an age of 10 Gyr (see \textcolor{black}{Figure \ref{fig:mr_curve_varyage}}). This is a larger difference than the 13.3\% found by \citet{LopezFortney2014} for an object with the same parameters. An exact replication is not anticipated, primarily because this work is a static interior structure model that does not explicitly evolve the planetary interior as done by \citet{LopezFortney2014} and secondarily because our studies use different atmospheric opacities, equations of state, and thermal profiles. Nonetheless, our overall agreement with \citet{LopezFortney2014} (especially in the nominal case of 4.5 Gyr ages, see Fig.~\ref{fig:mr_curve_varyage}) shows that our model is capable of accounting for the evolution of planetary radii with age.}

These fits are derived for a smaller parameter space than explored in this study ($1<\frac{M}{M_\oplus}<40$,  $w_\mathrm{H_2O} < 0.1$) so inaccuracies due to extrapolation may occur at high masses or water mass fractions. \citet{Mordasini2020}'s fits underestimate Jupiter's modern luminosity by a factor of $2.78$, indicating that even far outside the fitted parameter space the estimates are the correct order of magnitude. On the other end of the mass spectrum, \citet{Mordasini2020}'s estimate for modern Earth's luminosity is 0.6 times its actual value \citep{KamlandCollaborationGando2011}, indicating that these fits may not be accurate for old planets with low envelope masses. These planets have low internal luminosities, making the impact of this effect minor.   \citet{HaldemannDorn2024} find that variations in the parameters reported by \citet{Mordasini2020} tend to result in $\lesssim1\%$ changes in exoplanetary radii.

Additionally, our prescription for temperature jumps (see below) means that the thermal structures of rocky planet interiors are dominated by the melting temperatures of their outermost mantle rather than their equilibrium temperatures \textcolor{black}{(as long as their equilibrium temperature is below their mantle melting temperature)}, so small changes in the effective temperature of rocky planets without magma oceans have a vanishingly small impact (of order $10^{-9}$ for modern Earth).

We do not use the albedos calculated by the radiative \textcolor{black}{transfer} model of \citet{ParmentierGuillot2014,ParmentierGuillot2015} as they do not incorporate clouds, instead we assume an Earth-like $A_B$ of 0.3 \citep{StephensO'Brien2015}, which is also within 0.05 of the $A_B$ of Mars \citep{StatellaPina2015}, Titan \citep{LiNixon2011}, Uranus \citep{IrwinWenkert2025}, and Neptune \citep{PearlConrath1991}, while being consistent with ${\sim}$half of hot Jupiters for which measurements exist \citep{FortneyDawson2021}\footnote{As well as the out-of-date values for Jupiter by \citet{HanelConrath1981} and Saturn by \citet{HanelConrath1983} that have recently been revised upward by \citet{LiJiang2018} and \citet{WangLi2024}, respectively.}.

Usage of this internal luminosity prescription relies on the assumption that planets have mantle abundances of the key long-lived radiogenic isotopes \textsuperscript{40}K, \textsuperscript{238}U, and \textsuperscript{232}Th matching Earth's. Stellar abundances of \textsuperscript{232}Th vary by a factor of two in solar twins while the abundances of the other isotopes have yet to be directly measured, implying that radiogenic heating could differ by a factor of a few from this estimate, especially considering that K\textcolor{black}{'s abundance} on Earth is depleted by a factor of five relative to solar for unknown reasons \citep{UnterbornJohnson2015,BotelhoMilone2019,UnterbornSchaefer2020}. We emphasize that our results are relatively insensitive to a factor of a few difference in radiogenic heating.

\textcolor{black}{As a consequence of this temperature profile, we find that planets with outer boundary temperatures greater than $\sim300$ K, $w_\mathrm{H\textsubscript{2}O}>0$, and $w_\mathrm{H/He} = 0$ have extended steam atmospheres and no layer of liquid water\footnote{In some high-mass planets, our model recovers thin sheets of liquid water at high planetary radii immediately between an overlying steam layer and underlying steam layers. This is a consequence of our non-treatment of rain-out.}, in agreement with the calculated boundaries wherein the runaway greenhouse effect occurs and thus steam atmospheres \textcolor{black}{form} found by \citet{NakajimaHayashi1992,KopparapuRamirez2013a,KopparapuRamirez2013b,TurbetEhrenreich2019}.}

Within the interior layer of $\tau > 1000$, the temperature gradient $\nabla$ is calculated following the standard criterion
\begin{equation}
    \nabla=\max{\left( \frac{3}{16\pi ac G}\frac{\kappa L_\mathrm{int}P}{mT^4},\nabla_\mathrm{ad} \right)}\mathrm{,}
\end{equation}
where $a$ is the radiation constant, $c$ is the speed of light and $\nabla_\mathrm{ad}$ is the adiabatic temperature gradient (see \ref{sec:thermal_background}). The first term represents the radiative temperature gradient and the second term represents the adiabatic temperature gradient.

In the solid and liquid portion of the planet below these layers, we adopt an adiabatic temperature gradient. We impose a temperature jump to the mantle melting temperature at the top and bottom of the mantle (if the temperature would have otherwise been below that temperature) \textcolor{black}{ following the melting curves of Eq.~\ref{eq:melting_T} and Eq.~\ref{eq:MgSiO3_melting_T} for the planet's composition and local pressure}. This is because planetary interiors form molten and rapidly cool until the temperature of the outermost regions of a layer reach the melting temperature and solidify, making further cooling much more inefficient and thus leaving the temperature immediately below the boundary relatively constant at the melting temperature throughout time \citep{GaidosConrad2010,Stixrude2014}. This justification is only valid if the interior was above the melting temperature when it formed. Planetary interiors do cool with time, \textcolor{black}{so this prescription is only strictly valid for a particular moment in time. The detailed evolution of the interior temperature profile with time is beyond the scope of this work, as it requires the significant computational investment of calculating the interior profile for multiple time steps rather than at one time.} This temperature jump prescription means that the temperature profile below the atmosphere is nearly independent of the temperature profile of the atmosphere when the temperature at the base of the atmosphere is below the melting temperature of the mantle.

\subsection{Transit Radii}
\label{sec:transit}
We report our final planetary radii not as $r_\mathrm{out}$, corresponding to the outermost radial layer of a planet in our model \textcolor{black}{(where $P=100$ Pa)}, but rather as the radius at which a light ray traveling along a grazing chord would reach an optical depth of $\tau_\mathrm{chord}=\frac{2}{3}$. This approximates the radius actually obtained via transit exoplanet detections \citep{Guillot2010}. To do so, we follow \citet{Guillot2010} and use Eq.~\ref{eq:transit_radius}:
\begin{equation}
    \tau_\mathrm{chord}(r) = 2\int_0^\infty \rho \kappa\frac{z+r}{\sqrt{z^2+2rz}}dz\mathrm{,}
    \label{eq:transit_radius}
\end{equation}
where $r$ is the distance from the centre of the planet at which $\tau_\mathrm{chord}$ is measured and $z$ is a distance above $r$ with $r\parallel z$. 
For $r=r_\mathrm{trans}$, the transit radius of the planet, Eq.~\ref{eq:transit_radius} evaluates to $\frac{2}{3}$.
\begin{figure}
    \centering
    \includegraphics[width=1.\columnwidth]{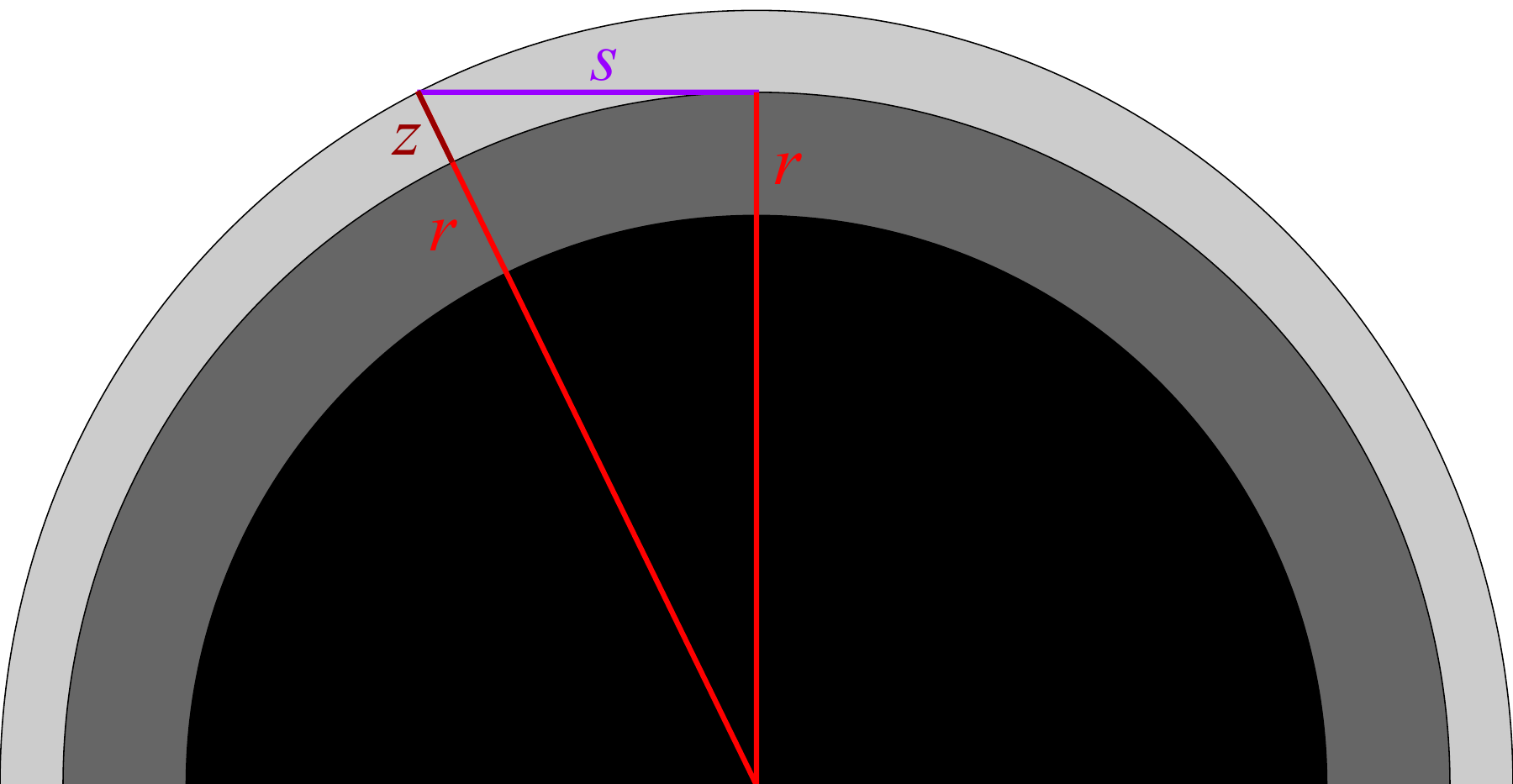}
    \vspace{-2em}\caption{A diagram showing the variables $r$ (light red lines) and $z$ (dark red line) in relation to $s$ (purple line), the length along which $\tau=\frac{2}{3}$. The darkest circle represents the solid surface of the planet, the lighter circle the transit radius of the planet, and the lightest circle the outermost layer of the atmosphere, defined here to be where $P=100$ Pa.}
    \vspace{-2em}\label{fig:geom_graph}
\end{figure}

We get $\rho(z)$ and $\kappa(z)$ from a valid solution to the interior structure equations \ref{eq:mass_cons}-\ref{eq:thermal_struct}. The chord over which the optical depth is measured, $s$, forms a right triangle with the radius of the planet at which the optical depth is measured, $r$, and $r+z$ is the hypotenuse of that triangle (see Fig.~\ref{fig:geom_graph}).

The iteration proceeds as follows: first guess a value $r_\mathrm{guess}=r$, then integrate Eq.~\ref{eq:transit_radius} from the outer pressure boundary, where $z$ is some high number approximating $\infty$, to the radius at which the total optical depth is $\frac{2}{3}$. When the total optical depth is $\frac{2}{3}$, $z$ should be $0$ and thus $r$ should be $r_\mathrm{trans}$. We thus minimize the $z$ for which $\tau_\mathrm{chord}=\frac{2}{3}$ via the Newton-Raphson method. We iterate guesses until $r_\mathrm{guess}$ and $r(\tau_\mathrm{chord}=\frac{2}{3})$ agree \textcolor{black}{to} within a factor of $10^{-4}$. \textcolor{black}{If $\tau_\mathrm{chord} < \frac{2}{3}$ (the atmosphere is found to be optically thin), the transit radius is instead the radius of the outermost layer of condensed matter in the planet.}

\textcolor{black}{A crucial caveat to this calculation of transit radii is that we assume an atmosphere that is free of clouds and hazes, whereas clouds and/or hazes are both theoretically predicted \citep[e.g][]{GaoZhang2020,EstrelaSwain2022} and commonly observed (with varying levels of degeneracy with metallicity) \citep[e.g.][]{KemptonZhang2023,RadicaCoulombe2024,ScarsdaleWogan2024,AlamGao2025} in the atmospheres of low-mass planets. Either clouds or hazes would result in a very high $\kappa$ surface high in the atmosphere, causing the transit radius to be much higher than that calculated here. To accommodate this fact, all mass-radius tables list not only the transit radius as calculated here and treated as fiducial throughout this work, but also \textcolor{black}{the radii at the pressures of} 100 and 2000 Pa, the radius at which $\tau=\frac{2}{3}$ for a chord of light going directly towards the planetary center (rather than grazing), and the radius at which the planet becomes condensed. Full interior structure profiles are provided to allow the extraction of the radii at any arbitrary pressure. If clouds and hazes are indeed a ubiquitous feature, the transit radii presented here are underestimates of the true radii for planets with extended atmospheres.}

\subsection{Rotation}\label{sec:rotation}
We do not assume that planets are perfectly spherical and employ an analytical approximation for the effects of planetary rotation by multiplying Eq.~\ref{eq:hydro_equil} by $f_P$ and Eq.~\ref{eq:thermal_struct} by $f_T/f_P$, where $f_P$ and $f_T$ are derived by \citet{PaxtonCantiello2013,PaxtonSmolec2019} for rigidly rotating bodies in hydrostatic equilibrium. \textcolor{black}{Although originally developed for stars, the assumptions underlying \citet{PaxtonCantiello2013,PaxtonSmolec2019}--hydrostatic equilibrium and rigid rotation--also hold for planets.} In this formalism, $r$ does not represent a physical radius and instead represents the radius of a sphere enclosing an equivalent volume to the volume within the physical mass enclosed $m$. $f_P$ and $f_T$ are functions solely of $\omega\equiv\Omega/\Omega_\mathrm{crit}$, where $\Omega$ is the rotation rate and $\Omega_\mathrm{crit}$ is the critical rotation rate. \textcolor{black}{$\omega$ is then solved for by plugging in the critical rotation rate.} The critical rotation rate depends on the equatorial rather than the equivalent volume radius and thus depends itself on $\omega$ via Eq.~\ref{eq:omega} \citep{PaxtonCantiello2013,PaxtonSmolec2019}, \textcolor{black}{where $r_e$ is the equatorial radius of the planet:}
\begin{equation}
    \omega=\Omega\sqrt{\frac{r_e^3}{Gm}}=\frac{\Omega r^\frac{3}{2}}{\sqrt{Gm}}(1+\frac{\omega^2}{6}-0.0002507\omega^4+0.06075\omega^6)^\frac{3}{2}\mathrm{.}
    \label{eq:omega}
\end{equation}
We solve Eq.~\ref{eq:omega} via rootfinding with the Newton-Raphson method. For a planetary rotation period of $\infty$ \textcolor{black}{($\omega=0$)}, $f_P=f_T=1$. \textcolor{black}{$f_P$ and $f_T$ monotonically decrease to $0$ at $\omega=1$, with $f_P$ smaller than $f_T$ at all times and decreasing more rapidly than $f_T$ for $\omega<0.916$.}
The effects of rotation are small but \textcolor{black}{noticeable}, varying a model Earth's \textcolor{black}{volume-equivalent} radii by ${\sim}0.02\%$ compared to a case where rotation is not taken into account.

\textcolor{black}{To verify the soundness of our methodology, we compared the rotationally-induced increase in equatorial radii in our model to that of \citet{HureNoe2025}. We found that the difference in equatorial radii between a 9.967$M_\oplus$, 42.3\% \textcolor{black}{mantle}, 57.7\% \textcolor{black}{water} (with 50\% MgO and 50\% SiO$_\mathrm{2}$) planet with a 5.05-hour rotation period and a 9.372$M_\oplus$, 43.8\% \textcolor{black}{mantle}, 56.2\% \textcolor{black}{water} (with 50\% MgO and 50\% SiO$_\mathrm{2}$) planet with an infinite rotation period was \textcolor{black}{4.7\%}, in contrast with the 6.3\% found by \citet{HureNoe2025}. The independent variables they use to define a planet are not the same as those used in our model, explaining the slightly varying mass and compositions of the 5.05-hour and infinite rotation period planets. Because of this, our smaller difference in equatorial radii is partially explainable by the EOS in our model having different compressibilities and thus reacting differently to a change in planetary mass than those used by \citet{HureNoe2025}. These different compressibilities lead to their work finding radii for planets of a given period \textcolor{black}{$\gtrsim$4}\% larger than ours due to differences in EOS.}

\textcolor{black}{The overall agreement of the effect of decreasing rotational periods coming from entirely different interior structure models lends credibility to our model's treatment of rotation. The smaller difference found in our model might be indicative of an underestimate of the effects of rotation in this work, but any such impact is convoluted with impact of our models using different EOS. Note that the impact of rotation on the volume-equivalent radius, $r$, is smaller than the impact of rotation on the equatorial radius, $r_e$, with the \textcolor{black}{4.7}\% increase in $r_e$ discussed above only corresponding to a \textcolor{black}{2.8}\% increase in $r$.}

\subsection{Brief Summary}
\label{sec:effects_summary}
We include much more physics than the basic model of a H/He envelope, a water layer, a MgSiO$_\mathrm{3}$ mantle, and a Fe core.
The inclusion of Fe in the planetary mantle systematically increases planetary densities, while the inclusion of S and O in the planetary core systematically decreases planetary densities. The inclusion of high\textcolor{black}{-}pressure phases in the mantle systematically increases densities for high\textcolor{black}{-}mass planets. The inclusion of temperature jumps at the core/mantle and mantle/core barriers, inclusion of melting in the mantle and core, and inclusion of thermal terms in our solid EOS all systematically decrease planetary densities. The new EOS we use for H/He and Fe are systematically denser than the majority of the literature. Directly calculating transit radii rather than the radii of a 100 Pa or 2000 Pa surface also systematically decreases planetary radii. \textcolor{black}{The inclusion of internal luminosity decreases planetary density \textcolor{black}{at} low ages while the inclusion of rotation decreases planetary density at short rotation periods.} The interplay of these systematic increases and decreases from different sources cannot be determined a priori, motivating the simultaneous inclusion of these effects in this model. 

\section{Model Validation} \label{sec:validation}
The internal composition of a body can be obtained via seismology by measuring the time that mechanical waves, such as those generated in quakes, take to move through \textcolor{black}{it} \citep{MoulikEkstrom2025a}. Seismology can also measure the eigenfrequencies of a planet (c.f spherical harmonics) \citep{MontagnerRoult2008}. These seismographic constraints as well as measured bulk properties such as masses, radii, moments of inertia, and love numbers can be used to constrain the interior compositions of planets \citep{GarciaKhan2019,MoulikEkstrom2025a}.

Seismometers have been installed on three bodies in the solar system, allowing for the detailed determination of their interiors: Earth, Mars, and the Moon. Three additional solar system bodies have non-seismographic but relatively strong constraints on their interior structures from their moments of inertia and additional geophysical arguments: Venus, Mercury, and Europa. Here we demonstrate our model's acceptable treatment of all of these bodies. We discuss Saturn, which has a H/He mass fraction greater than \textcolor{black}{that} of interest in this study, in appendix \ref{sec:saturn}.

We particularly note the importance of replicating planetary Moment of Inertia (MoI) coefficients ($C$, as appear\textcolor{black}{s} in $I=CMR^2$), which are sensitive to the mass distribution within planetary interiors. Solely focusing on reproducing planetary radii can lead to cases wherein a model that systematically overpredicts densities in one region and underpredicts them in another \textcolor{black}{appears} to perform well overall. However, such a model will not extrapolate well beyond the mass regimes of the planets that it was validated on (typically Earth).  

We derive our model $C$ for planets by numerically integrating Eq.~\ref{eq:MoI},
\begin{equation}
    C=\frac{8\pi}{3M_pR_p^2}\int_0^{R_{p}}r^4\rho(r)dr\mathrm{,}
    \label{eq:MoI}
\end{equation}
(derived by combining the differential MoI of a spherical shell and Eq.~\ref{eq:mass_cons}) with \texttt{scipy.integrate.trapezoid} using $\rho(r)$ from our calculated interior structure profile.
The smaller the value of $C$ is, the more mass is concentrated toward the planetary centre and thus the more differentiated the planet is. In the edge case of a constant-density sphere, $C=0.4$.

MoI and thus $C$ can be attained without \textcolor{black}{installing a seismometer on a body} either by \textcolor{black}{(1)} precisely measuring its precession (as for Venus \citep{MargotCampbell2021} and Mars \citep{KonoplivPark2020}), \textcolor{black}{(2)} using spacecraft-derived constants related to the shape of its gravitational field and measuring its libration ($C_{22}$ and $J_2$, as for the Moon \citep{WilliamsKonopliv2014}), or \textcolor{black}{(3)} by using the Darwin–Radau equation to infer $C$ from a spacecraft-derived constant related to the shape of its gravitational field and the relationships between MoI about different axis (as for Europa \citep{AndersonSchubert1998}; the constant is $C_{22}$; requires that the object is in the Cassini state, where the spin axis, orbit precessional axis, and normal to its orbital plane are all in the same plane \citep{PealePhillips2002}; the MoI about different axes are determined via flattening as in \citet{AndersonSchubert1998} or from obliquity as in \citet{GenovaGoossens2019}).

Throughout this section, we will be using the notation X (Y\%; Z) to notate a value of X with an error of Y from the real value ($Y\%=100(X_\mathrm{Model}-X_\mathrm{Observed})/{X_\mathrm{Observed}}$) that is Z $\sigma$ from the real value. We only list Z if the uncertainty in the true value is high, as is the case for the latter three bodies in our validation sample but not the first three. We emphasize that although we engage in detailed analysis throughout this section of any discrepancies between our model and reality, our model's overall high accuracy is attested to by the low values of these errors. The sources of the basic parameters of objects in our validation sample (mass, radius, moment of inertia coefficient, and rotational period) are discussed in appendix \ref{sec:constants}.
 
\subsection{Earth} \label{sec:earth}

\begin{figure*}
\centering     \includegraphics[width=\textwidth]{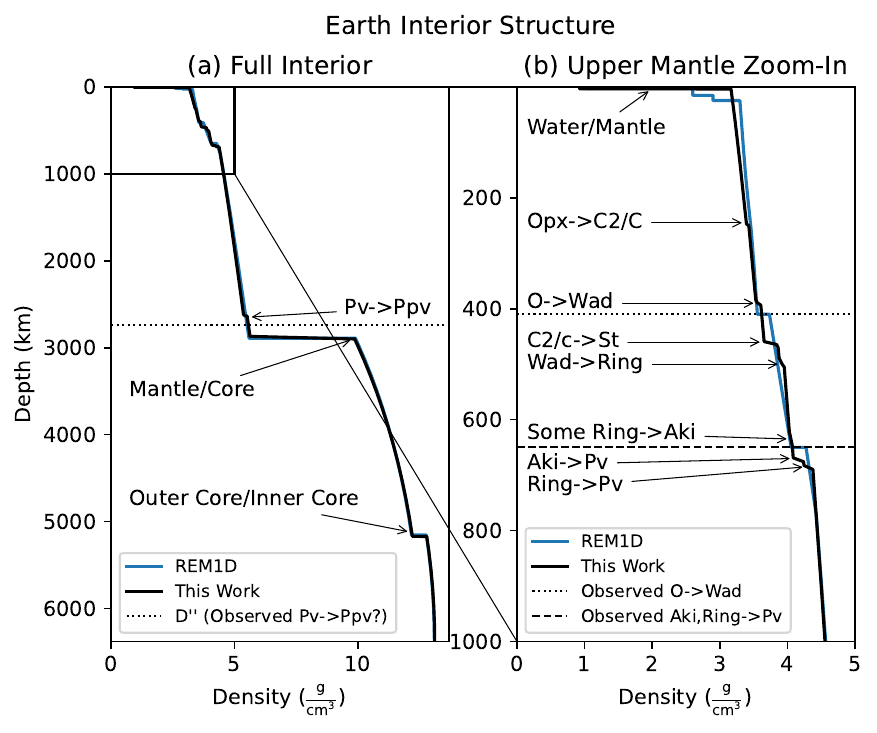}
\vspace{-2em}\caption{The internal r-$\rho$ profile of Earth as determined by the model in this work as compared to REM1D, a reference model for the Earth derived from seismology and other constraints \textcolor{black}{\citep{MoulikEkstrom2025a,MoulikEkstrom2025b}}. The box in the upper left of panel a is shown in more detail in panel b. Horizontal lines indicate the observed locations of phase transitions in Earth's interior (although note that the association of $D^{\prime\prime}$ with the Pv->Ppv transition is not universally accepted). More controversially observed phase transitions are discussed in the text. Annotations indicate phase transitions and boundaries between layers. For phase transitions, the format is lower pressure phase -> higher pressure phase. For boundaries between layers, the format is lower pressure layer/higher pressure layer. Multiple phases coexist throughout the mantle, the labelled phases represent a transition only for some subset of mantle material. Opx is short for Orthopyroxene. C2/c is short for Clinopyroxene. O is short for Olivine. Wad is short for Wadsleysite. St is short for Stishovite. Ring is short for Ringwoodite. Aki is short for Akimotoite. Pv is short for Perovskite. Ppv is short for Post-Perovskite. Note that this is for a spherically averaged Earth and both the density profile and mineralogy may change with latitude and longitude. All mantle EOS above the Pv->Ppv transition are from HeFESTo solved using Perple\_X \citep{Connolly2009,StixrudeLithgow-Bertelloni2024}. See text for in-depth discussion.}
\vspace{-2em}\label{fig:earth_structure}
\end{figure*}

The bulk composition of the silicate Earth (core+mantle)--also known as the primitive mantle (primitive referring to the mantle composition before the crust is differentiated)--is not exactly constrained \citep{PalmeOneill2014}. We follow \citet{PalmeOneill2014}'s literature review, which constrains $x_\mathrm{FeO}$, $x_\mathrm{MgO}$, and $x_\mathrm{SiO_2}$ from measurements of rocks from the upper mantle as well as cosmochemical abundances. We account for our non-inclusion of Ca, Al, and other elements by forcing the molar ratios of Mg/Si and Mg/Fe to reflect reality. The resulting fractions are $x_\mathrm{MgO}=0.512$, $x_\mathrm{SiO_2}=0.425$, and $x_\mathrm{FeO}=0.063$ (c.f $x_\mathrm{MgO}=0.521$, $x_\mathrm{SiO_2}=0.417$, and $x_\mathrm{FeO}=0.062$ from \citet{McDonoughSun1995}; $x_\mathrm{MgO}=0.528$, $x_\mathrm{SiO_2}=0.409$, and $x_\mathrm{FeO}=0.063$ from \citet{WorkmanHart2005}; $x_\mathrm{MgO}=0.519$, $x_\mathrm{SiO_2}=0.423$, and $x_\mathrm{FeO}=0.058$ from \citet{HaldemannDorn2024}).

We select the elemental abundances in the core to reflect the observed density profile, resulting in $x^\mathrm{Liquid}_\mathrm{O}=0.08$ and $x^\mathrm{Liquid}_\mathrm{S}=0.09$ in the liquid core and $x^\mathrm{Solid}_\mathrm{S}=0.115$ in the solid core\textcolor{black}{, with $x^\mathrm{Solid}_\mathrm{O}=0$ assumed to reduce the number of free parameters} (see section \ref{sec:core} for further discussion \textcolor{black}{and comparison to literature values}).

In Fig.~\ref{fig:earth_structure} we compare our model constructed using Earth's known composition to the interior of the Earth as reported in the one-dimensional Reference Earth Model (REM1D), a model representing the consensus of the seismographic community \citep{MoulikEkstrom2025a,MoulikEkstrom2025b}. REM1D is a spherically-averaged model of the interior structure of Earth derived from wave travel times, Earth's normal modes, and its bulk properties (e.g mass, radius, moment of inertia) \citep{MoulikEkstrom2025a,MoulikEkstrom2025b}. REM1D is an update to PREM by \citet{DziewonskiAnderson1981}. PREM has found extremely wide use in the geophysics community \citep{StaceyDavis2008}, with extrapolations from it being used in previous super-Earth mass-radius relations \citep{ZengSasselov2016,ZengJacobsen2019}. The $\frac{\chi^2}{N}$ (where $N$ is the number of parameters in the model) of REM1D's model Earth mass, moment of inertia, and $C$ compared to observations is 0.12, making it the only reference Earth model consistent with these parameters at the 1$\sigma$ level (c.f PREM 21 \citep{DziewonskiAnderson1981}, AK135 117 \citep{KennettEngdahl1995}, AK135F 195 \citep{MontagnerKennett1996}), motivating our adoption of it \citep{MoulikEkstrom2025b}. The difference between the densities of REM1D and PREM is small but nonzero (a maximum of $0.5$\%, generally much lower) \citep{MoulikEkstrom2025b}. REM1D reports the density of the Earth as a series of piecewise functions between several discontinuities (discussed below) \citep{MoulikEkstrom2025a}.

We find that our model Earth has a radius of 1.0015 $R_\oplus$ (0.15\% error v. actual Earth). \textcolor{black}{To more directly compare our results to the literature, we also consider the case of an Earth with simplified parameters wherein $w_\mathrm{H\textsubscript{2}O}=0$, $w_\mathrm{core}=0.325$, and rotation is neglected. In this case,} we get an Earth radius of \textcolor{black}{1.0008} $R_\oplus$ (0.0\textcolor{black}{8}\% error v. actual Earth). To the best of our knowledge, this is the closest replication of Earth's radius in any interior structure model built for application to exoplanets that is not explicitly constructed using Earth's \textcolor{black}{mantle density profile}\footnote{\citet{ZhangRogers2022} report a $<0.1\%$ discrepancy and cannot be correctly compared to this study as no exact number is quoted.} (see section \ref{sec:literature_comp} for more comparison to the literature). 

\textcolor{black}{We emphasize that although our model's core composition is constructed by direct comparison with seismology out of necessity,  all other parameters of this model were selected completely blind to Earth's seismographically-derived profile and thus this model's replication of Earth's interior is independent of the seismological models we compare it to. As the core is volumetrically much smaller than the mantle, the impact of mantle composition is much more significant tha\textcolor{black}{n} the impact of core composition. Our replication of Earth's radius thus arises from the state-of-the-art physics within our model rather than any attempt to fit for it.} \textcolor{black}{In testing the accuracy that this model could achieve if it were to explicitly attempt to replicate $R_\oplus$, we have found that varying the core compositional values within reported uncertainties in the literature (see section \ref{sec:core}) can result in values of $R_p$ arbitrarily close to $R_\oplus$.} It is thus clear that our model's replication of Earth's radius is nearing the level of accuracy at which uncertainty in its input parameters dominate\textcolor{black}{s}.

We also find that our model gives $C=0.33056$, a tiny \textcolor{black}{-0.048}\% error compared to Earth \citep{MoulikEkstrom2025a}. If we used the \textcolor{black}{simplified} values typical of the literature of $w_\mathrm{core}=0.325$ and $w_{H_2O}=0$ with no effect of rotation, we would have a model Earth MoIC of $0.33083$ (\textcolor{black}{0.035}\%). For comparison, the Preliminary Earth Reference Model (PREM) has an Earth $C=0.33090$ (0.055\%) \citep{DziewonskiAnderson1981}. 

Our MoIC is thus of comparable accuracy to that derived from seismology. This is important because our forward model parameters were constructed independently of seismology wherever possible. The uncertainty in the measurement of Earth's MoIC itself is $0.007\%$, dominated by uncertainty in the value of G \citep{MoulikEkstrom2025a}. The error in our replication of the distribution of Earth's interior structure is thus only an order of magnitude from its theoretical floor. We emphasize that this was achieved with no explicit attempts to fit this variable.

We now compare the density jumps in our model to those observed \textcolor{black}{i}n Earth. Density jumps arise from phase and composition transitions.  Rapid increases in density in REM1D at less than 25 km correspond to the transition from the water layer to the mantle, modeled here, and the crust, not modeled here. The largest density jump corresponds to the transition between Earth's mantle and core and great agreement can be observed between our model and REM1D. The second largest density jump at a greater depth corresponds to the transition between Earth's liquid outer core and solid inner core, where we again find agreement between our model and REM1D (although note that our melting temperature of Fe was selected for this purpose, see section \ref{sec:core}).

Earth's mantle has two widely-accepted density jumps included in REM1D \citep{MoulikEkstrom2025a,MoulikEkstrom2025b}. The first is at a depth of 410 km and corresponds to the transition between (Mg,Fe)$_2$SiO$_4$ olivine and (Mg,Fe)$_2$SiO$_4$ wadsleyite \citep{Helffrich2000,CormierLithgow-Bertelloni2023,MoulikEkstrom2025b}. The second is at a depth of 650 km and corresponds to the transition between between (Mg,Fe)$_2$SiO$_4$ ringwoodite (wadsleyite transitions to ringwoodite at intermediate depths but this does not cause a large density jump, see below) and (Mg,Fe)SiO$_3$ perovskite \citet{Helffrich2000,ShimDuffy2001,CormierLithgow-Bertelloni2023,MoulikEkstrom2025b}. Both of these transitions are recovered in our model, with the 410 km jump at 390 km and the 650 km jump at 685 km.

Between these two jumps, our model finds three additional mantle density jumps not included in REM1D. The shallowest and largest at 460 km is caused by SiO\textsubscript{2} transitioning from the coesite to stishovite phase. \textcolor{black}{This transition has been identified as the possible cause of a seismic discontinuity} observed at varying depths around 300 km in some locations known as the X-discontinuity \citep{RevenaughJordan1991,KempJenkins2019,SrinuKumar2021}. We invoke our model's non-treatment of the Earth's crust, our simplified chemical inventory, and Earth's heterogeneous interior as potential explanations for the 160 km difference between our model's coesite to stishovite transition and the X-discontinuity. However, a small difference in density over ${\sim}2.5\%$ of Earth's radius is negligible.

The next deepest upper mantle density jump at 500 km is associated in our model with the phase transition from wadsleyite to ringwoodite, which must occur between the 410 km olivine to wadsleyite transition and 650 km ringwoodite to perovskite transition to supply the ringwoodite for that transition. A jump at this depth is found in some regions on Earth \citep{Shearer1996,CormierTian2020,TianLv2020,ZhangSchmandt2022}.

The deepest of these jumps occurs at a depth of 635 km and arises in our model as a consequence of the appearance of (Mg,Fe)SiO$_3$ akimotoite. \citet{CormierLithgow-Bertelloni2023} have found that heterogeneities in Earth's mantle peak at around this depth and attribute this to the topography of a phase transition (i.e waves are bouncing off \textcolor{black}{the} surface of \textcolor{black}{a} phase transition with a depth that varies with latitude and longitude due to varying composition and temperature profiles).

Our model contains one mantle density jump above the 410 km jump at 245 km due to the transition of (Mg,Fe)SiO$_3$ from the opx to C2/c phase \citep{Woodland1998,StixrudeLithgow-Bertelloni2005}. This location is consistent with the ${\sim}$300 km X-discontinuity observed at varying depths at some locations, however, the opx to C2/c transition is unlikely to be detectable with current technology \citep{KempJenkins2019}. Additionally, the depth variability of the X-discontinuity indicates that it is likely caused either by a non-phase-transition mechanism or the transition of SiO\textsubscript{2} from the coesite to stishovite phase (see above), so this is likely coincidental \citep{RevenaughJordan1991,KempJenkins2019,SrinuKumar2021}.

Our deep mantle contains a density jump at 2650 km associated with the transition of (Mg,Fe)SiO$_3$ perovskite to (Mg,Fe)SiO$_3$ post-perovskite that is not included in REM1D. Although no 2650 km density jump appears to exist within Earth's interior, this depth is near the top of the well-attested $D^{\prime\prime}$ layer, a region where the slope of wave velocities with depth abruptly changes \citep{MoulikEkstrom2025b}. This change in velocities has been associated with the perovskite to post-perovskite phase transition, in agreement with our results \citep{TsuchiyaTsuchiya2004}. However, it also has been argued that this change could also result from a magma ocean at the base of Earth's mantle that either still exists or has since solidified and left behind a layer enriched in light elements \citep{LabrosseHernlund2007,HuDeng2024}.

The lack of a\textcolor{black}{n observed density jump corresponding to the $D^{\prime\prime}$ layer} despite the phase transition between perovskite and post-perovskite could be the result of chemical stratification (whose existence is evidenced by the lowermost mantle's changing value of $K^\prime$) that depresses densities in Earth's lowermost mantle \citep{MoulikEkstrom2025b}, a phenomenon not captured by our model's uniform elemental (not not species) composition. Post-perovskite could also not be a universal feature in the lowermost mantle and only appear at certain locations with the right temperatures and compositions \citep{Houser2007,MoulikEkstrom2025b}.

It is also possible that post-perovskite does not form in Earth's lowermost mantle or forms less readily than our model predicts for one of two reasons. First, elements such as \textcolor{black}{aluminum} that are not considered here could push the pressure at which a system becomes pure post-perovskite sufficiently high that the transition does not occur in Earth's interior \citep{CatalliShim2009,StixrudeLithgow-Bertelloni2024}. Second, the increase in temperature in our model between the mantle and the core happens instantaneously at the top of the core, whereas in reality the lowermost regions of the mantle have a super-adiabatic temperature curve. This makes it more difficult for post-perovskite to form, potentially causing a double transition in which post-perovskite forms at some depth and then returns to perovskite at even greater depths \citep{HernlundLabrosse2007}.

To conclude, our model replicates Earth's radius and $C$ from its known mass and composition with extreme accuracy (<0.2\%). It also includes jumps in density at the correct locations for Earth's mantle-outer core and inner core-outer core boundary as well as the well-attested phase transition depths between olivine and wadsleysite, and ringwoodite and perovskite. Additional phase transitions that have been reported (but are not confirmed) within Earth's interior--such as between wadsleysite and ringwoodite, and perovskite and post-perovskite--are predicted in our model at locations similar to those predicted by seismology.

\subsection{Mars} \label{sec:mars}
\begin{figure*}
\centering
\includegraphics[width=\textwidth]{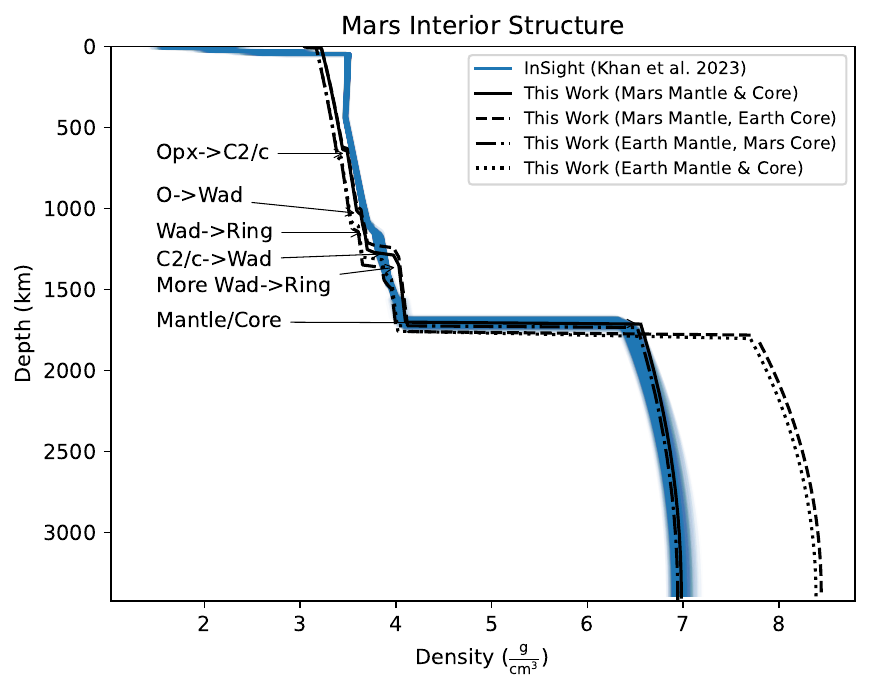}
\vspace{-2em}\caption{The internal r-$\rho$ profile of Mars as determined by the model in this work compared to the inversion of InSight seismic data from \citet{KhanHuang2023}. \citet{KhanHuang2023} provide 1000 models consistent with Martian observations, all of which are plotted with some transparency such that darker shades of blue represent regions where more models agree. We plot four Mars models with each possible combination of Mars/Earth Mantle/Core \textcolor{black}{elemental} abundances (see text). See Fig.~\ref{fig:earth_structure} caption for abbreviations and formatting of annotations. Annotations are for the fiducial Mars Mantle \& Core Case (Case 1).}
\vspace{-2em}\label{fig:mars_structure}
\end{figure*}

We compare our model Mars to Mars' \textcolor{black}{density} profile derived from the InSight seismometer--which landed on the Martian surface in 2018--in Fig.~\ref{fig:mars_structure} \citep{BanerdtSmrekar2020,KhanHuang2023}. We present a profile (1) using a mantle mineralogy with a composition derived from Martian meteorites following \citet{YoshizakiMcDonough2020} ($x_\mathrm{MgO}=0.444$, $x_\mathrm{SiO_2}=0.438$, and $x_\mathrm{FeO}=0.118$, broadly consistent with other estimates of Mars' mantle composition such as \citet{KhanSossi2022}, see review by \citet{KuskovKronrod2024}) and a Martian core with abundances selected to match InSight's results ($x_\mathrm{S}^\mathrm{Liquid}=0.03$, $x_\mathrm{O}^\mathrm{Liquid}=0.42$)\footnote{As our EOS for liquid FeS is for Fe\textsubscript{81}S\textsubscript{19}, high values of $x_\mathrm{S}^\mathrm{Liquid}$ deplete all free Fe. Thus, for planets with high \textcolor{black}{core light element abundances} like Mars, $x_\mathrm{S}^\mathrm{Liquid}$ must be lowered to accommodate high $x_\mathrm{O}^\mathrm{Liquid}$.}; (2) using a Martian mantle composition and core abundances fit to Earth's core; (3) using Earth's mantle composition and core abundances fit to Mars' core; and (4) using Earth's mantle composition and core abundances. In all cases, we use a Martian core mass fraction from \citet{KhanHuang2023} of 0.21. This is lower than the previously\textcolor{black}{-}accepted value of 0.25 as in \citet{KhanSossi2022,LeMaistreRivoldini2023} due to \citet{KhanHuang2023}'s discovery of a molten layer at the base of the Martian mantle that was previously identified with Mars' liquid outer core.

We calculate Martian radii of 0.5333$R_\oplus$ (0.24\%), 0.5288$R_\oplus$ (-0.60\%), 0.5371$R_\oplus$ (0.95\%), and 0.5327$R_\oplus$ (0.13\%), respectively. We get Martian Moment of Inertia coefficients of \textcolor{black}{0.36209 (-0.361\%), 0.35505 (-2.296\%), 0.36212 (-0.353\%), and 0.35549 (-2.176\%)}, respectively. In all cases, we find an entirely liquid core--as is currently understood to be the case for Mars \citep{KhanHuang2023,LeMaistreRivoldini2023}--and a core radius near the results from InSight (for the fiducial case 1, it is fully consistent with seismographic results). Mars' lack of a magnetic field despite its entirely liquid core has been explained by a lack of convection within its interior and a stratification between H-rich and S-rich layers of the core, effects not included within our model \citep{HemingwayDriscoll2021,YokooHirose2022,YokooHirose2024}. We do not recover a molten liquid mantle layer as our model's simplified prescription for the temperature jump at the mantle-core boundary does not increase temperatures in the lowermost mantle as occurs in reality.

It is of interest that using the chemical composition of Earth reproduces Mars' radius better than using the chemical composition of Mars. An investigation of the MoIC reveals that using Mars' chemical composition yields an order of magnitude more accurate result, indicating that the replication of Mars' mantle when using Earth's composition is the result of a systematic underprediction in mantle density being counteracted by a systematic overprediction in core density.

The light element fractions we require in the Martian core correspond to a core Fe weight abundance of $80\%$, ${\sim}5\%$ lower than the lower bound provided by \citet{KhanHuang2023}. This is likely a product of our limited chemical inventory as well as a simplified mantle structure leading to an inaccurate outer pressure boundary condition for the core. The density of our model's core moving with increasing depth from being on the denser to lighter end of \citet{KhanHuang2023}'s retrievals is a potential sign of our inaccurate elemental abundances.

Our model's systematic underprediction of density in the upper ${\sim}$500 km of the Martian interior corresponds to the Martian lithosphere, which has a conductive profile starting from the temperature of the Martian surface rather than an adiabatic profile starting at the melting temperature of the Martian mantle and is thus much colder (and therefore denser) in reality than in our model \citep{KhanSossi2022,KhanHuang2023}. As planets increase in mass, their Rayleigh numbers increase, shrinking the size of their lithospheres \citep{Howard1966,ValenciaSasselov2007,FoleyHouser2020}, so the inaccuracy of our simplified temperature prescription becomes increasingly small. The cancelling out of errors when assuming a\textcolor{black}{n Earth-like chemical composition for Mars' mantle and core} thus disappears, illustrating the importance of constructing models that replicate planetary interior structures broadly rather than just their radii.

\subsection{The Moon} \label{sec:luna}
\begin{figure*}
\centering     \includegraphics[width=\textwidth]{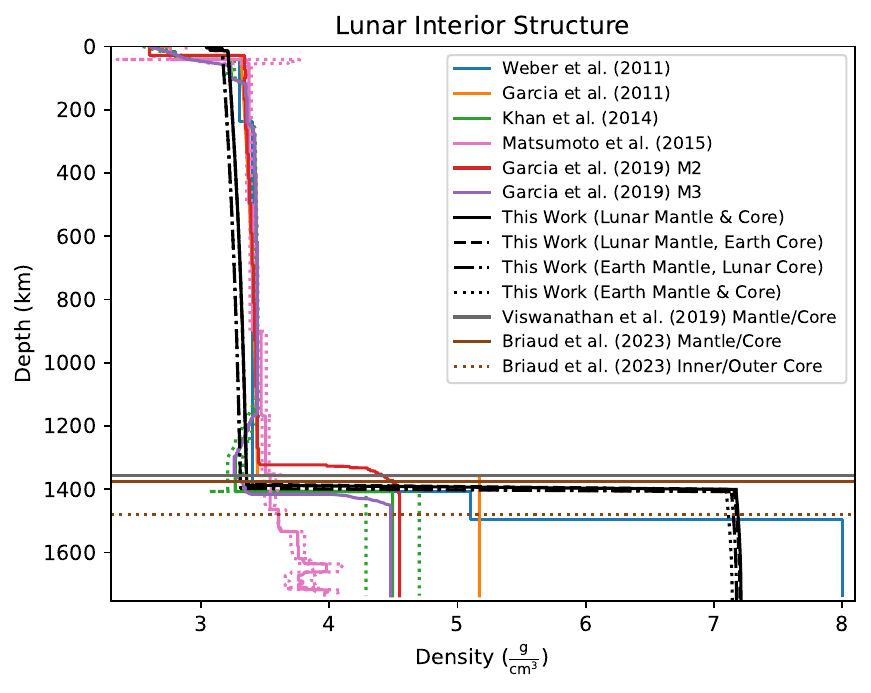}
\vspace{-2em}\caption{The internal r-$\rho$ profile of the Moon as determined by the model in this work and found in various models of the lunar interior informed by seismology. Dotted lines indicate errors where applicable. Horizontal lines correspond to lunar core radii from other considerations (\citet{ViswanathanRambaux2019}: the oblateness of the lunar core, \citet{BriaudGanino2023}: lunar tidal deformation). The format of boundaries between layers is lower pressure layer/higher pressure layer. Colours are arbitrary.}
\vspace{-2em}\label{fig:luna_structure}
\end{figure*}
Seismometers installed on the lunar surface by astronauts during the Apollo space program collected seismic data through 1977 \citep{NunnGarcia2020}. A new seismometer was recently placed on the Moon as part of the Chandrayaan-3 mission but has not yet collected adequate signals to construct an interior structure model \citep{JohnThamarai2024}. The Moon has no widely-accepted inter\textcolor{black}{ior} structure model like PREM or REM1D and thus we compare our model to a range of lunar interior structure models in Fig.~\ref{fig:luna_structure}. We plot the models of \citet{GarciaGagnepain-Beyneix2011,WeberLin2011,KhanConnolly2014,MatsumotoYamada2015,GarciaKhan2019}, each model differing in data analysis and input data, with M1 and M2 in the model of \citet{GarciaKhan2019} corresponding to imposing and not imposing a discontinuity at the crust-mantle boundary\textcolor{black}{, respectively}. We note that \citet{WeberLin2011} does not include data on the bulk Moon and thus we disfavor it, however its inclusion of a solid core is supported by \citet{BriaudGanino2023}, who argue that a solid inner core is needed to reproduce the \textcolor{black}{M}oon's observed tidal deformation. Constraining the lunar interior at depths greater than ${\sim}$1200 km with seismology is difficult, hence the divergence of our reference models at this depth \citep{NunnGarcia2020}. We take the Moon's core mass fraction as 1.68\%, the median of values reported by \citet{ViswanathanRambaux2019} derived from the oblateness of the Moon's core inferred from lunar laser ranging (precisely measuring distances to locations on the lunar surface using laser round-trip travel times). This value is slightly above the 1.5\% core mass fraction that had been reported as an upper bound from analysis of data from the GRAIL spacecraft \citep{WilliamsKonopliv2014}.

We obtain the bulk FeO and MgO of The Moon from \citet{TaylorTaylor2006}'s estimates from geochemistry and the seismographically-derived lunar mantle density. We then assume that all non-Fe\textcolor{black}{O} non-MgO material is SiO\textsubscript{2} as the Mg/Si of the Moon is poorly constrained \citep{TaylorTaylor2006}. This results in $x_\mathrm{MgO}=0.468$, $x_\mathrm{SiO_2}=0.428$, and $x_\mathrm{FeO}=0.104$. These values are broadly consistent with other geophysical as well as geochemical lunar mantle composition estimates \citep{KhanConnolly2014,KuskovKronrod2024,SossiNakajima2024}, with the exception of $x_\mathrm{SiO_2}=0.428$, which is an overestimate due to it representing all non-MgO-non-FeO materials.

In Fig.~\ref{fig:luna_structure}, we present lunar profiles (1) using the lunar mantle mineralogy of \citet{TaylorTaylor2006} and a lunar core with Earth-like abundances in the liquid \textcolor{black}{phase} and a sufficiently high $x_\mathrm{S}^\mathrm{Solid}$ to be entirely liquid ($x_\mathrm{S}^\mathrm{Solid}=0.3$, \textcolor{black}{$x_\mathrm{S}^\mathrm{Liquid}=0.09$, $x_\mathrm{O}^\mathrm{Liquid}=0.08$)}; (2) using the lunar mantle composition and core abundances fit to Earth's core; (3) using Earth's mantle composition and core abundances fit to the lunar core as in case 1; and (4) using \textcolor{black}{Earth's} mantle composition and core abundances. We get lunar radii of 0.2740$R_\oplus$ (0.49\%), 0.2740$R_\oplus$ (0.49\%), 0.2752$R_\oplus$ (0.94\%), and 0.2752$R_\oplus$ (0.94\%), respectively. We get lunar Moment of Inertia coefficients of \textcolor{black}{0.39427 (0.294\%), 0.39428 (0.296\%), 0.39443 (0.336\%), and 0.39445 (0.341\%)}, respectively.

We set core elemental abundances to obtain \textcolor{black}{a melting curve that results in} an outer liquid core, as included in all models we consult. We do not attempt to recreate a solid inner core and thus an outer/inner core transition because the slope of the temperature in the core is generally steeper than the slope of the melting curve. This would indicate that our model \textcolor{black}{M}oon would have an outer solid core and liquid inner core. In our model, this would result in a higher-density layer overlying a lower-density layer, an unphysical situation that we avoid, so we restrict $x^\mathrm{Solid}_\mathrm{S}$ to values where this does not occur; however, this result does have physical meaning that we discuss in appendix \ref{sec:iron_snow}. If a solid lunar inner core does exist (its detection in seismographic data is not universally agreed, see Fig.~\ref{fig:luna_structure}), our inability to recreate it owes to one or several of the following simplifications in our model: (1) alloy melting temperatures following Eq.~\ref{eq:melting_T}, (2) a lack of a compositional gradient in the core, and (3) no core thermal evolution \textcolor{black}{(see section \ref{sec:thermal}}\textcolor{black}{)}.

Our model underpredicts densities throughout the lunar mantle even when accounting for its higher Fe content than the Earth, indicating either an inaccurate temperature profile as for Mars, the use of an inaccurate composition of the mantle, or the mantle being in the wrong phase, potentially due to elements not included in our formalism. The solutions with lunar cores have liquid cores and the solutions with Earth cores have solid cores. The extremely close densities of the liquid and solid phases likely indicates that \citet{IchikawaTsuchiya2020}'s EOS for liquid FeS and FeO do not extrapolate down to the few GPa pressures of the Moon's mantle well.

We recover a core radius of 345 km, which is only $-1.15\sigma$
from the 362 km radius reported by \citet{BriaudGanino2023}, and is thus consistent with the true lunar core radii, if not in the middle of plausible values.

\subsection{Venus} \label{sec:venus}
Current constraints on Venus' moment of inertia \textcolor{black}{coefficient}, $C$\textcolor{black}{,} are consistent with Earth's but too weak to further constrain interior structure models \citep{MargotCampbell2021,ShahHelled2022}. We thus assume a composition of Venus' mantle and core identical Earth's, but a slightly lower $w_\mathrm{Core}$ of $0.3$ near the median of the distribution reported by \citet{ShahHelled2022}. We calculate a Venus radius of 0.9497$R_\oplus$ (-0.016\%) and a Venus MoIC of \textcolor{black}{0.33491 (-\textcolor{black}{0.62}\%); \textcolor{black}{-0.087$\sigma$}}). As \citet{ShahHelled2022} derived Venus' interior from Venus' MoI, this good fit is not the result of an independent constraint as is the case for planets with seismographic constraints. Following this concern, we also tested Venus as having the exact same parameters as Earth besides its rotational period and mass. This resulted in a radius of 0.9447$R_\oplus$ (-0.55\%) and MoIC of \textcolor{black}{0.33270 (-1.28\%; -0.18$\sigma$)}. 

The comparison of these two answers serves as an important constraint: Venus is the planet with the most Earth-like composition for which we have in situ measurements that is not itself the Earth. Venus is thus a test case for the broad extrapolatory power of the Earth. Although our results assuming an Earth-like composition for Venus are close to reality, it should be kept in mind that the smallest compositional difference between planets for which we have in situ measurements represents a radius change at the half-\% level. As Earth and Venus orbit the same star, this reflects a minimum scatter in radius arising from concerns beyond the host star and planetary mass.

Better measurements of Venus' MoIC from the EnVision and VERITAS missions--with planned launches between 2028 and 2031--are expected to provide stronger constraints on how well our model replicates Venus and how close Venus' composition is to Earth's \citep{CascioliHensley2021,RosenblattDumoulin2021,WidemannGhail2022}.

\subsection{Mercury} \label{sec:mercury}
\begin{figure}
\centering     \includegraphics[width=1.\columnwidth]{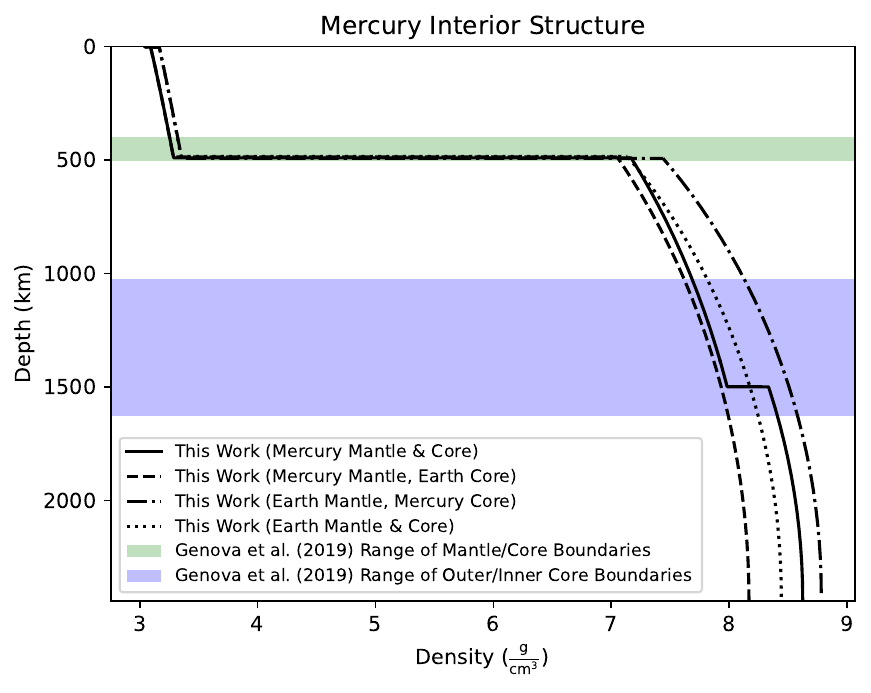}
\vspace{-2em}\caption{The internal r-$\rho$ profile of Mercury as determined by the model in this work compared to the plausible locations of Mercury's mantle/core and inner/outer core transitions from \citet{GenovaGoossens2019}, indicated by horizontal shaded regions. The format of boundaries between layers is lower pressure layer/higher pressure layer. The inner/outer core transition boundary is 0.4-0.7 times the core radius following \citet{GenovaGoossens2019}, the core radius comes from their fig.~S9d, the\textcolor{black}{ir} closest model to ours. The shaded region indicates a 3$\sigma$ spread. Colors are arbitrary.}
\vspace{-2em}\label{fig:mercury_structure}
\end{figure}
We compare our results to constraints on Mercury's interior from the literature in Fig.~\ref{fig:mercury_structure}. Although no seismic data exists for Mercury, Mercury's core does not librate due to solar torques while Mercury's mantle does, allowing MoIC for the mantle and core to be determined separately, giving a constraint on its interior \citep{PealePhillips2002,MargotPeale2012}. Additionally, surface features on Mercury such as ridges indicate that the planet has contracted \citep{ByrneKlimczak2014}. These pieces of evidence and Mercury's low MoIC provide evidence for a a solid inner core \citep{GenovaGoossens2019}.

We take $x_\mathrm{MgO}=0.532$, $x_\mathrm{SiO\textsubscript{2}}=0.466$, and $x_\mathrm{FeO}=0.002$ following \citet{FischerParman2025}, which assumes that Mercury has a composition similar to \textcolor{black}{e}nstatite chondrite but with somewhat lower Si fractions.

We take the core mass fraction of Mercury as $0.7136$, resulting from taking a Mercury core radius of $1975$ km and bulk density of $7300$ kg/m\textsuperscript{3}. The core radius is that inferred by \citet{GenovaGoossens2019}'s interior structure models built to match Mecrury's MoIC and fractional MoIC from the core. The core density is within the plausible range retrieved by \citet{GenovaGoossens2019}.

We present density profiles of Mercury (1) using the Mercury mantle mineralogy of \citet{FischerParman2025} and a core with the right composition ($x^\mathrm{Solid}_\mathrm{S}=0.06$, $x^\mathrm{Liquid}_\mathrm{S}=0.19$, indicative of extreme partitioning\textcolor{black}{, with $x^\mathrm{Liquid}_\mathrm{O}=x^\mathrm{Solid}_\mathrm{O}=0$ to reduce the number of free parameters}) to result in a liquid/solid core transition within the radii predicted by \citet{GenovaGoossens2019} (0.4-0.7 times the favored outer core radius of ${\sim}$1988 km from their supplemental fig.~S9d which most directly matches our interior model, a slightly larger radius than their overall preferred radius); (2) using Mercury's mantle composition and core abundances fit to Earth's core; (3) using Earth's mantle composition and core abundances fit to Mercury's core; and (4) using \textcolor{black}{Earth's} mantle composition and core abundances. 

We find Mercurian radii of 0.3822$R_\oplus$ (-0.17\%), 0.3833$R_\oplus$ (0.11\%), 0.3779$R_\oplus$ (-1.31\%), and 0.3805$R_\oplus$ (-0.62\%), respectively. We get Mercurian Moment of Inertia coefficients of 0.33575 (0.827\%; 1.65$\sigma$), 0.33744 (1.334\%; 2.665$\sigma$), 0.33393 (0.281\%; 0.561$\sigma$), and 0.33691 (1.173\%; 2.344$\sigma$), respectively.

Experiments at conditions similar to Mercury's core indicate that nearly all S partitions into the liquid core while Si partitions nearly evenly into the liquid and solid cores \citep{TaoFei2021,Pommier2025}, indicating that our model's S in the solid core is actually representing another light element. This light element is likely not Si, as Si reduces core melting temperatures inefficiently and our required light element abundance arises from the need to reduce the core melting temperature \citep{KnibbeRivoldini2025}.

Our finding of a partitioning between Mercury's inner and outer cores is in apparent contradiction with Mercury's observed magnetic field, which gives no hint of the compositional convection that would be anticipated from such a scenario \citep{ManglikWicht2010,TakahashiShimizu2019,KnibbeRivoldini2025}. \textcolor{black}{This could be more evidence that our assumption that the light elements in Mercury's core are \textcolor{black}{sulphur} and oxygen is inaccurate.}

The BepiColombo spacecraft, which is scheduled to arrive in orbit around Mercury in November 2026 \citep{Sanchez-CanoHadid2025}, will provide a stronger constraint on Mercury's MoIC \citep{BenkhoffMurakami2021}, helping to shine light on the exact makeup of Mercury's interior and advance the understanding of its unique-in-the-solar-system high core mass fraction regime.

\subsection{Europa} \label{sec:europa}
Although no seismic data exists for Europa, measurements of its moment of inertia from the Galileo spacecraft coupled with its low gravity leading to a relatively constant density water (solid and liquid) layer allows some constraints to be placed on its interior composition \citep{GomezCasajusZannoni2021,PetriccaCastillo-Rogez2025}. Europa's small inferred core (see below) also reduces the available parameter space, allowing compositional inferences from \textcolor{black}{its} MoIC \citep{PetriccaCastillo-Rogez2025}. We follow the interior structure model results of \citet{PetriccaCastillo-Rogez2025} and adopt for Europa $w_\mathrm{H_2O}=0.074$, $x_\mathrm{MgO}=0.375$, $x_\mathrm{SiO_2}=0.353$, and $x_\mathrm{FeO}=0.272$. This results in a very low MgSiO$_\mathrm{3}$ fraction that causes Eq.~\ref{eq:melting_T} to break, so we set our melting temperatures to what they would be for an Earth-like mantle composition. \citet{PetriccaCastillo-Rogez2025} do not find a strong constraint on $w_\mathrm{core}$, we take their approximate median core radius and density to find $w_\mathrm{core}=0.0045$ and use Earth's core's chemical composition. Using Europa's measured $A_B$ of 0.68 \citep{GrundyBuratti2007}, we find a Europa radius of 0.2486$R_\oplus$ (1.48\%) and MoIC of 0.35624 (0.434\%; 0.641$\sigma$) when calculating Europa's surface temperature from its solar instellation (as would be the case for an exoplanet). 

\textcolor{black}{Our model by default does} not produce a liquid ocean due to \textcolor{black}{its} neglect of tidal heating, assuming a purely convective thermal structure in the ice layer, and ignoring the impact of any impurities on the melting temperature. However, if we instead use the insight that Europa's ice shell \textcolor{black}{likely} has a radius of $\lesssim25$ km \citep{NimmoGiese2005,SchenkMatsuyama2008,CoxBauer2015,VilellaChoblet2020}, comparable to the radius of Earth's neglected crust, and set the outer boundary of the water layer \textcolor{black}{to a temperature of 273.15 K (and a \textcolor{black}{p}ressure of $10^{5}$ Pa) to ensure a liquid state}--similar to how we treat the mantle--we get a radius of 0.2473$R_\oplus$ (0.95\%) and MoIC of 0.35973 (1.419\%; 2.097$\sigma$). We take the latter case as fiducial while emphasizing that our failure to replicate a subsurface ocean in the first case indicates that our model is not applicable for detailed characterization of tidally-heated icy moons. The relatively high error even when forcing the water layer into the right state is also likely a reflection of our inability to calculate a proper melting temperature, potentially leading to an interior that is far too hot.

The Europa Clipper mission will reach an orbit around Jupiter in April 2030 and will collect measurements of Europa's induced magnetic field and love number \citep{RobertsMcKinnon2023,PappalardoBuratti2024}. Love numbers determine the response of a body to tidal forcing and are thus particularly effective for Europa due to Jupiter's extreme tidal forcing \citep{PappalardoBuratti2024}. These will provide better constraints on Europa's interior that could be compared to \textcolor{black}{the density and temperature profiles predicted by this model}.

\subsection{Summary} \label{sec:val_sum}
Our model generates radii and moment of inertia coefficients within 0.5\% or 2$\sigma$ of the true values for Earth, Mars, the Moon, \textcolor{black}{Venus,} and Mercury. We replicate Europa's radius and moment of inertia \textcolor{black}{coefficient} to within 1\% or 3$\sigma$. In addition, the core radii of Earth, Mars, the Moon, and Mercury are completely consistent with reality, only off by $1.16\sigma$ from other estimates at worst. The outer/inner core transitions of Earth and Mercury (but not of the Moon) are replicated. Both unambiguous density jumps within Earth's mantle are replicated to within 35 km, and several other density jumps consistent with at least some theories of Earth's interior are present. Where our model is discrepant with seismographic constraints, the chief explanation tends to relate to physics and geophysics occurring at masses below Earth's, \textcolor{black}{so our model is expected to extrapolate well into the exoplanetary mass regime}.

Having demonstrated our model's accuracy, we apply it to calculate mass-radius relations for exoplanetary populations.

\section{Mass-Radius Curves} \label{sec:mr}
\subsection{New Mass-Radius Relations} \label{sec:new_mr_relations}
\begin{figure*}
\centering     \includegraphics[width=\textwidth]{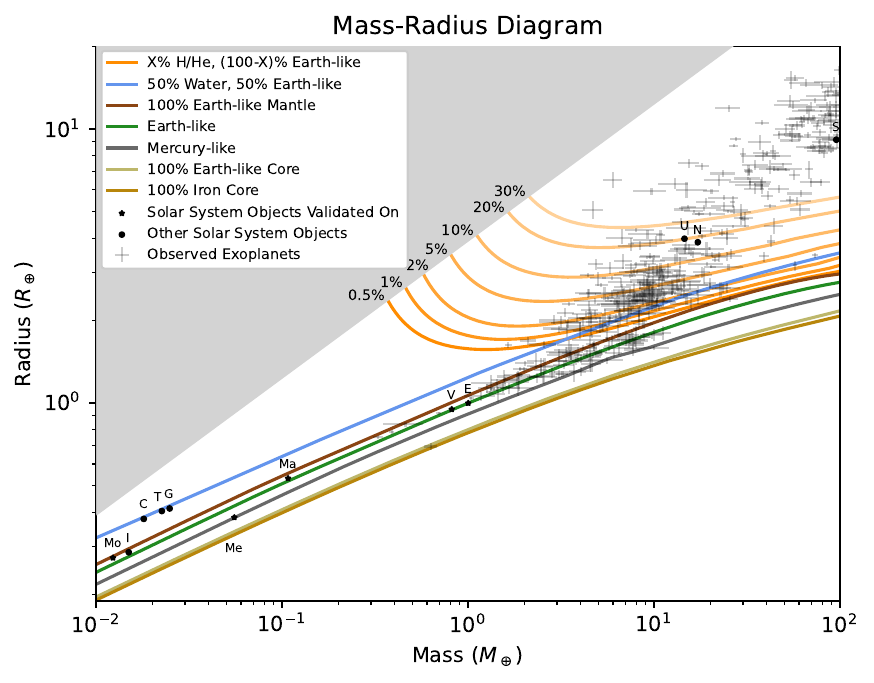}
\vspace{-2em}\caption{The isocomposition curves in mass-radius space for several planetary compositions. \textcolor{black}{Multiple H/He-rich compositions are present, higher transparency curves have higher H/He fractions and the H/He fraction by mass is labelled on the left of each curve. See text for further elaboration on compositions.} The shaded region corresponds to planets with gravities at their transit radius outside the bounds of the analytical radiative transfer model fits used by this study (see section \ref{sec:thermal}). The background error bars are the observed exoplanet population \textcolor{black}{via the PlanetS catalog \citep{OtegiBouchy2020,ParcBouchy2024} as of 30 March 2026}. The dots are solar system objects (both moons and planets), with stars being the six objects within our validation sample (see section \ref{sec:validation}). From lowest to highest mass, the solar system objects are: the Moon (see section \ref{sec:luna}), Io, Callisto, Titan, Ganymede, Mercury (see section \ref{sec:mercury}), Mars (see section \ref{sec:mars}), Venus (see section \ref{sec:venus}), Earth (see section \ref{sec:earth}), Uranus, Neptune, and Saturn (see appendix \ref{sec:saturn}). Even though not included in the validation sample, the agreement of Callisto, Titan, and Ganymede's properties with the half water isocomposition curve indicates the applicability of this model to water-rich bodies. Note a discrepancy: solar system planetary radii are actual radii while exoplanet radii are transit radii (see section \ref{sec:transit}). No de-biasing is applied. Radii and mass uncertainties restricted to below 8\% and 25\%, respectively. Note that $M$ is plotted on the x axis even though we fit slopes of $M$ in terms of $R$ in section \ref{sec:power_law}, this is a matter of convention. All of our model data used to make this figure is available at \url{https://github.com/Bennett-Skinner/SkinnerPudritzCloutier2026-MR-curves/} and \url{https://doi.org/10.5281/zenodo.20382154}.}
\vspace{-2em}\label{fig:overall_mr}
\end{figure*}

The isocomposition mass-radius (M\textcolor{black}{-}R) relations from our new interior structure model are presented and compared to the exoplanetary population in Fig.~\ref{fig:overall_mr} \textcolor{black}{(see appendix \ref{sec:more_mr_curves} for more curves: Fig.~\ref{fig:mr_curve_varyage} for varied age, and Fig.~\ref{fig:mr_curve_varyrot} for varied rotation period)}. We \textcolor{black}{display twelve} compositions, from least dense to most dense for Earth-like masses and temperatures: H/He-enveloped \textcolor{black}{(X\% H/He+(100-X)\% Earth-like with X=0.5\%-30\%)}, Water world (50\% H$_\mathrm{2}$O+50\% Earth-like), pure mantle (100\% mantle with Earth-like abundances), Earth-like, Mercury-like (see section \ref{sec:mercury}), Earth-like core (100\% core mass fraction with Earth-like abundances\textcolor{black}{)}, and a 100\% pure iron core. \textcolor{black}{All other variables are set following Table \ref{tab:free_params}.} Curves for different temperatures \textcolor{black}{and varied water and core mass fractions} are provided below. All mass-radius data points are publicly available at \url{https://github.com/Bennett-Skinner/SkinnerPudritzCloutier2026-MR-curves/} and \url{https://doi.org/10.5281/zenodo.20382154}.

\textcolor{black}{W}e generated 200 model planets for these compositions log-uniformly spaced between 0.01 and 100 $M_\oplus$ and thus our mass resolution is only as high as any two model planets are far apart in mass space (0.04$M_\oplus$ at the low end of this mass range and 0.37$M_\oplus$ at the high end of this mass range). We consider these compositions to bracket the parameter space of super-Earths and sub-Neptunes. We include higher-mass versions of terrestrial objects only for Earth and Mercury because those are the only objects with constrained liquid and solid core compositions, whereas for most solar system objects only liquid core compositions are known. 

We immediately note that--as demonstrated in section \ref{sec:validation}--the Earth-like compositional curve passes very close to Earth \textcolor{black}{and} Venus, the Mercury-like compositional curve passes very close to Mercury, and the pure mantle curve passes nearby the Moon, as expected from its very low core mass fraction. Beyond the validation sample, Io is consistent with a rocky interior, while Callisto, Titan, and Ganymede are consistent with half-water interiors, in agreement with observation and with theories of Galilean moon formation \citep{HellerPudritz2015,ShibaikeOrmel2019}. \textcolor{black}{Finally, Mars lies close to but noticeably above the Earth-like compositional curve, indicating a low core mass fraction and light element enhancement in the core. The overlap of Uranus and Neptune with the 20\% H/He isocomposition curve should not be taken as a prediction of their composition because these curves are generated at Earth's equilibrium temperature.}

Broadly, isocomposition curves for predominantly solid bodies follow power laws \textcolor{black}{$M=bR^a$} at masses a few times Earth's. These curves are roughly parallel. As mass increases, compression within the planetary interior becomes increasingly important, resulting in an increasingly flat power law. In other words, the $M$-$R$ relationship is concave down in log-log space when $M$ is used as the x axis. All power-law $M$-$R$ curves thus eventually degrade. Planets with significant gaseous envelopes--be they H/He or steam--experience a flattening of the $M$-$R$ relationship at masses $\lesssim2M_\oplus$ as the total mass of the planet becomes inadequate to hold onto the surrounding tenuous atmosphere. Planets with masses and radii consistent with being in this flattening region are rare (and none meet the mass and radius precision cutoff we set in our figures), indicating that such a scenario is not viable long-term, with either such tenuous materials never being accreted, rapidly being lost, or the planet cooling to sufficient temperatures \textcolor{black}{to shrink its envelope size}. Planets with masses above ${\sim} 10$--$20\, M_\oplus$ always have densities less than a planet of that mass made of pure Earth composition mantle material, indicating that they are highly likely to be volatile-enhanced. This is in agreement with predictions of gas accretion and pebble isolation masses from planet formation theory and conform\textcolor{black}{s} to the apparent masses of the cores of Jupiter and Saturn \citep{WahlHubbard2017,BitschIzidoro2019,LambrechtsMorbidelli2019,OtegiBouchy2020,MankovichFuller2021,HowardGuillot2023}.

\begin{figure}
\centering
\includegraphics[width=1.\columnwidth]{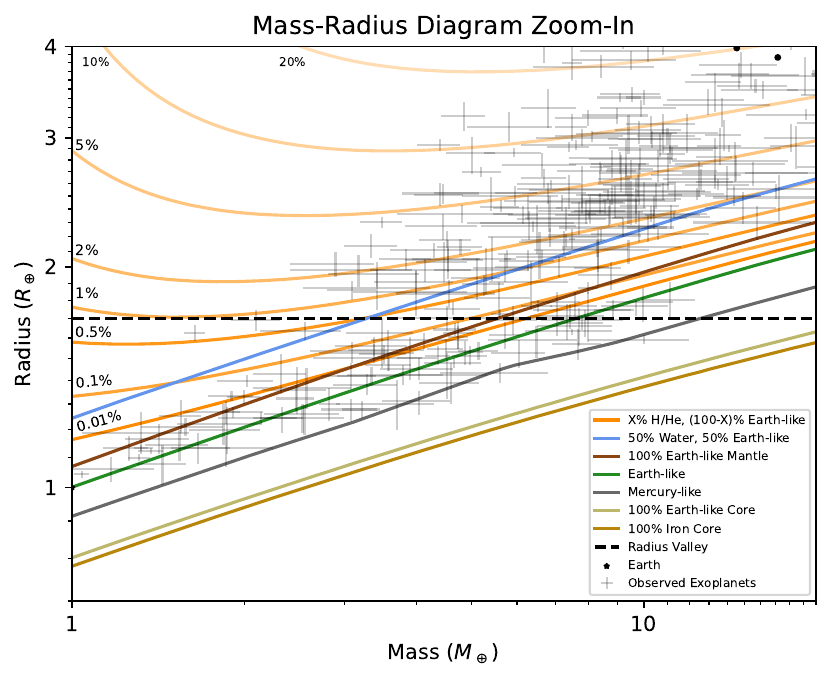}
\vspace{-2em}\caption{The same as Fig.~\ref{fig:overall_mr} but zoomed in on the parameter space of super-Earths and sub-Neptunes. The kink in the Mercury-like curve is due to Mercury's core becoming increasingly liquid with mass until it is fully liquid at $5.672M_\oplus$ before becoming increasingly solid with mass (see section \ref{sec:power_law}).}
\vspace{-2em}\label{fig:overall_mr_zoomed}
\end{figure}

In Fig.~\ref{fig:overall_mr_zoomed}, we zoom into masses between 1 and 20 $M_\oplus$, the super-Earth to sub-Neptune regime. The radius valley, located at ${\sim}$1.7 $R_\oplus$ (with variance with instellation and stellar type), is indicated with a horizontal dashed line \citep[e.g.][]{FultonPetigura2017,FultonPetigura2018,CloutierMenou2020}. At masses less than ${\sim}8M_\oplus$, a radius of 1.7$R_\oplus$ (the nominal centre of the radius valley) is greater than that of a planet with an Earth-like composition, indicating either a core mass fraction much lower than Earth's or some amount of H/He (\textcolor{black}{$<1\%$}) or H$_\mathrm{2}$O (up to 50\%). For these low masses, lying above the radius valley thus requires a significant amount of H/He (\textcolor{black}{$\geq1\%$}) or an extended steam atmosphere. Below the intersection of the half-water, half-Earth-like curve with the radius valley at ${\sim}3.7M_\oplus$, all but \textcolor{black}{\textcolor{black}{two planets with precise} mass and radii measurements have \textcolor{black}{densities} consistent with being between a pure iron core and a water world, indicating a relative lack of volatile-enriched low-mass planets.} The\textcolor{black}{re are two exceptions:} L98-59 \textcolor{black}{d} \citep{CadieuxL'Heureux2025}\textcolor{black}{--}which observations with JWST NIRSPec and the IRGINS high-resolution spectrograph have found likely has a thick sulphur-rich secondary atmosphere \citep{CheverallMadhusudhan2026,BanerjeeBarstow2024,GressierEspinoza2024}\textcolor{black}{--and Kepler-54 c, whose mass varies by a factor of a few between literature sources \citep{OfirYoffe2025,LeleuDelisle2023,HaddenLithwick2014}.}

\begin{figure*}
\centering
\includegraphics[width=\textwidth]{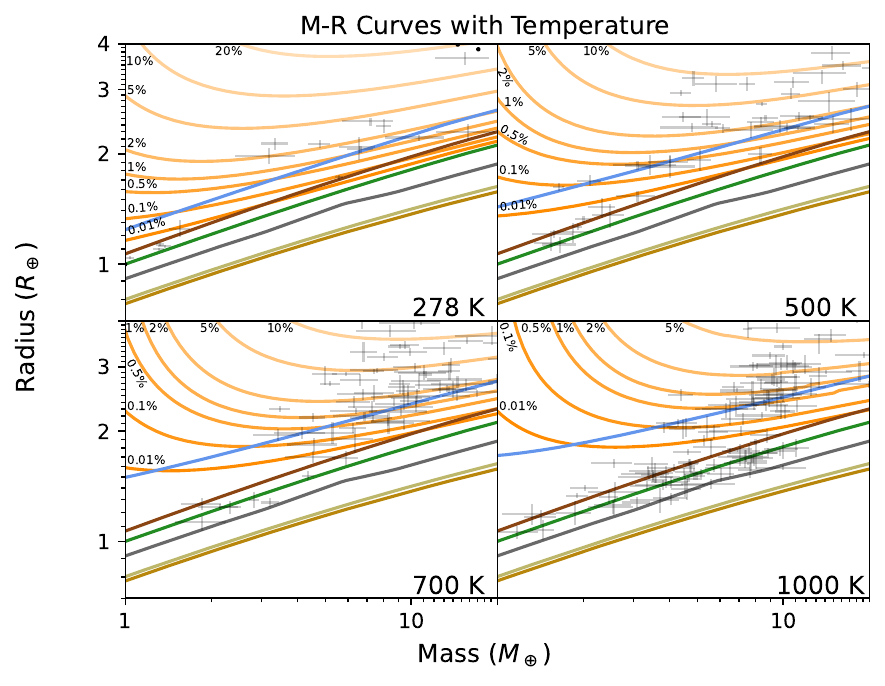}
\vspace{-2em}\caption{The same as Fig.~\ref{fig:overall_mr_zoomed} but with planetary equilibrium temperatures (as defined for $A_B=0$, note that our planets have an assumed $A_B=0.3$, so \textcolor{black}{their effective temperatures are approximately} $(1-0.3)^{0.25}$ times the listed temperature) varied. Curve colours are the same as Fig.~\ref{fig:overall_mr}. Background exoplanets are sorted into the panel for an equilibrium temperature closest to their temperature reported \textcolor{black}{in the PlanetS catalog}. The noticeable kink\textcolor{black}{s at high \textcolor{black}{masses} for H/He-enveloped planets correspond to the total mass fraction of the mantle that is melting from the thermal blanketing of the overlying envelope increasing.}}
\vspace{-2em}\label{fig:hot_planets}
\end{figure*}

Mass and radius are not the only available parameters, as most exoplanets have \textcolor{black}{a} known $T_\mathrm{eq0}$ as well. In Fig.~\ref{fig:hot_planets}, we plot mass-radius curves for varying \textcolor{black}{zero-albedo} equilibrium temperatures and compare to the exoplanet population at those temperatures. Higher temperatures increase planetary radii, with the effect more pronounced the more extended the planetary atmosphere. This amounts to a $>10\%$ increase in radius as equilibrium temperature increases from 278 to 1000 K in the ${\sim}10M_\oplus$ mass regime for planets with a 1\% H/He mass fraction and several $\%$ in the same regime for planets with a 50\% water mass fraction. This means that planets with a H/He envelope have radii more sensitive to stellar instellation than water-rich planets, causing the mass at which a planet with a 50\% water, 50\% Earth-like \textcolor{black}{composition} and a \textcolor{black}{planet with a} 1\% H/He, 99\% Earth-like \textcolor{black}{composition} have the same radius to increase with stellar instellation.

\begin{figure}
\centering
\includegraphics[width=1.\columnwidth]{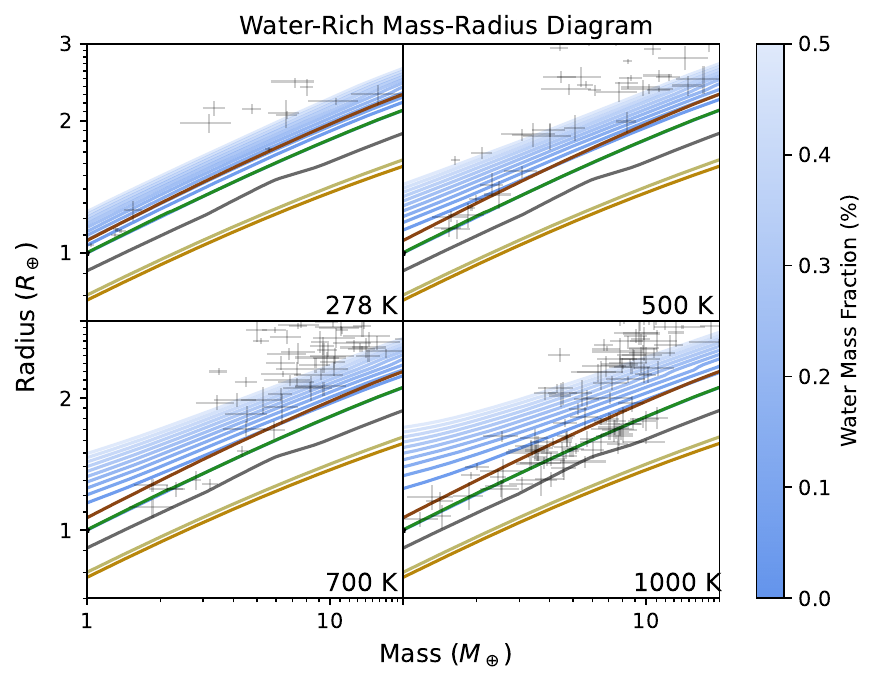}
\vspace{-2em}\caption{\textcolor{black}{The same as Fig.~\ref{fig:hot_planets} but with many intermediate water compositions included. Each water curve has a composition of X\% water, (1-X)\% Earth-like, with X varying from 5 to 50\% in 5\% increments.}}
\vspace{-2em}\label{fig:mr_curve_tpanels_water}
\end{figure}

\textcolor{black}{
M-R curves with water mass fraction varied at a high resolution are shown in Fig.~\ref{fig:mr_curve_tpanels_water}. As water mass fraction and temperature increase, planetary radii increase. For low-temperature planets whose water is entirely in condensed form (top-left panel), the isocomposition curves are relatively parallel to each other and to rocky bodies, but as temperature increases and a steam atmosphere makes up an increasing fraction of the total planetary radius, a flattening of the mass-radius relation at lower masses becomes increasingly apparent. At $1.02M_\oplus$, increasing the zero-albedo equilibrium temperature from 278 K to 1000 K \textcolor{black}{increases the radius of a 50\% water, 50\% Earth-like planet by 37.4\%}. Accounting for thermal effects is thus necessary in the study of hot water worlds ($T_\mathrm{eq}>273.15$, note that reported temperatures are $T_\mathrm{eq0}$, not $T_\mathrm{eq}$). The distance between subsequent isocomposition curves shrinks with increasing water mass fraction, indicating that a linear equation is insufficient to relate radius and water mass fraction and motivating our use of a quadratic function in section \ref{sec:power_law}. The kinks in radius at high masses are caused by the melting of the mantle caused by the thermal blanketing of the overlying water layer.}

\begin{figure}
\centering
\includegraphics[width=1.\columnwidth]{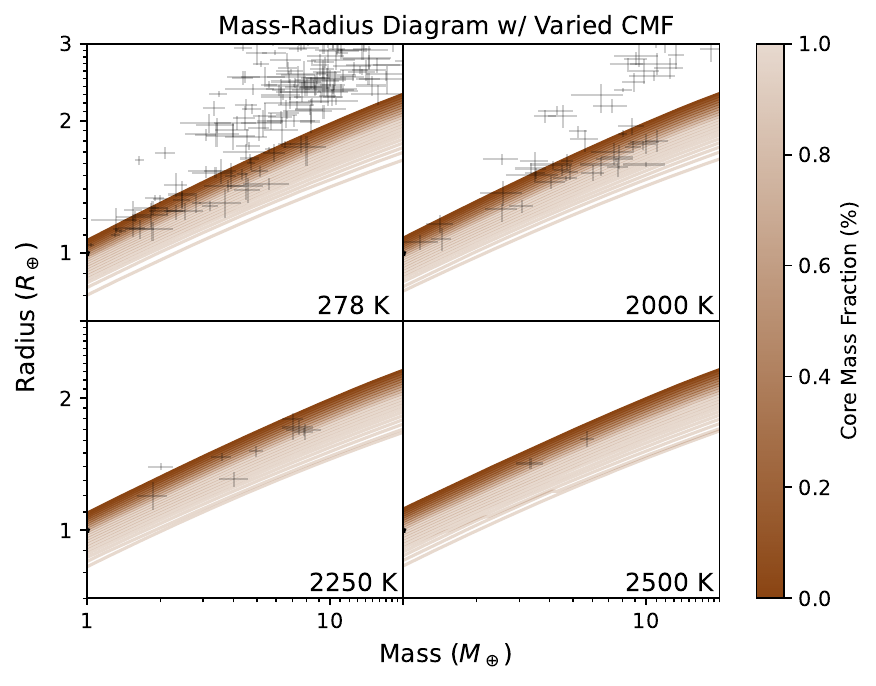}
\vspace{-2em}\caption{\textcolor{black}{The same as Fig.~\ref{fig:hot_planets} but showing planets with core mass fractions varied between 0\% and 100\% in 5\% increments. The mantle and core elemental compositions are Earth-like.}}
\vspace{-2em}\label{fig:mr_curve_tpanels_cmf}
\end{figure}

\textcolor{black}{
M-R curves with core mass fraction varied at a high resolution are shown in Fig.~\ref{fig:mr_curve_tpanels_cmf}. The isocomposition curves are relatively parallel to each other. The distance between subsequent isocomposition curves shrinks with decreasing core mass fraction (increasing mantle mass fraction), indicating that a linear equation is insufficient to relate radius and core mass fraction and motivating our use of a quadratic function in section \ref{sec:power_law}. The kinks in radius at high masses are due to the \textcolor{black}{solidifying of the core}. At an Earth mass, the mantle is fully solid at Earth temperatures, partially molten at 2000 K and 2250 K, and fully molten at 2500 K. At masses immediately above the Earth's, the bottom of the mantle becomes solid due to high pressures, even at 2500 K, reducing the impact of temperature on planetary radii. For a planet with a 0\% core mass fraction, raising the zero-albedo equilibrium temperature from 278 K to 2500 K causes an increase in radius by 4.6\% at a mass of $M_\oplus$ and 1.5\% at a mass of $\textcolor{black}{10}M_\oplus$. For a planet with a 100\% core mass fraction, these values are \textcolor{black}{3.5}\% and \textcolor{black}{3.1}\%, respectively. For intermediate compositions and masses, the increase in planetary radii with temperature is comparable \textcolor{black}{($\sim1-4\%$).}}

\subsection{Power-Law Fits to the M-R Relation }\label{sec:power_law}
\textcolor{black}{We fit $R(M,w_\mathrm{H_2O})$ over the mass range $0.93<M/M_\oplus<3.57$ (corresponding to Earth-like planets with mantles whose highest-pressure phase is post-perovskite) to M-R curves with water mass fractions between 0 and 50\% (see Fig.~\ref{fig:mr_curve_tpanels_water}) and find that
\begin{equation}
    R/R_\oplus=1.010(1+0.643w_\mathrm{H_2O}-0.377w_\mathrm{H_2O}^2)(M/M_\oplus)^{(1/3.748)}
\end{equation}
for $0<w_\mathrm{H_2O}<0.5$.}

\textcolor{black}{We also fit $R(M,w_\mathrm{core})$ to Earth-like M-R curves with core mass fractions between $0$ and $1$ (see Fig.~\ref{fig:mr_curve_tpanels_cmf}) and find that}
\begin{equation}
    R/R_\oplus=1.075(1-0.178w_\mathrm{core}-0.061w_\mathrm{core}^2)(M/M_\oplus)^{(1/3.762)}\mathrm{.}
\end{equation}

\textcolor{black}{We caution that the many phase transitions in both the mantle and the core occurring in this mass regime mean that both of these fits relating composition to radius can have errors compared to our data of up to $0.02R_\oplus$, even within the regime over which they were fit.}

\textcolor{black}{To counter this, we also} present $M=bR^a$ curves for fits to the condensed worlds in our dataset obtained via \texttt{scipy.optimize.curve\_fit} \textcolor{black}{for a subset of compositions} in Table \ref{tab:results}.

\begin{figure}
\centering
\includegraphics[width=1.\columnwidth]{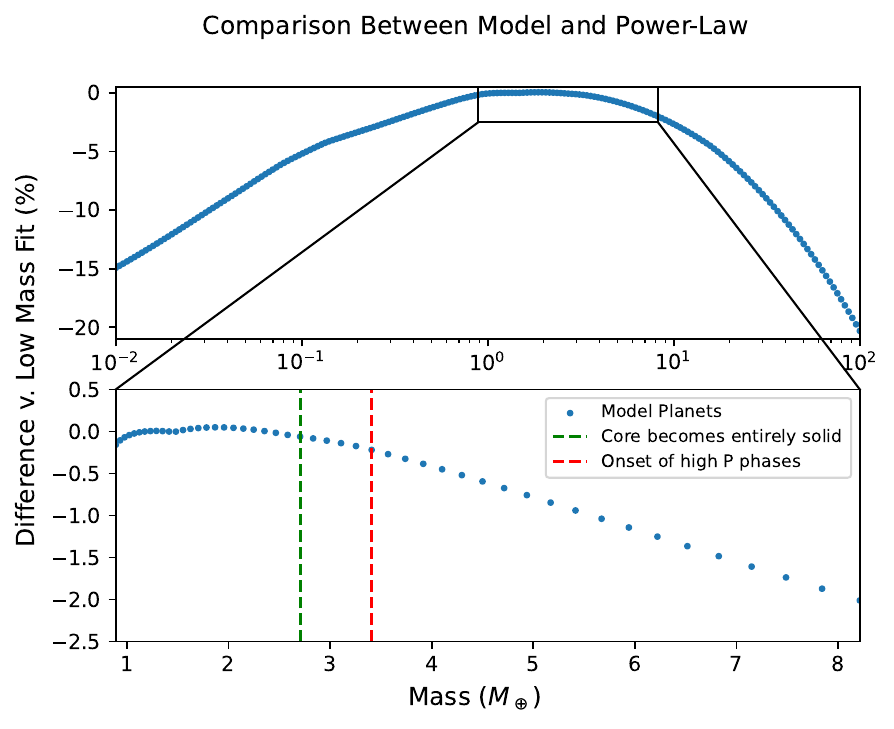}
\vspace{-2em}\caption{The relative difference between our calculated planetary radii for an Earth-like composition and the $M=0.994R^{3.701}$ power law that best fits it for masses between \textcolor{black}{0.93} and \textcolor{black}{2.58} Earth masses (see Table \ref{tab:results}). Two kinks corresponding to changes in the planetary interior structure are labelled with vertical lines. We emphasize that the plotted difference is solely due to a power law fit not capturing the full complexity of the mass-radius relationship within our model and is not a result of differences between our model and physics. (Top) Masses between 0.01 and 100 $M_\oplus$ on a log scale (Bottom) Masses between \textcolor{black}{0.93} and 8.21 $M_\oplus$ on a linear scale.}
\vspace{-2em}\label{fig:diff_v_lowPcurve}
\end{figure}

\textcolor{black}{W}e emphasize that these fits are only valid for a limited range of masses as there is no physical reason to expect a universal power-law mass-radius relation. We thus strongly emphasize the importance of using our full tables rather than these fits (see Fig.~\ref{fig:diff_v_lowPcurve} and discussion below).

We consider Earth-like and Mercury-like compositions because these are the two solar system bodies whose solid core compositions are constrained. We model all curves as piecewise functions with breaks at the masses at which various new phases appear. Pure core objects are always fully solid as the mantle material melting point at zero pressure is much lower than the core material melting point at zero pressure, resulting in no phase transitions. For these pure core objects, we provide a fit over the entire M-R space of interest and between the boundaries for an Earth-like object. The increasing value of the power law $a$ with mass, even without phase transitions, is an indication that compression cannot be explained entirely by a power law.

We only consider masses between a low-mass limit of \textcolor{black}{0.93}$M_\oplus$ (the mass at which post-perovskite appears in an Earth-like composition) and a high-mass limit of 8.21$M_\oplus$\footnote{Rounded up from 8$M_\oplus$ to include a model planet at the mass of 8.21$M_\oplus$} in these fits, similar to the 1--8$M_\oplus$ range used by \citet{ZengSasselov2016} and approximately following the mass range where an Earth-like power-law fit is appropriate.

The power\textcolor{black}{-law} index $a$ is always greater than 3, indicating masses greater than \textcolor{black}{in the case where} all objects \textcolor{black}{have} the same density. This is because compression causes higher-mass objects to have higher bulk densities, with \textcolor{black}{$a$} encoding the strength of this compression effect. Inspection of Table \ref{tab:results} reveals that $a$ generally increases with increasing planetary mass and increasing planetary core mass fraction, both of which cause greater pressures to be reached in the interior and \textcolor{black}{thus} more extreme compressions. The factor $b$ decreases with increasing mass as a result of the bulk density not increasing as quickly as implied by the local value of $a$ alone (the $M$-$R$ relationship can be re-parameterized to $\rho\propto bR^{a-3}$). Notably, the power laws of water worlds and Earth-like planets are within a few $\%$ at masses where both are expected to have the same state of core; similarly, the power laws of Earth-like and iron cores are within 1\% of each other. However, we emphasize that these power laws are not parallel, and interpolated compositions encoding such an assumption are flawed.

The Mercury-like curve is a notable exception to these general trends, owing to the state of the core. The extremely low iron fraction in the mantle coupled with Eq.~\ref{eq:melting_T} causes a high mantle melting temperature and thus high temperature at the top of the planetary core (see section \ref{sec:thermal}) which causes the planetary core temperature to be much closer to the iron melting temperature at higher masses than is the case for other compositions. This results in the unique phenomenon of the core becoming increasingly liquid up to a mass of $5.672M_\oplus$--when the core becomes \textcolor{black}{almost} fully liquid (only the innermost 1.5\% of the total planetary mass is \textcolor{black}{solid})--before becoming increasingly solid up to $9.438M_\oplus$--when the core becomes fully solid. The increasing prominence of lighter liquid iron at greater masses below $5.672M_\oplus$ causes a shallower-than-otherwise $M$-$R$ relationship, while the solidification of the core above that mass causes a steeper-than-otherwise $M$-$R$ relationship. As our temperature profile arises from a simple prescription and does not include the thermal or dynamical evolution of the interior (see appendix \ref{sec:iron_snow}), we do not anticipate these changes in slope are reflected in the planetary population, but we emphasize that these changes in slope are a physical result.

In Fig.~\ref{fig:diff_v_lowPcurve}, we compare the model radii of Earth-like planets with the power-law fit for masses between \textcolor{black}{0.93} and \textcolor{black}{2.58} $M_\oplus$ from Table \ref{tab:results}. We observe that applying the Earth-like mass power-law fit outside of its intended mass range results in a discrepancy compared to our mass-radius curves reaching 10s of \% due to the non-constant value of the power-law $a$ with mass. We also observe that kinks appear in the mass-radius curve at the locations at which the state of the interior changes, indicating that the power law is no longer appropriate, justifying our prescription of changing power laws at these masses. 

These points illustrate that it is not in general possible to capture planetary mass-radi\textcolor{black}{us} relations with a simple power law. We have estimated the errors that arise if one makes such an assumption.  Although the inaccuracy from doing so is $\lesssim2\%$ for planets in the \textcolor{black}{0.93}--8.21 $M_\oplus$ range--which may be acceptable in some circumstances--it is not generally correct. In other circumstances the error arising from its use is comparable to the level of observational uncertainty. The errors are nearly universally negative, indicating that the power-law fits provide an upper limit on the \textcolor{black}{planetary} radius when extrapolated beyond their intended regime.

\begin{table*}
    \caption{Power-Law fits to the isocomposition curves of \textcolor{black}{a selection of consolidated compositions} considered in our sample. $M$ and $R$ are in Earth units. We encourage the use of our $M$-$R$ tables rather than these fits.}
    \begin{center}
    \centerline{
    \begin{threeparttable}
\begin{tabular}{ccc}
    Mass Range ($M_\oplus$) & Change to the interior at minimum of mass range & Best-Fit Equation\\\hline
    \multicolumn{3}{c}{Water World}\\
    \hline
        \textcolor{black}{0.93}--1.18 & - & $M=0.456R^{3.604}$\\
        1.18--1.63 & Post-Perovskite appears & $M=0.439R^{3.754}$\\
        1.63--\textcolor{black}{2.83} & Ice X appears & $M=0.426R^{3.835}$\\
        \textcolor{black}{2.83}--\textcolor{black}{3.41} & Core becomes purely solid & $M=0.417R^{3.878}$\\
        \textcolor{black}{3.41}--8.21 & High-pressure mantle phases appear & $M=0.398R^{3.981}$\\
        \hline\multicolumn{3}{c}{Earth-Like Mantle}\\\hline
        \textcolor{black}{0.93}--2.97 & - & $M=0.783R^{3.608}$\\
        2.97--8.21 & High-pressure mantle phases appear & $M=0.691R^{3.926}$\\
        \hline\multicolumn{3}{c}{Earth-Like}\\\hline
        \textcolor{black}{0.93}--\textcolor{black}{2.58} & Post-perovskite appears & $M=0.994R^{3.701}$\\
        \textcolor{black}{2.58}--\textcolor{black}{3.57} & Core becomes purely solid & $M=0.969R^{3.808}$\\
        \textcolor{black}{3.57}--8.21 & High-pressure mantle phases appear & $M=0.885R^{4.047}$\\
        \hline\multicolumn{3}{c}{Mercury-Like}\\\hline
        \textcolor{black}{0.93}--1.96 & - & $M=1.402R^{3.745}$\\
        1.96--5.67 & Post-Perovskite Appears & $M=1.442R^{3.675}$\\
        5.67--6.83 & Core becomes \textcolor{black}{maximally} liquid & $M=0.610R^{5.921}$\\
        6.83--8.21 & High-pressure mantle phases appear & $M=0.667R^{5.707}$\\
        \hline\multicolumn{3}{c}{Earth-Like Core}\\\hline
        \textcolor{black}{0.93}--8.21 & - & $M=2.301R^{4.131}$\\
        \textcolor{black}{0.93}--\textcolor{black}{2.58} & - & $M=2.301R^{3.806}$\\
        \textcolor{black}{2.58}--\textcolor{black}{3.57} & - & $M=2.294R^{4.025}$\\
        \textcolor{black}{3.57}--8.21 & - & $M=2.196R^{4.324}$\\
        \hline\multicolumn{3}{c}{Pure Iron Core}\\\hline
        \textcolor{black}{0.93}--8.21 & - & \textcolor{black}{$M=2.570R^{4.161}$}\\
        \textcolor{black}{0.93}--\textcolor{black}{2.58} & - & $M=2.547R^{3.807}$\\
        \textcolor{black}{2.58}--\textcolor{black}{3.57} & - & $M=2.554R^{4.041}$\\
        \textcolor{black}{3.57}--8.21 & - & $M=2.454R^{4.378}$\\
    \end{tabular}
    \end{threeparttable}}
    \end{center}
    \label{tab:results}
\end{table*}

\section{Discussion} \label{sec:discussion}
\subsection{Comparison to Empirical Fits}\label{sec:emp_fit_comp}
Power law fits of the $M$-$R$ relation to empirical data are widely used. As noted above, our power laws can be used and compared with empirical $M$-$R$ relations in the terrestrial planet regime. As an example, it is consistent with the empirical power laws from \citet{ChenKipping2017}--whose fit is dominated by solar system objects--and \citet{MullerBaron2024}--whose fit includes no solar system objects, indicating that up to ${\sim}2$ $M_\oplus$, planets behave much like Earth and Venus. Our fit is slightly higher than that reported by \citet{OtegiBouchy2020}. Beyond this mass, our $R$ exponent increases beyond the empirical mass-radius relation. However, the empirical mass-radius relation beyond 2 $M_\oplus$ in \citet{ChenKipping2017} and 4.4 $M_\oplus$ in \citet{MullerBaron2024} becomes contaminated with volatile-enriched worlds, so a conclusion cannot currently be drawn on whether this divergence truly represents diverging behavior in the interiors of rocky worlds. Nonetheless, a search for a kink in the empirical exoplanet mass-radius relation at \textcolor{black}{${\sim}3.6$} $M_\oplus$ could be a powerful observational test for high pressure phase transitions in the mantle.

However, there is another reason that our mass-radius relations diverge from the empirical relations for super-Earths for reasons unrelated to the phase changes considered here. Hot super-Earths may sequester a significant portion of their primordial H/He atmospheres, resulting in a higher density deficit in the core and thus mass increasing less strongly with radius than it otherwise would \citep{SchlichtingYoung2022}.  Although this is beyond the scope of this paper, we will address it in a subsequent publication.

\subsection{Comparison to the Literature}
\label{sec:literature_comp}
Having presented our model, validated it, and presented its results, we compare it with the results in the literature and evaluate the relative merits of different models.

\begin{figure}
\centering     \includegraphics[width=1.\columnwidth]{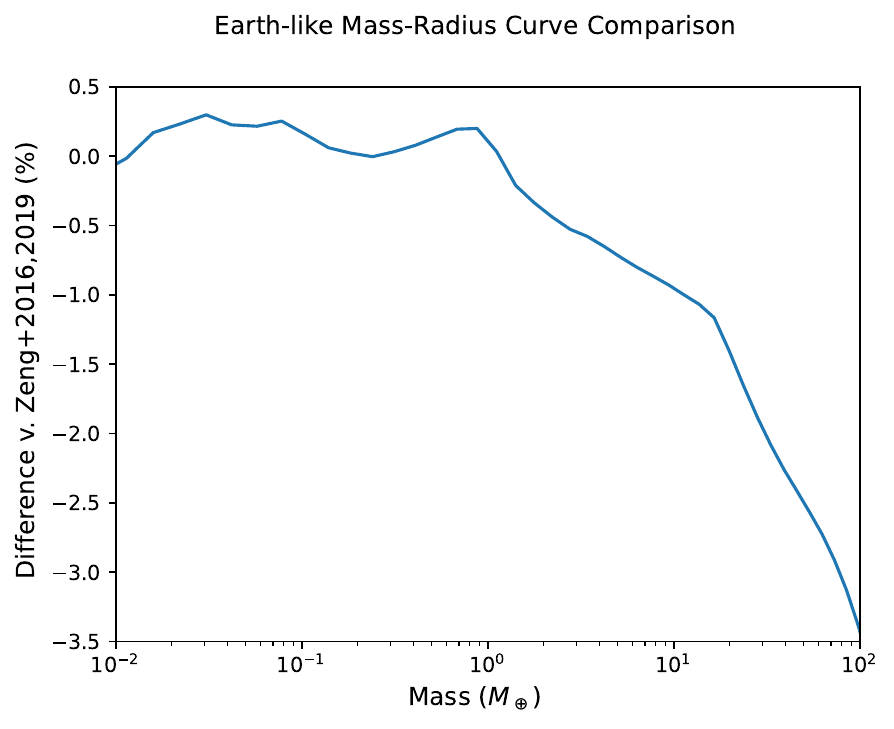}
\vspace{-2em}\caption{The difference between the radii of Earth-like planets calculated by this work and that of \citet{ZengSasselov2016,ZengJacobsen2019}, calculated via $\mathrm{\%}=(R_\mathrm{This Work}-R_\mathrm{Zeng})/R_\mathrm{This Work}*100$. Note the systematic downwards trend starting at ${\sim}1M_\oplus$ and accelerating at ${\sim}17M_\oplus$.}
\vspace{-2em}\label{fig:earthlike_zeng}
\end{figure}

\textcolor{black}{In Fig.~\ref{fig:earthlike_zeng}, we show the difference between our model and that of \citet{ZengSasselov2016,ZengJacobsen2019} for an Earth-like mass-radius curve for the entire mass regime considered in this work. The two models generally agree for masses ${\lesssim}M_\oplus$. However, \citet{ZengSasselov2016,ZengJacobsen2019} predict increasingly smaller radii at higher masses because their models were constructed \textcolor{black}{using EOS derived by fitting the density profile of Earth's interior}. For ${\lesssim}M_\oplus$ planets with pressures probed by Earth's interior, their methodology is largely consistent with a full physics-based approach as in this paper. However, at higher masses their extrapolation from Earth's interior leads to systematically underdense results due to the non-inclusion of phase transitions. This trend begins at masses immediately greater than Earth's due to the increasing fraction of the mantle composed of post-perovskite, and then accelerates at masses $\gtrsim16.5M_\oplus$ due to the increasing fraction of the mantle composed of high-pressure oxides (with the most relevant dense phase being Fe\textsubscript{2}P-type SiO\textsubscript{2}).}

\textcolor{black}{In the super-Earth (${<20M_\oplus}$) regime wherein our model radii diverge from simpler models previously published in the literature by 2\%, \textcolor{black}{$\sim$}10\% of recently-published planetary radii have observational precisions 2\% or better \citep{ChristiansenMcElroy2025}--motivating the necessity of the state-of-the-art updates to interior structure modeling presented here for interpretation of observational data. As one moves away from Earth in composition space and temperature space, extrapolation from Earth's interior becomes increasingly invalid and the difference between this work and \citet{ZengSasselov2016,ZengJacobsen2019} increases.} 

\begin{figure}
\centering     \includegraphics[width=1.\columnwidth]{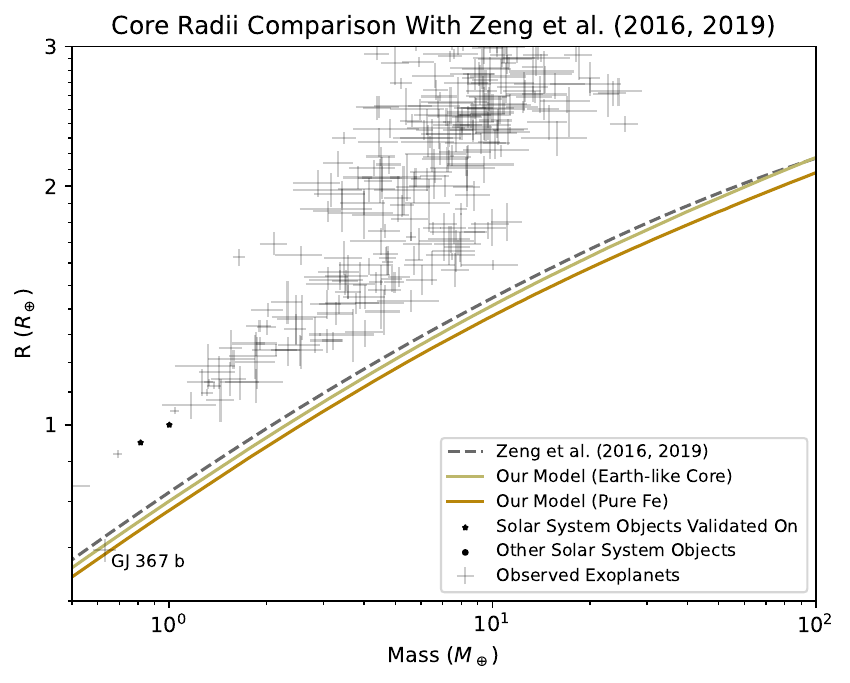}
\vspace{-2em}\caption{The mass-radius relations for planets with a 100\% core mass fraction in our model, both composed of pure \textcolor{black}{iron} and an Earth-like \textcolor{black}{core composition}, compared to the findings of \citet{ZengSasselov2016} \textcolor{black}{for pure iron}. Scatterpoints are the same as in Fig.~\ref{fig:overall_mr}. Note that several additional planets have been observed in this parameter space but do not meet the accuracy criteria in mass and radius for consideration in this publication.}
\vspace{-2em}\label{fig:fe_zeng}
\end{figure}

In Fig.~\ref{fig:fe_zeng}, we compare the radii for our pure-core planets (both with elemental abundances following Earth's and a pure iron composition) \textcolor{black}{with \citet{ZengSasselov2016,ZengJacobsen2019}}, which have radii several $\%$ larger than ours. \textcolor{black}{GJ 367 b has an observed density higher than a pure iron planet in the model of \citet{ZengSasselov2016,ZengJacobsen2019} \citep{GoffoGandolfi2023}, whereas it is compatible with an extremely iron-rich composition in our model.} This suggests  that our model \textcolor{black}{is more physical}. \textcolor{black}{There are two reasons for this. First, \cite{ZengSasselov2016} assume purely liquid cores following Earth's core state. However, our model finds that super-Earths generally have purely solid cores due to the high pressures reached in their interiors. Second, they use a poorly-extrapolating BM2 EOS (Eq.~\ref{eq:BM3} with $K_0^\prime=4$) for iron in the core.} They justify the usage of this EOS by its good extrapolation to ultrahigh pressures (${\sim}12$ TPa), but these pressures are only achieved at ${\sim}100$ $M_\oplus$, a mass beyond the parameter space of interest for \textcolor{black}{iron}-rich planets. \textcolor{black}{\citet{UnterbornDesch2023} also find that \citet{ZengSasselov2016} systematically underestimates the radii of iron-rich bodies compared to their model which assumes purely liquid cores. This confirms that the disagreement between our model and \citet{ZengSasselov2016} is not solely due to a different assumed phase.}

The curves of \citet{ZengSasselov2016,ZengJacobsen2019} are thus not appropriate for application to objects with high core mass fractions as they tend to  overestimate their radii by several \%\textcolor{black}{, comparable to observational errors for the best-characterized planets.}

\begin{figure}
\centering     \includegraphics[width=1.\columnwidth]{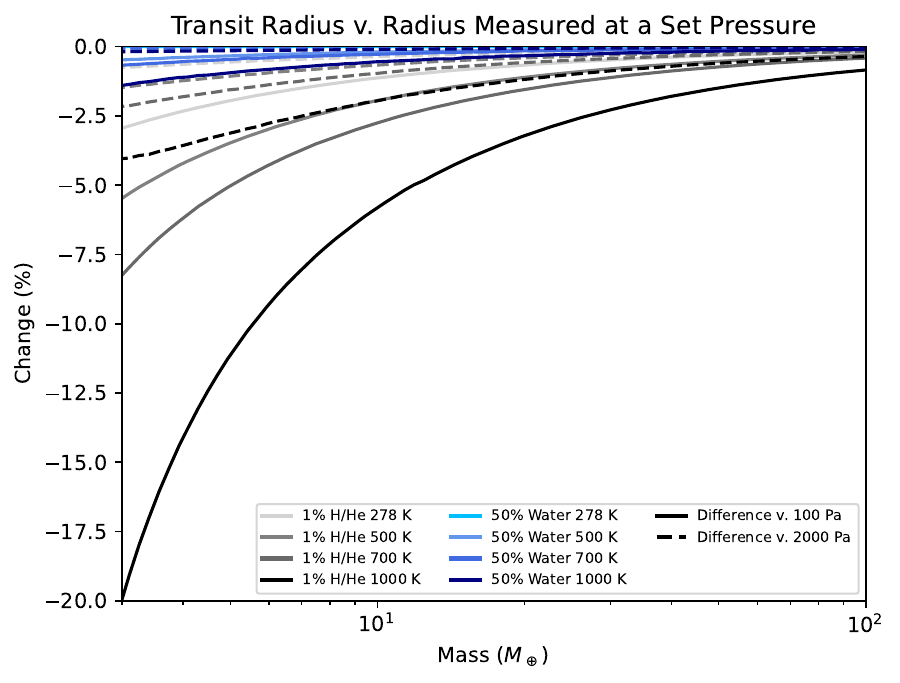}
\vspace{-2em}\caption{The radii of planets in our model as calculated for a transit radius (see section \ref{sec:transit}) compared to the radii of planets in our model at a specific outer boundary pressure. We compare to outer boundary pressures of 100 Pa \textcolor{black}{(solid lines)} and 2000 Pa \textcolor{black}{(dashed lines)}, both of which are common in the literature.}
\vspace{-2em}\label{fig:trans_r}
\end{figure}
In Fig.~\ref{fig:trans_r}, we compare the transit radii of planets calculated in our model to the radii of the planets in our model at set outer boundary pressures of 100 Pa and 2000 Pa (see \ref{sec:transit}). The vast majority of the literature uses radii at set outer boundary pressures (\citet{HaldemannDorn2024} is a notable exception); for example, \citet{ZengJacobsen2019} use an outer boundary pressure of 100 Pa and \citet{LopezFortney2014} use an outer boundary pressure of 2000 Pa. We see that the use of these boundary conditions alone can result in radii several \% larger than the transit radii probed by observations, with the error much more extreme at higher temperatures \textcolor{black}{and lower masses}. The literature thus systematically overestimates planetary radii as compared to observations \textcolor{black}{(unless clouds significantly raise the transit radius, see section \ref{sec:transit})}.

The BICEPS model of \citet{HaldemannDorn2024} is the most similar work in the literature to ours, including mantle Gibbs free energy minimization, a correction for transit radii, an irradiated atmospheric temperature profile using the same prescriptions as here, and using approximately half of the same EOS used here. In contrast with BICEPS, our model uniquely includes more transitions in the high-pressure mantle, preferential partitioning of light elements into the liquid core, and rotation; on the other hand, BICEPS uniquely includes H$\mathrm{_2}$O-H/He mixing (among many other smaller differences). These effects are minor (${\sim}0.5\%$) for our validation sample and thus it is likely that the BICEPS model would reproduce the radii that we validate on to within $<1\%$. However, these effects are not necessarily small in parts of parameter space not probed in the solar system. \citet{HaldemannDorn2024} likely slightly overestimate super-Earth radii by not including the high-pressure phase transitions considered here, while this work moderately underestimates planetary radii for planets with significant H$\mathrm{_2}$O and H/He, if the two are indeed miscible.

The other modern model that includes Gibbs free energy minimization of the mantle using HeFESTo is ExoPlex \citep{UnterbornDesch2018,UnterbornHinkel2018,UnterbornPanero2019,SchulzeWang2021,UnterbornDesch2023}. ExoPlex assumes a fully liquid core and thus systematically underestimates planetary densities (overestimates planetary radii) for planets with solid cores, especially above the mass at which planetary cores may be entirely solid ($\gtrsim2.70M_\oplus$ in this model for Earth-like compositions). ExoPLEX does not currently include ice X and thus is not appropriate for use on planets with very high water content, however this issue has already been identified and will be remedied in future ExoPLEX releases \citep{UnterbornDesch2023}. Unlike our model and BICEPS \citep{HaldemannDorn2024}, ExoPLEX does include Al and Ca in the planetary mantle and thus is the planetary interior structure model that currently models Earth-like mantles most accurately. This is reflected in ExoPLEX's 0.1\% (-0.22\% in an earlier version \citep{UnterbornDesch2018}) error on their model Earth radii, only slightly larger than that found by our model \citep{UnterbornDesch2023}.

The model of \citet{DornKhan2015,DornVenturini2017} includes Gibbs free energy minimization of the mantle with Al and Ca as well. It also solves for an atmosphere of H, He, C, and O \textcolor{black}{in chemical equilibrium} \citep{DornKhan2015,DornVenturini2017}, giving it the most expansive non-core chemical inventory in the literature we have reviewed. However, it does not include light elements in the core, melting, or the newer EOS released since its publication \citep{DornKhan2015,DornVenturini2017}.

ExoPie does not include Gibbs free energy minimization in the mantle but does include the EOS for olivine, wadsleysite, ringwoodite, perovskite, and post-perovskite, as well as the ability to put Fe in the mantle \citep{ValenciaO'Connell2006,ValenciaSasselov2007,PlotnykovValencia2020,PlotnykovValencia2024}. ExoPie does include Ni in the core and uses Si instead of S as the dominant light element--which likely reflects reality, see section \ref{sec:earth} \citep{PlotnykovValencia2020}. ExoPie's model Earth radius has an error of ${\sim}$-0.44\% (from their \textcolor{black}{fig.~2}).

The model of \citet{AcunaDeleuil2021} \textcolor{black}{(an extension of \citet{SotinGrasset2007,BruggerMousis2016,BruggerMousis2017} to include low-temperature ice phases)} also does not include Gibbs free energy minimization, instead including olivine, enstatite, perovskite, and the ability to put Fe in the mantle. This is a slightly more limited inventory than the other mantle models, but the effect of the transition zone (in Earth, the mantle at depths between 420 and 650 km, see Fig.~\ref{fig:earth_structure}) is relatively minor. More significantly, the exclusion of post-perovskite leads to a systematic (if minor, post-perovskite is only ${\sim}$1.5\% denser than perovskite in mantle conditions) underestimation of planetary densities (and thus overestimation of their radii). \citet{AcunaDeleuil2021} do not include a H/He envelope but do include a coupled atmospheric model for which radiative transfer profiles are computed that includes CO$_\mathrm{2}$, a potential advance upon our model for planets with secondary atmospheres \citep{Marcq2012,MarcqSalvador2017}. \citet{BruggerMousis2017}'s model Earth has a radius error of $-0.8\%$.

\citet{AguichineMousis2021} \textcolor{black}{(an extension of \citet{SotinGrasset2007,BruggerMousis2016,BruggerMousis2017} coupling the atmosphere and interior)} also does not include Gibbs free energy minimization, but include an upper mantle made of olivine and enstatite and a lower mantle made of MgSiO$_\mathrm{3}$ and MgO. \citet{AguichineMousis2021} also include S in the planetary core but assume a purely solid core. \citet{AguichineMousis2021}'s mode\textcolor{black}{l} Earth has a radius of 0.992 $R_\oplus$. Unlike this model, \citet{AguichineMousis2021,AguichineBatalha2025} is part of a planetary evolutionary model and thus explicitly accounts for the thermal history of a planet. It thus may be more appropriate for young planets than our model as it treats their thermal structures self-consistently \citep{AguichineMousis2021,AguichineBatalha2025}.

\citet{LopezFortney2014} also present a planetary evolution model, albeit focused on H/He-enveloped worlds rather than the water worlds of \citet{AguichineMousis2021}. \citet{LopezFortney2014} use the older H/He EOS of \citet{SaumonChabrier1995} \textcolor{black}{and do not include advances in H/He EOS after 2014, such as the non-ideal effects published by \citet{HowardGuillot2023} \textcolor{black}{or recent high-pressure simulations of \citet{ChabrierMazevet2019}}}, resulting in systematically lower densities and thus larger radii. \citet{LopezFortney2014} also approximate planetary cores with EOS for olivine and iron. The simplified nature of the\textcolor{black}{ir} prescription for the core and mantle makes the\textcolor{black}{ir} model unfit for application to Earth-like planets.

The MAGRATHEA model of \citet{HuangRice2022,RiceHuang2025} also does not include Gibbs free energy minimization, instead including olivine, enstatite, perovskite, and a silicate melt. Crucially, MAGRATHEA does not include Fe in the mantle, although \citet{RiceHuang2025} does discuss a basic prescription to account for this by modifying the MgSiO$_\mathrm{3}$ EOS that is not in the default model. MAGRATHEA also does not include light elements in the core by default, although \citet{RiceHuang2025} discuss the addition of a light element in the core. \citet{HuangRice2022}'s model Earth has a radius of 0.967$R_\oplus$ by default and 0.9884 $R_\oplus$ if temperature jumps are included.

The model of \citet{vandenBergYuen2019} neglects low-pressure phases in the mantle but does include the high-pressure phase transitions of \citet{UmemotoWentzcovitch2017} as in our model. \citet{vandenBergYuen2019} do not include Fe in the mantle or light elements in the core. These simplifications are reflected in \citet{vandenBergYuen2019}'s $-1.30\%$ error model Earth radius, although we do note that \citet{vandenBergYuen2019} is the only other publication to include \citet{UmemotoWentzcovitch2017}'s high-pressure mantle phase transitions and thus is not subject to the systematic overprediction of super-Earth radii elsewhere in the literature. \citet{vandenBergYuen2019} also numerically model convection, allowing a more realistic interior temperature evolution than modeled here.

\citet{BoujibarDriscoll2020} do not include Fe in the mantle, but do include a mantle prescription composed of peridotite, perovskite, post-perovskite, and the fictitious phases post-post-perovskite 1 and 2, which have EOS calculated \textcolor{black}{by} assuming that the EOS jumps between peroskite and post-perovskite and replicated once and twice, respectively. \citet{BoujibarDriscoll2020} do not include Fe in the mantle or light elements in the core but do allow for both a solid and liquid core.

The APPLE model of \citet{SurSu2024} is built for application to giant planets and thus has a simplified rock prescription that assumes pure post-perovskite\textcolor{black}{, in contrast with our comprehensive mineralogy}. Unlike our model, APPLE includes semi-convection, helium rain, and metallicity gradients in envelopes \citep{SurSu2024}, all of which are crucial to the interior structures of gas giants. APPLE is thus appropriate for application to planets either above our 100 $M_\oplus$ cutoff or that are majority H/He such as Saturn \citep{SurSu2024,TejadaArevaloSu2024,SurTejadaArevalo2025}.

% extrapolating the density profile of Earth from PREM to pressure regimes beyond the Earth

Finally, we return to the interior structure \textcolor{black}{model} with the broadest use in the community due to the ease of access to its mass-radius curves, that of \citet{ZengSasselov2016,ZengJacobsen2019}. It is constructed by \textcolor{black}{generating EOS from a fit to Earth's interior profile from PREM} and thus has no explicit treatment of mineral physics, making a direct comparison difficult as a model Earth's radius has no error by construction. This has the effect of avoiding the systematic mantle density underprediction (radius overprediction) from excluding iron in the mantle, but comes at the cost of not including any high-pressure phase transitions. Additionally, as planets around different stars (and even different planets around our own sun) have different compositions, extrapolations assuming Earth-like compositions are not strictly valid.

In sum, our model recovers the Earth's radius more accurately than any other model in the literature because it includes more core and mantle phases and more up-to-date EOS than any other model in the literature (to the author's knowledge). The other models that include Gibbs free energy minimization and light elements in the core, BICEPS and ExoPlex, have comparable model Earth radii but do not include high-pressure mantle EOS and thus grow less reliable with increasing planet mass \citep{UnterbornDesch2023,HaldemannDorn2024}. The most commonly used model in the literature, \citet{ZengJacobsen2019}, is inappropriate for planets with significant atmospheres, large cores, significantly non-Earth-like chemical compositions, or with large enough masses that an extrapolation from Earth's conditions misses additional phase transitions.

\subsection{Water Outside the Water Layer}
\label{sec:water_outside_water}
\subsubsection{Water in the interior}
Our model does not (yet) include volatile sequestration within planetary interiors, which can significantly reduce the radii of planets ($3$--$25\%$) \citep{DornLichtenberg2021,ShahAlibert2021,LuoDorn2024}. This effect is particularly pronounced for planets with magma oceans, into which water partitions much more readily than solids \citep{DornLichtenberg2021,Elkins-Tanton2008}. The details of where water can be partitioned relies on the pressures and thus depths at which water equilibrated with the surrounding material, requiring a planetary evolutionary model rather than a static model as presented here \citep{LuoDorn2024}. A particular process of interest is chemical equilibration with the atmosphere, which can destroy the vast majority (up to ${\sim}95\%$) of accreted water \citep{WerlenDorn2025}. The interplay of these processes as well as atmospheric escape is necessary to fully capture the distribution of water within the planet population. 

That said, our model serves as a useful representative of planets without water sequestration. If sequestration induces a systematic reduction of planetary radii, it would appear as a discrepancy between our model's predicted and the actual radius, with ours being larger than the planet's true radius. \textcolor{black}{Thus, if observations could constrain some of the free parameters of Table \ref{tab:free_params} independent of interior structure models--such as by assuming that planets inherit elemental abundance ratios from their stars to set core mass fractions and mantle abundance ratios--and the results were input into our interior structure model, the resultant radius would be lower} than observed. This discrepancy could be used \textcolor{black}{as evidence for} water sequestration, as has been done (using a different interior structure model) \textcolor{black}{by \citet{WeissermanGromek2026}}.

\subsubsection{Water in the H/He envelope}
Our model currently does not address the miscibility of H/He and water.  This can produce a mixed H/He-water layer as is included \textcolor{black}{by} \citet{HaldemannDorn2024}. \citet{BurnMordasini2024b} find that this effect is crucial in replicating the radius valley with their model of planet formation.  Limited experimental studies have been conducted on such miscibility, but only at pressures up to a few GPa. At very low ($<0.2$ GPa) pressures, \citet{SewardFranck1981} find H/He and water \textcolor{black}{miscible at temperatures above $650$ K}. At ${\sim}$ GPa pressures, \cite{BaliAudétat2013,VlasovAudetat2023} find that H/He and water are miscible within silicate material at temperatures above ${\sim}900$ K, with the temperature at which miscibility is possible moving to higher temperatures with higher pressures, implying high-pressure immiscibility of H/He and water, although such an extrapolation cannot continue indefinitely \citep{BergermannFrench2024}. 

Given the uncertainty of extrapolations due to the lack of direct high-pressure probing by experimentation, the high-pressure miscibility of H/He and water is determined by simulation. \citet{SoubiranMilitzer2015} find H/He and water are always miscible between 2 and 70 GPa and 1000 and 6000 K, while \citet{BergermannFrench2024} use a similar methodology but include nuclear quantum effects and find that at pressures above 10 GPa, H/He and water become immiscible for $T<2000$ K \citep{BethkenhagenFrench2013}. \textcolor{black}{\citet{GuptaStixrude2025} find a similar result, with the boundary between miscibility and immiscibility scaling from ${\sim}$1000 K at ${\sim}$1 GPa to ${\sim}$6000 K at ${\sim}$3000 GPa.} Some reasonable planet parameters within our model (e.g 1\% H/He 99\% Earth-like, $10M_\oplus$, \textcolor{black}{$T_\mathrm{eq0}$} of 300 K) have temperature profiles lying on the immiscible side of this line while others (e.g 1\% H/He 99\% Earth-like, $10M_\oplus$, \textcolor{black}{$T_\mathrm{eq0}$} of 1000 K) have temperature profiles that cross into the miscible region. \textcolor{black}{\citet{Piaulet-GhorayebThorngren2025} use these curves and an interior structure model (with a significantly less complete set of EOS than used here) to find that sub-Neptunes with an equilibrium temperature of 350 K and atmospheric metallicities above $\sim 100 \times$ solar likely have atmospheric regions that cool below the miscibility threshold, resulting in demixing and a gradient of water content in the atmosphere rather than the extreme end members of either a fully-mixed envelope or separate H/He and water layers.}

We do not include any H/He mixing with water due to the sparse sampling in P-T space of miscibility studies, the complex consequences of the process of demixing\textcolor{black}{, and the incompatibility of a compositional gradient with this model's assumption of a layered interior}. If H/He-water mixing is indeed significant\textcolor{black}{--which several JWST observations have found tentative evidence for \citep[e.g.][]{BennekeRoy2024,Piaulet-GhorayebBenneke2024,AhrerRadica2025,CoulombeBenneke2025}--}our model underestimates radii by placing water at systematically higher pressures than it would be in if it was mixed \citep{BurnMordasini2024b}. In any case, our model can be taken as a no H/He-water mixing end member \textcolor{black}{and will serve as the basis for future work incorporating H/He-water mixing}.

\section{Conclusion} \label{sec:conclusion}
\subsection{Summary} \label{sec:sum}
We have constructed a new interior structure model that covers the entire parameter space of exoplanets besides gas giants. This interior structure model improves upon the literature by incorporating the following physics simultaneously:
\begin{itemize}
    \item A Gibbs free energy minimization equilibrium mineralogy inside the upper mantle
    \item High-pressure phases of mantle material
    \item Low-pressure phases of iron
    \item Light elements in the liquid and solid core and partitioning between the two
    \item Temperature jumps at the top and bottom of the mantle
    \item A prescription to calculate transit radii instead of radii at a set pressure surface
    \item An upper atmospheric temperature profile from radiative transfer equations
    \item Thermal terms in all equations of state
    \item H/He non-ideal mixing
    \item Equations of state for iron that extrapolate well to high pressures
    \item Rotation
\end{itemize}

We validate our work by creating forward models for bodies in the solar system using parameters derived--wherever possible--from constraints independent of interior structure models (such as seismology and meteoritics) in the literature. Our validation samples and results are:
\begin{itemize}
    \item Earth: We find radii and moment of inertia coefficients within 0.2\% of measurements. Our model also reproduces the locations of phase transitions in the mantle, the mantle-core boundary, and the outer-inner core boundary to extreme accuracy (see Fig.~\ref{fig:earth_structure}). To our knowledge, this is the best replication of Earth for an exoplanetary interior structure model in the literature.
    \item Mars and the Moon: These are the only two extraterrestrial bodies for which seismological data are available. For these worlds, we find radii and moment of inertia coefficients within 0.5\% of reality. 
    \item Venus, Mercury, and Europa: For these bodies, we compute radii and moment of inertia coefficients that are within 1\% or 3$\sigma$ of observations.
\end{itemize}
     
We encourage authors of interior structure model publications targeting Earth-like planets to include their model Earth's $r-\rho$ profile, radius and moment of inertia coefficient to allow assessment \textcolor{black}{their model's} accuracy.

\textcolor{black}{We generate radii for 32,97\textcolor{black}{5} model planets with masses spanning $0.01M_\oplus$ to $100M_\oplus$, H/He mass fractions spanning 0\% to 30\%, water mass fractions spanning 0 to 50\%, core mass fractions spanning 0 to 100\%, and zero-albedo equilibrium temperatures spanning 278 K to 2500 K. We find:}

\begin{itemize}
       \item The M-R relation is more general than can be described by power law fits, with planets of all compositions getting denser with mass than can be described by a simple power law (see Fig.~\ref{fig:overall_mr}) owing to compression and phase transitions in the interior
       \textcolor{black}{\item Within a restricted range of terrestrial planet masses, we fit new $M=bR^a$ relations to these curves as piecewise functions}
       \textcolor{black}{\item Non-ideal mixing and new high-pressure EOS create systematically denser planets in the low-to-intermediate mass regime than previous studies}
       \textcolor{black}{\item Steam atmospheres can increase the radii of water-rich worlds compared to the case in which all their water is ice by tens of \%}
       \textcolor{black}{\item The melting of planetary mantles and cores can increase their radii by several \%}
       \item The usage of outer boundary pressure radii rather than transit radii to define model radii can result in several \% systematic overestimates of planetary radii
       \item The literature systematically underestimates the radii of high-mass planets by neglecting high-pressure phase transitions in the mantle
       \item The iron-rich EOS of \citet{ZengSasselov2016} extrapolates poorly to the planetary core pressure regime, leading to systematic overestimates of planetary radii for iron-rich bodies
       \textcolor{black}{\item All of these systematic effects can cause changes in radii larger than observational radii uncertainties}
\end{itemize}

Our mass-radius relations are available at \url{https://github.com/Bennett-Skinner/SkinnerPudritzCloutier2026-MR-curves/} and \url{https://doi.org/10.5281/zenodo.20382154} and we strongly encourage their use throughout the community. As an example of work to come, appendix \ref{sec:example_planet_profiles} provides the interior structure profile of a model 20$M_\oplus$ super-Earth.
 
\subsection{Future Work} \label{sec:future_work}
Future improvements to the model will include an expansion of the chemical inventory in the planetary mantle and core, the sequestration of water within the planetary interior, \textcolor{black}{the mixing of H/He and water in planetary atmospheres,} and the increasingly curious question of carbon\textcolor{black}{, which could present a significant component of sub-Neptune interiors} \citep{HaldemannDorn2024,LiBergin2026}. Our model will also move beyond the simple melting curve of Eq.~\ref{eq:melting_T} and incorporate the experimentally-derived melting temperatures of mixtures.

\textcolor{black}{This model has already been used in the work of \citet{OsbornCloutier2026} to infer the composition of the hot super-Neptune TOI-672 b, whose position in the mass-radius diagram is poorly sampled by existing mass-radius relations. Future work will use this model to infer the compositions of more individual planets via the generation of additional mass-radius curves. New mass-radius curves using this model tuned to specific systems can be generated upon correspondence with the lead author.}

Ultimately, our forward model will serve as the basis for a code for retrieving planetary compositions from measured masses and radii in a forthcoming publication. The use of this model to recalculate radii from the planet population synthesis model of \citet{AlessiPudritz2017,AlessiPudritz2018,AlessiInglis2020,AlessiPudritz2022} for comparison to observations is also forthcoming.

\section*{Acknowledgements}
BNS led the programming, analysis, writing, \textcolor{black}{editing,} and interpretation. REP and RC contributed to the analysis, editing, and interpretation, while overseeing the project.

We are indebted to Artyom Aguichine, Komal Bali, Remo Burn, Caroline Dorn, Jonas Haldemann, \textcolor{black}{David Rice,} Laura Schaefer, Ankan Sur, and Emerson Tao for stimulating discussions. \textcolor{black}{We are also indebted to Xinran Liu for sharing computing resources.} \textcolor{black}{Finally, we thank the referee for a thoughtful report that thas improved the quality of the manuscript.} BNS is supported by McMaster University. REP and RC are supported by the Natural Sciences and Engineering Council of Canada (NSERC) through the Discovery Grants program.

This work makes use of the Exoplanet Archive (DOI: 10.26133/NEA12) \citep{ChristiansenMcElroy2025}, \texttt{NumPy} \citep{HarrisMillman2020}, \texttt{pandas} \citep{Pandas2025}, \texttt{Matplotlib} \citep{Hunter2007}, and \texttt{SciPy} \citep{VirtanenGommers2020}.

\section*{Data Availability}
The data underlying this article is available in a GitHub repository at the following link: \url{https://github.com/Bennett-Skinner/SkinnerPudritzCloutier2026-MR-curves/}. It is also available at a Zenodo repository at the following link: \url{https://doi.org/10.5281/zenodo.20382154}. Due to file size limitations, the full interior profile of objects in our sample is only available on Zenodo. Due to file number limitations, the files available from Zenodo are provided zipped, whereas the files available from GitHub can be directly downloaded.

\bibliography{main,nonadssources}{}
\bibliographystyle{mnras}

\appendix
\section{Numerical Details}\label{sec:num_details}
Our solver is multi-threaded, solving five problems at once on different threads. Two threads integrate inward from the surface, one for the guessed radius $r$ and another for a slightly higher value of the radius, $r(1+d)$. Three threads integrate outward from the core, one for the guessed pressure and temperature $P,T$, one for a slightly higher value of pressure $P(1+d),T$, and one for a slightly higher value of temperature $P,T(1+d)$. The integrations at slightly different values are calculated to perform numerical derivation for the Newton-Raphson method. We find \textcolor{black}{universal convergence in our sample when using a value of $d$ of $10^{-3}$, reduced to $10^{-5}$ if melting occurs in the core or mantle and $10^{-6}$ if the phase of the core is different at the fitting mass for different threads.}

The \textcolor{black}{fitting mass at which our outwards and inwards integrations meet is by default the mass of the core, but becomes a quarter the mass of the core if the phase of the core is different at the fitting mass for different threads. If the core mass is 100\% of the planet's total mass, the fitting mass is set to 40\% of the planet's total mass. If the core mass is 0\% of the planet's total mass, the fitting mass is set to the lowest value of (1) the envelope-water boundary, (2) the water-mantle boundary, and (3) 40\% of the planet's total mass.}

\textcolor{black}{We impose agreement at the fitting mass } of $r$, $P$, and $T$ \textcolor{black}{by factors of $10^{-4}$, $10^{-3}$, and $10^{-3}$, respectively.} \textcolor{black}{Our chosen threshold for relative error in Cash-Karp steps is $10^{-5}$. If we significantly increase ou\textcolor{black}{r} numerical accuracy by imposing an agreement of $10^{-5}$ in all dimensions and using a relative error threshold of $10^{-7}$, our radii do not change to the third decimal place, justifying that our numerical parameters are sufficient for the level of accuracy in this publication. However, planetary moment of inertia coefficients can vary at the third decimal place and thus for our validation sample \textcolor{black}{we} set the agreement threshold to $10^{-5}$ and the relative error threshold to $10^{-7}$. If melting occurs in the core or mantle, we set our relative error to $10^{-7}$, and if that melting occurs at the fitting mass, we set our relative error to $10^{-8}$.} \textcolor{black}{To avoid proximity to the divergence of $r$, we treat $R/\max{(R_p,m_p^{1/3.7})}=10^{-6}$ as the core of the planet. Lowering this value to $10^{-7}$ results in numerical instabilities due to the proximity to the divergence at the core. The max function in the denominator is used so that in planets where an envelope makes up the majority of the planetary radius, the innermost radius of the planet is still well within the planetary core.}

We need reasonable values of the pressure and temperature of the planetary \textcolor{black}{centre} to begin iterating. We attain these by integrating from the outer boundary condition to the inner boundary condition and refining our guesses via the Newton-Raphson method. In other words, we perform single-shooting integration to seed our initial guess for double-shooting integration.

Whe\textcolor{black}{n} we construct our own EOS table\textcolor{black}{s}, they are uniformly spaced in log space such that $\frac{P_{i+1}}{P_i}=\frac{T_{i+1}}{T_i}=1.01$. For the HeFESTo tables generated by Perple\_X, we use the highest possible number of datapoints and construct square tables. This corresponds to spacings of $P_{i+1}/P_i\lesssim1.012$ and $T_{i+1}-T_i\lesssim3$ K (the maximum number of datapoints permitted depends on composition). We interpolate in log space for all tables besides HeFESTo, where we interpolate in linear space. This is because the parameter space of HeFESTo is the smallest of our dataset. \textcolor{black}{Our generated tables only go up to a temperature of $10^5$ K and thus all temperatures above $10^5$ K are treated as $10^5$ K when calculating EOS. Expressed mathematically, $\rho(P,T>10^5\textrm{ K})=\rho(P,10^5\textrm{ K})$.}

%As $r$ and $m$ are not identically zero at our inner boundary condition, calculations of the local critical rotation can be unreliable for guesses too far from the solution. We thus set $\omega = \Omega f(m_\mathrm{p})/\sqrt{Gm_\mathrm{p}}$ when single shooting, where $f(m_\mathrm{p})$ is a rough estimate of the planetary radius from its mass. We found that the $\frac{M}{M_\oplus}=(1+0.55w_{\mathrm{H\textsubscript{2}O}}-0.14w_{\mathrm{H\textsubscript{2}O}}^2)(\frac{R}{R_\oplus})^{\frac{1}{3.7}}$ relation of \citet{ZengJacobsen2019}, multiplied by 1.5 for planets with envelope mass fractions above $10^{-2}$ and $2$ for planets with equilibrium temperatures above or equal to 700 K (if they have a non-condensed surface) was sufficient for this purpose.

Even for guesses close to the solution, the calculation of $\omega$ for $m/m_\mathrm{p}<10^{-4}$ or $r/r_\mathrm{out}<10^{-4}$ is unreliable. We thus set $r_e=r$ at these small mass \textcolor{black}{coordinates close to the \textcolor{black}{centre}}.

\textcolor{black}{The Newton-Raphson method formally requires a smooth function, which is not the case when melting occurs in the mantle and/or core. We address this complication by smoothing our EOS over the melting curve with a tanh function.  Specifically we calculate the melt volume fraction (note this is a volume fraction rather than a mass fraction) as:
\begin{equation}
    X_\mathrm{Melt}=0.5\Big(1+\tanh{\left(\frac{T-T_\mathrm{Melt}(P)}{T_\mathrm{Smooth}}\right)\Big)}\mathrm{,}
    \label{eq:smoothing}
\end{equation}
where $T_\mathrm{Smooth}$ parameterizes the range of temperatures over which smoothing occurs. We find that $T_\mathrm{Smooth}=\max{(20\text{ K},0.005T)}$ leads to universal convergence within our sample. Hyperbolic tangents are commonly used to smooth over melting in the literature \citep[see e.g.][]{DannbergGassmoller2022}, although it is typically done with pressure as the independent variable rather than temperature. We formulate it this way because all \textcolor{black}{of} our melting curves are in the form $T_\mathrm{Melt}(P)$ rather than $P_\mathrm{Melt}(T)$. Eq.~\ref{eq:smoothing} can be interpreted as representing a melt mixture, however we emphasize that $T_\mathrm{Smooth}$ is chosen solely for the purpose of convergence. Our implementation is thus not expected to reflect the real temperature range over which melt mixtures are present. This prescription is not required for the convergence of our validation sample and thus is not used in our validation sample, the impact on the final radii is below the precision that we report in this publication, although moment of inertia coefficients can be \textcolor{black}{a}ffected by up to \textcolor{black}{$\sim$0.1\%.}}

\textcolor{black}{Having calculated $X_\mathrm{Melt}$, we use the additive volume law to calculate $\rho=X_\mathrm{Melt}\rho_\mathrm{Liquid}+(1-X_\mathrm{Melt})\rho_\mathrm{Solid}$, where $\rho_\mathrm{Liquid}$ and $\rho_\mathrm{Solid}$ are the densities of the liquid and solid phases, respectively. We also calculate $\nabla_\mathrm{ad}=X_\mathrm{Melt}\nabla_\mathrm{ad,Liquid}+(1-X_\mathrm{Melt})\nabla_\mathrm{ad,Solid}$, which is not thermodynamically consistent as $\nabla_\mathrm{ad}$ is not extensive. We cannot use a thermodynamically consistent framework (i.e. using method 1 or method 2 in section \ref{sec:ad_grad_discussion} and combining the thermodynamic variables following section \ref{sec:combination}) because different thermodynamic frameworks are used in different pressure regimes (i.e. method 1 and method 2 use different thermodynamic parameters).}

\section{Stoichiometry}\label{sec:stoichiometry}
Our liquid mantle mantle \textcolor{black}{has} species abundances calculated identically to \citet{HaldemannDorn2024}. \textcolor{black}{The stoichiometry of the solid mantle that includes post-perovskite and lower-pressure phases or is completely dissociated is identical to \citet{HaldemannDorn2024}, while the stoichiometry at intermediate pressures is determined by Equations ~\ref{eq:mgo_step_1}-\ref{eq:sio2_step_3} in section \ref{sec:mantle}.}

\textcolor{black}{For the solid core, the total number of Fe, FeS, and FeO are given by Equations \ref{eq:N_Fe_Solid}-\ref{eq:N_FeO_Solid} \textcolor{black}{(where here and throughout this appendix $\tilde{N}$ represents the number of a species and $N$ represents the number of an element while a $s$ superscript means solid and an $l$ superscript means liquid),}}
\begin{equation}
    \tilde{N}^s_\mathrm{Fe} = N^s_\mathrm{Fe}-N^s_\mathrm{S}-N^s_\mathrm{O}
    \label{eq:N_Fe_Solid}
\end{equation}
\begin{equation}
    \tilde{N}^s_\mathrm{FeS} = N^s_\mathrm{S}
    \label{eq:N_FeS_Solid}
\end{equation}
\begin{equation}
    \tilde{N}^s_\mathrm{FeO} = N^s_\mathrm{O}
    \label{eq:N_FeO_Solid}
\end{equation}
\textcolor{black}{derived by the constraints that each FeS depletes Fe and S at a 1:1 rate and each FeO depletes Fe and O at a 1:1 rate.
When calculating molar fractions, we need to calculate total molar mass to obtain the denominator, yielding Eq.~\ref{eq:solid_denominator}:}
\begin{equation}
    N^s_\mathrm{Fe}-N^s_\mathrm{S}-N^s_\mathrm{O}+N^s_\mathrm{S}+N^s_\mathrm{O}=N^s_\mathrm{Fe}\mathrm{.}
    \label{eq:solid_denominator}
\end{equation}
\textcolor{black}{Dividing Equations \ref{eq:N_Fe_Solid}-\ref{eq:N_FeO_Solid} by Eq.~\ref{eq:solid_denominator} and utilizing the fact that number fractions are proportional to molar fractions yields Equations \ref{eq:x_Fe_solid}-\ref{eq:x_FeO_solid}:}
\begin{equation}
\tilde{x}^s_\mathrm{Fe} = 1-\frac{x^S_\mathrm{S}+x^S_\mathrm{O}}{x^S_\mathrm{Fe}}
\label{eq:x_Fe_solid}
\end{equation}
\begin{equation}
\tilde{x}^s_\mathrm{FeS} = \frac{x^S_\mathrm{S}}{x^S_\mathrm{Fe}}
\label{eq:x_FeS_solid}
\end{equation}
\begin{equation}
\tilde{x}^{s}_\mathrm{FeO} = \frac{x^S_\mathrm{O}}{x^S_\mathrm{Fe}}
\label{eq:x_FeO_solid}
\end{equation}

For the liquid core, the total number of Fe, FeS, and FeO are given by Equations \ref{eq:N_Fe}-\ref{eq:N_FeO},
\begin{equation}
    \tilde{N}^l_\mathrm{Fe} = N^l_\mathrm{Fe}-\frac{81}{19}N^l_\mathrm{S}-N^l_\mathrm{O}\mathrm{,}
    \label{eq:N_Fe}
\end{equation}
\begin{equation}
    \tilde{N}^l_\mathrm{FeS} = N^l_\mathrm{S}\mathrm{,}
    \label{eq:N_FeS}
\end{equation}
\begin{equation}
    \tilde{N}^l_\mathrm{FeO} = N^l_\mathrm{O}\mathrm{,}
    \label{eq:N_FeO}
\end{equation}
derived by the constraints that each Fe$_\mathrm{0.81}$S$_\mathrm{0.19}$ depletes 81 times as many free Fe atoms as free S atoms, each FeO deplets Fe and O at a 1:1 rate, and no free S or O are allowed. Note that $\tilde{N}_\mathrm{Fe}$ represents the number of free Fe atoms while $N_\mathrm{Fe}$ represents the total number of Fe atoms \textcolor{black}{(free or otherwise)}.

When calculating molar fractions, we need to calculate total molar mass to obtain the denominator, yielding Eq.~\ref{eq:denominator}:
\begin{equation}
    N^l_\mathrm{Fe}-\frac{81}{19}N^l_\mathrm{S}-N^l_\mathrm{O}+N^l_\mathrm{S}+N^l_\mathrm{O}=N^l_\mathrm{Fe}+\textcolor{black}{(1-\frac{81}{19})}N^l_\mathrm{S}\mathrm{.}
    \label{eq:denominator}
\end{equation}

Dividing Equations \ref{eq:N_Fe}-\ref{eq:N_FeO} by Eq.~\ref{eq:denominator} and utilizing the fact that number fractions are proportional to molar fractions yields Equations \ref{eq:x_Fe}-\ref{eq:x_FeO}:
\begin{equation}
    \tilde{x}^l_\mathrm{Fe}=\frac{x^l_\mathrm{Fe}-\frac{81}{19}x^l_\mathrm{S}-x^l_\mathrm{O}}{x^l_\mathrm{Fe}+(1-\frac{81}{19})x^l_S}\mathrm{,}
    \label{eq:x_Fe}
\end{equation}
\begin{equation}
    \tilde{x}^l_\mathrm{FeS}=\frac{x^l_\mathrm{S}}{x^l_\mathrm{Fe}+(1-\frac{81}{19})x^l_S}\mathrm{,}
    \label{eq:x_FeS}
\end{equation}
\begin{equation}
    \tilde{x}^l_\mathrm{FeO}=\frac{x^l_\mathrm{O}}{x^l_\mathrm{Fe}+(1-\frac{81}{19})x^l_\mathrm{S}}\mathrm{.}
    \label{eq:x_FeO}
\end{equation}

\textcolor{black}{Setting $x^l_\mathrm{O}$ or $x^S_\mathrm{O}$ to zero recovers the equations of \citet{HaldemannDorn2024}.}

\section{Additional Mass-Radius Curves}\label{sec:more_mr_curves}
\begin{figure}
\centering
\includegraphics[width=1.\columnwidth]{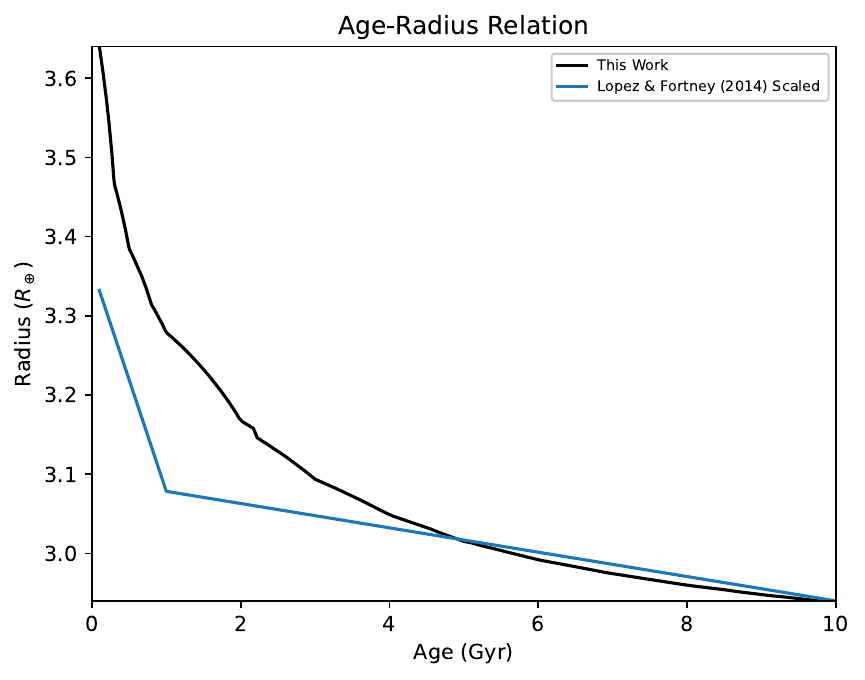}
\vspace{-2em}\caption{\textcolor{black}{The radius of a planet with age in this work and in \citet{LopezFortney2014}, with the radius from \citet{LopezFortney2014} scaled such that it is equal to this work at 10 Gyr. The planet is $13M_\oplus$ and has a 99\% Earth-like, 1\% H/He composition with an instellation of $1000F_\oplus$, corresponding to an equilibrium temperature of $(1-A_B)T_\mathrm{eq}=1565$ K.}}
\vspace{-2em}\label{fig:mr_curve_varyage}
\end{figure}
\textcolor{black}{The effect of age on a planet's radius for an example case with comparison to \citet{LopezFortney2014} is shown in Fig.~\ref{fig:mr_curve_varyage}. \textcolor{black}{The drop in radius at 2.2 Gyr corresponds to the solidification of the core as the planet cools.} The \textcolor{black}{remaining }non-smoothness of our curve is because our internal luminosity is calculated by interpolating over the low-resolution table of \citet{Mordasini2020}. Note the relative agreement at ages of 4.5 Gyr and greater. The shrinking of radius with age is not entirely due to the contraction of the envelope--the underlying mantle and core reduce in radius by \textcolor{black}{2.9\%} due to the higher pressures and lower temperatures at the envelope-mantle interface with age, shrinking the total planetary radius by \textcolor{black}{1.6\%}.}

\begin{figure}
\centering
\includegraphics[width=1.\columnwidth]{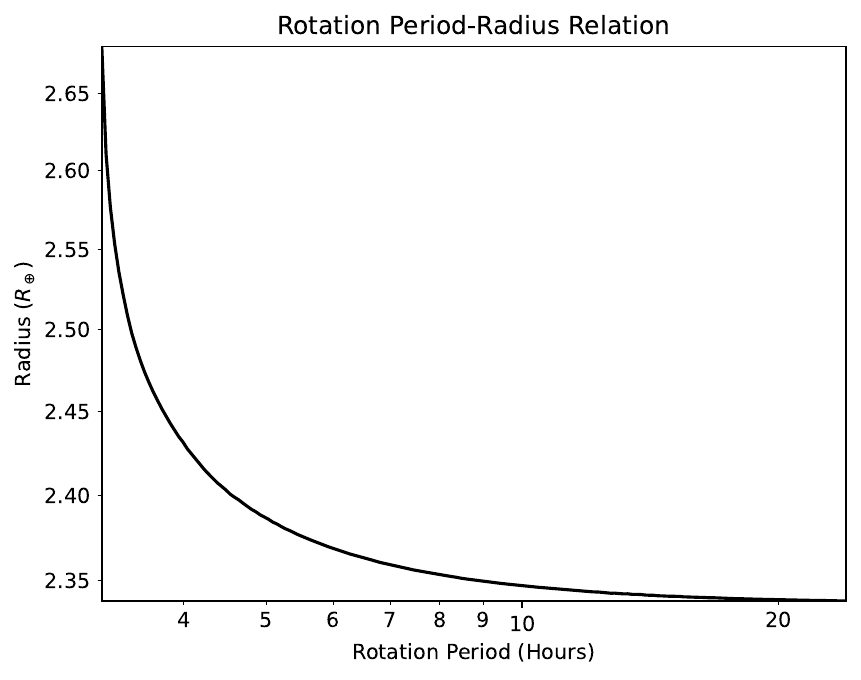}
\vspace{-2em}\caption{\textcolor{black}{The radius of a planet with rotation period. Note log scale on x axis. The planet is $10M_\oplus$ and has a 99\% Earth-like, 1\% H/He composition with a zero-albedo equilibrium temperature of 700 K.}}
\vspace{-2em}\label{fig:mr_curve_varyrot}
\end{figure}

\textcolor{black}{The effect of rotation period on a planet's radius for an example case is shown in Fig.~\ref{fig:mr_curve_varyrot}. Periods are from \textcolor{black}{3.21} to 24 hours, corresponding to \textcolor{black}{$\omega=0.78$} to $0.07$. Over this range, a faster-rotating planet has a radius \textcolor{black}{14.6\%} larger than a slower-rotating planet. Shorter periods than those shown are numerically unstable. The radius asymptotes at a rotation period of less than a day, indicating that the likely tidally-locked population probed by most exoplanet observations is not significantly impacted by rotation. However, short rotation periods are expected in the immediate aftermath of giant impacts such as the one that formed the Moon \citep[e.g.][]{Canup2012}.}

\section{Saturn}
\label{sec:saturn}
Our model does not include physics critical for gas giant interior models such as helium rain, diffuse cores, or conductive temperature gradients \citep{MankovichFortney2020,MankovichFuller2021,PreisingFrench2023,SurSu2024,SurTejadaArevalo2025}. Additionally, Saturn's equilibrium temperature is outside of the boundaries of the radiative transfer temperature profiles from \cite{ParmentierGuillot2015} and thus our temperature profile relies on potentially inaccurate extrapolation.

Nevertheless, Saturn has a mass $<100M_\oplus$ and serves as a useful edge case test that our prescription for H/He and high-density matter is sensible, bolstering confidence in the applicability of our model to sub-Neptunes with few $\%$ H/He envelopes that are within the scope of our model but do not exist in the solar system and thus for which we have no direct validation.

We take Saturn as having its reported albdeo of 0.41 \citep{WangLi2024}, $Z_\mathrm{atm}=3.16$ roughly following the upper atmospher\textcolor{black}{ic} value of \textcolor{black}{$\sim$5} found in \citet{SurTejadaArevalo2025}, solar $w_\mathrm{H}$ and $w_\mathrm{He}$, and a non-H/He centre that is 50\% water and 50\% Earthlike. If we take the centre as having the mass of Saturn's core found in the model of \citet{SurTejadaArevalo2025} that replicates Saturn's current properties well (4$M_\oplus$), we get a radius of 9.\textcolor{black}{253}$R_\oplus$ (\textcolor{black}{1.23}\%) and a MoIC of 0.25\textcolor{black}{64}\footnote{As is standard for Saturn \citep{SaillenfestLari2021}, we normalize Saturn's MoI using our model's equatorial radius in Eq.~\ref{eq:MoI} rather than using Saturn's volume-equivalent radius. Note that Eq.~\ref{eq:MoI} assumes spherical symmetry, we do not account for the flattening of objects in our MoI calculations.} (1\textcolor{black}{6.54}\%). If we take the total mass of Saturn's fuzzy core as 17$M_\oplus$--as derived by observations of pulsation modes in Saturn's rings originating from Saturn's interior \citep{MankovichFuller2021}--we get a radius of 8.\textcolor{black}{477}$R_\oplus$ (-\textcolor{black}{7.26}\%) and a MoIC of 0.2\textcolor{black}{296} (4.\textcolor{black}{37}\%). Although high, our trend in errors is directionally correct: adding more material to the centre decreases the planetary radi\textcolor{black}{us}, indicating a higher average density, and decreases the planetary MoIC, indicating a greater degree of mass concentration towards the planet's center.

The most prominent reason for our error is our simplified temperature structure. The combined effects of Helium rain and conduction cause our temperature at $\frac{m}{M_\mathrm{pl}}=0.2$ to be ${\sim}$12,000 K lower than \citet{SurTejadaArevalo2025}'s model (15,000 K v. 3,000 K). This nearly order of magnitude lower temperature leads to systematically higher densities, explaining our systematic underprediction of Saturn's radius \textcolor{black}{when using a reasonable total metallic mass.} \textcolor{black}{Our temperature jump at the top of the mantle remedies this discrepancy, so the unrealistically low temperatures are only present in the lower envelope and water layer.} This discrepancy starts to occur under a total H/He envelope mass of ${\sim}25M_\oplus$, so we estimate that our model still applies well to planets with $w_\mathrm{H/He}\lesssim20\%$.

Another possible reason for our model's underprediction of Saturn's radius is that our EOS predict higher densities at high pressures compared to the literature due to \textcolor{black}{the inclusion of additional} phase transitions and using a Holzapfel EOS for iron (as opposed to assuming silicates are in the post-perovskite phase and using a Keane EOS for iron as in \citet{ZhangRogers2022}). If this were the case, a version of our model including more physics relevant to gas giants would still predict smaller Saturn radii than the literature when using the same input parameters, indicating that the literature systematically overestimates Saturn's core mass fraction.

Replicating Saturn's radius is difficult even for interior structure models more suited to it: \citet{SurTejadaArevalo2025} find a radius of Saturn 1.18$\%$ smaller than reality (comparable to our \textcolor{black}{0.76}\% smaller than reality) while reproducing Saturn's surface temperature and $J_2$ gravitational moment to within $0.13\%$ and $0.4\%$ of reality, respectively.

\section{Constants}
\label{sec:constants}
We take molar masses for all elements from NIST\footnote{\url{https://webbook.nist.gov/}}. We take the value for G from NIST of $6.67430*10^{-11}\frac{\mathrm{m\textsuperscript{3}}}{\mathrm{kgs\textsuperscript{2}}}$\footnote{\url{https://physics.nist.gov/cuu/pdf/wall_2022.pdf}}.

We take the mass of Earth as $5.9723651*10^{24}$ kg from \citet{MoulikEkstrom2025a} (atmospheric mass included) and the radius of Earth as $6.371000*10^{6}$ m from \citet{MoulikEkstrom2025a}. We take Earth's rotational period as 0.997270 days following \citet{MoulikEkstrom2025a}'s value of Earth's angular velocity. We take Earth's irradiance as 1360.8 $\frac{\mathrm{W}}{\mathrm{m\textsuperscript{2}}}$ following \citet{KoppLean2011}, which we note is from the 2008 solar minimum and thus a lower bound.

We take the mass of the \textcolor{black}{M}oon as $7.3463*10^{22}$ kg, radius of the \textcolor{black}{M}oon as $1.737151*10^{6}$ m, and MoIC of the \textcolor{black}{M}oon as $0.393112\pm0.000012$ from \citet{WilliamsKonopliv2014}. We take the rotation period of the \textcolor{black}{M}oon as 27.321661 days following \citet{Kuiper1954}.

We take the mass of Mars $6.417*10^{23}$ kg, the radius of Mars as $3389.5$ km, and the MoIC of Mars as $0.3634$ following \citet{ArchinalActon2018,KonoplivPark2020,KhanSossi2022}. We take the rotation period of Mars as $1.025957$ days following \citet{Bakich2000}.

We take the mass of Venus as $4.8673*10^{24}$ kg following \citet{SalibyFienga2023}. We take the radius of Venus as 6051.8 km following \citet{ArchinalActon2018}. We take the MoIC of Venus as $0.337\pm0.024$ and rotation period of Venus as $243.0226$ days following \citet{MargotCampbell2021}.

We take the mass of Europa as $4.79982*10^{22}$ kg following \citet{AndersonSchubert1998} and the radius of Europa as $1560.8$ km following \citet{ArchinalActon2018}. We use the instellation of Europa of 51 $\frac{\mathrm{W}}{\mathrm{m\textsuperscript{2}}}$ and rotational period of Europa of 3.547 days from \citet{Ashkenazy2016}. We take the MoIC of Europa as $0.3547\pm0.0024$ following \citet{GomezCasajusZannoni2021}, higher than the previously-accepted value of $0.346\pm0.005$ following \citet{AndersonSchubert1998}.

We take the radius of Mercury as 2439.4 km following \citet{ArchinalActon2018}. We take the mass of Mercury as $3.3009999*10^{23}$ kg and MoIC of Mercury as $0.333\pm0.005$ following \citet{GenovaGoossens2019}. We obtain a Mercury rotation period of 58.646146 days from \citet{MazaricoGenova2014}.

We take the mass of Saturn as $5.6845789*10^{26}$ kg following \citet{JacobsonAntreasian2006}. We take the radius of Saturn as 58232 km following \citet{ArchinalActon2018}. We take the semi-major axis of Saturn as 9.5826 AU (to calculate its instellation) following \citet{TuryshevToth2022}. This is at a pressure of $10^5$ Pa, we report Saturn radii at $10^5$ Pa rather than the $100$ Pa outer boundary to account for this. We take the bulk rotation period of Saturn as 11.18 hours following \citet{MankovichMarley2019}. We take the Moment of Inertia \textcolor{black}{Coefficient} of Saturn as 0.22$\pm0.022$ following \citet{FortneyHelled2018,Helled2019}, although note that \citet{MilitzerHubbard2023} report a much more constrained value on the lower end of 0.2181$\pm$0.0002 using an interior model of Saturn.

For all solar system bodies beyond those specified, we take masses from JPL\footnote{\url{https://ssd.jpl.nasa.gov/planets/phys_par.html} for planets, \url{https://ssd.jpl.nasa.gov/sats/phys_par/} for moons}.

In the body text, we reduce the number of significant figures of the pressure changes in Eq.~\ref{eq:Fe_melting}. We use values of 82.65823 GPa and 409.76935 GPa in our code.

\section{Iron Snow}\label{sec:iron_snow}
Our model lunar core's adiabatic temperature profile is slightly less steep ($\frac{\partial T}{\partial P}$) than the Fe melting curve at the same pressures. For the right $x_\mathrm{S}^\mathrm{Solid}$, this means that the core could be solid at the core-mantle boundary but then become liquid at higher pressures. This creates an unphysical scenario in our model wherein a denser layer overlies a less-dense layer. However, if such a situation were to occur in reality, this unphysicality would be remedied by the dynamical motion of the denser layer. This phenomenon is known as iron snow and it has been proposed to describe the present state of Ganymede's core \citep{HauckAurnou2006}, the future state of Mars' core \citep{StewartSchmidt2007}, and possibly the current state of Mercury's core \citep{ChenLi2008,DumberryRivoldini2015}. Alternatively, the sinking iron could re-melt then float back upwards, resulting in compositional convection that could explain the modern-day dynamo of Ganymede \citep{Christensen2015a,Christensen2015b,RuckriemenBreuer2015,BruggerBurn2020}. Essential to the phenomenon of iron snow is the relation of the core composition to the eutectic, the lowest melting temperature achieved in an alloy, which cannot be accommodated for by our monotonic function for melting temperature of Eq.~\ref{eq:melting_T} \citep{BruggerBurn2020,UnterbornSchaefer2020}.

\textcolor{black}{We obtain a core with solid iron overlaying liquid iron for a Mercury composition planet with a mass of 0.0304$M_\oplus$. In this case,} there ultimately exists a solid planetary core and thus we are not capturing outside-in crystallization but rather a limited region in which iron snow could form. Our model Mercury's temperature is within $165$ K of the melting curve throughout the entire core, indicating that even small changes to our model--such as the inclusion of a eutectic--could dramatically change the masses and compositions where this phenomenon occurs.

\citet{GaidosConrad2010} proposed that iron snow could occur in super-Earths, however, the newer iron melting curves used in this work from \citet{Gonzalez-CataldoMilitzer2023,DongMardaru2025,DongFischer2025} are sufficiently shallow that we found no super-Earths with iron snow.

The appearance of iron snow in our model, even if it is not consistently handled, is a testament to our model's ability to capture physics over a wide range of planetary parameters.

\section{Example Super-Earth Profile}\label{sec:example_planet_profiles}
\label{lastpage}
\begin{figure}
\centering     \includegraphics[width=1.\columnwidth]{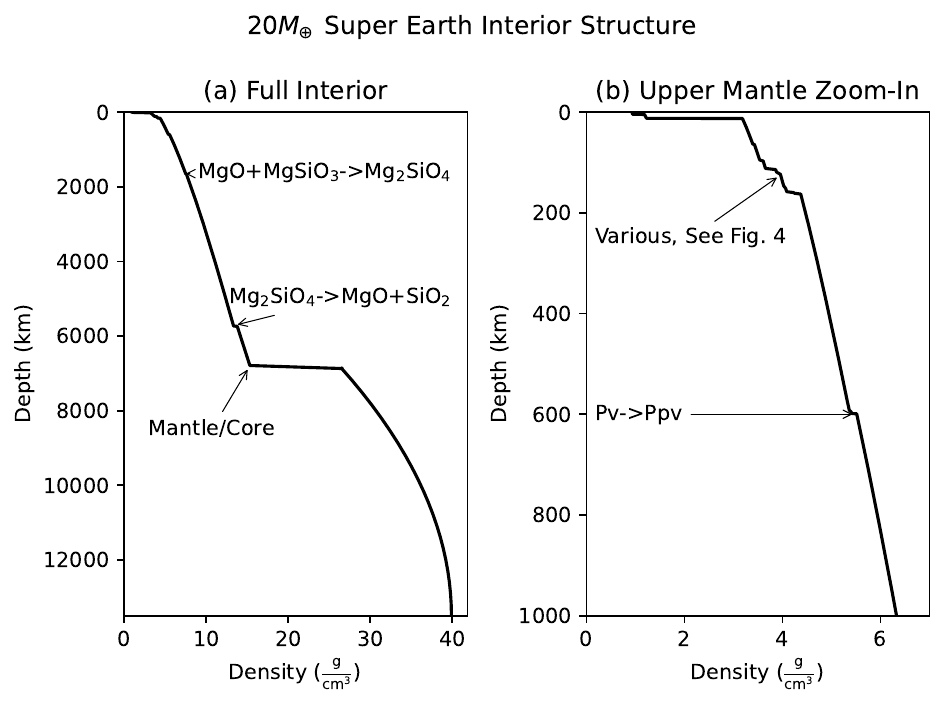}
\vspace{-2em}\caption{An example profile of a 20$M_\oplus$ super-Earth. Transition and boundary notation is the same as in Fig.~\ref{fig:earth_structure}. The uppermost mantle contains many phase transitions in close proximity in the same order as in Fig.~\ref{fig:earth_structure}.}
\vspace{-2em}\label{fig:super_earth_structure}
\end{figure}
In Fig.~\ref{fig:super_earth_structure} we plot the interior structure of a 20$M_\oplus$ super-Earth. We choose this high mass--nearing the highest mass super-Earths detected--to show all mantle transitions in our model. Note the relatively small jump caused by the recombination of MgO and MgSiO$_\mathrm{3}$ into Mg$_\mathrm{2}$SiO$_\mathrm{4}$ and large jump caused by the dissociation of Mg$_\mathrm{2}$SiO$_\mathrm{4}$ into MgO and SiO$_\mathrm{2}$. The non-inclusion of these density jumps in planetary interior structure models in the literature results in systematic underestimates of planetary densities. Also note how the higher surface gravity of the planet causes mantle features to move closer to each other and the surface as compared to Fig.~\ref{fig:earth_structure}.
\end{document}